\documentclass{aa}  

\usepackage{graphicx}

\usepackage{natbib, twoopt}
\usepackage[varg]{txfonts}

\begin{document} 

\title{The imprint of satellite accretion on the chemical and dynamical properties of disc galaxies}
\titlerunning{The imprint of satellite accretion on disc galaxies}

\author{T. Ruiz-Lara, \inst{1, 2} C. G. Few, \inst{3} B. K. Gibson, \inst{4} I. P\'erez, \inst{1, 2} E. Florido, \inst{1, 2} I. Minchev, \inst{5} \and P. S\'anchez-Bl\'azquez \inst{6}}

\authorrunning{T. Ruiz-Lara et~al.}

\institute{\inst{1} Departamento de F\'isica Te\'orica y del Cosmos, Universidad de Granada, Campus de Fuentenueva, E-18071 Granada, Spain \\
\email{ruizlara@ugr.es} \\
\inst{2} Instituto Carlos I de F\'isica Te\'orica y computacional, Universidad de Granada, E-18071 Granada, Spain \\ 
\inst{3} School of Physics, University of Exeter, Stocker Road, Exeter EX4 4QL, UK \\
\inst{4} E.A. Milne Centre for Astrophysics, Deptartment of Physics \& Mathematics, University of Hull, Hull, HU6 7RX, UK \\
\inst{5} Leibniz-Institut für Astrophysik Potsdam (AIP), An der Sternwarte 16, 14482 Potsdam, Germany \\
\inst{6} Departamento de F\'isica Te\'orica, Universidad Aut\'onoma de Madrid, E-28049 Cantoblanco, Spain \\ }

\date{Received ---; accepted ---}

\abstract
{}
{We study the effects of the cosmological assembly history on the 
chemical and dynamical properties of the discs of spiral galaxies as a 
function of radius.}
{We make use of the simulated Milky-Way mass, fully-cosmological discs, 
from {\tt RaDES} (Ramses Disc Environment Study). We analyse their 
assembly history by examining the proximity of satellites to the 
galactic disc, instead of their merger trees, to better gauge which 
satellites impact the disc. We present stellar age and metallicity profiles, Age-Metallicity Relation 
(AMR), Age-Velocity dispersion Relation (AVR), and Stellar Age Distribution (SAD) 
in several radial bins for the simulated galaxies.}
{Assembly histories can be divided into three different stages: i) a 
merger dominated phase, when a large number of mergers with mass ratios 
of $\sim$1:1 take place (lasting $\sim$3.2$\pm$0.4~Gyr on average); ii) 
a quieter phase, when $\sim$1:10 mergers take place (lasting 
$\sim$4.4$\pm$2.0~Gyr) - these two phases are able to kinematically heat 
the disc and produce a disc that is chemically mixed over its entire 
radial extension; 
iii) a ``secular'' phase where the few mergers that 
take place have mass ratios below 1:100, and which do not affect the disc 
properties (lasting $\sim$5.5$\pm$2.0~Gyr). Phase ii ends with a final 
merger event (at time $t_\mathrm{jump}$) marking the onset of important radial 
differences in the AMR, AVR, and SAD.}
{Inverted AMR trends in the outer parts of discs, for stars younger than 
$t_\mathrm{jump}$, are found as the combined effect of radial motions and star 
formation in satellites temporarily located in these outer parts. 
``U-shaped'' stellar age profiles change to an old plateau 
($\sim$10~Gyr) in the outer discs for the entire {\tt RaDES} sample. 
This shape is a consequence of inside-out growth of the disc, radial 
motions of disc stars (inwards and outwards), and the accretion of old 
stars from satellites. We see comparable age profiles even when ignoring 
the influence of stellar migration due to the presence of early {\it 
in-situ} star formation in the outer regions of the galaxy.}


\maketitle


\section{Introduction}

The accepted scenario of galaxy formation involves an extended process 
of hierarchical merging of structures until systems settle into the 
entities we observe today \citep[e.g.][]{1978MNRAS.183..341W, 
1993MNRAS.262..627L}. During these mergers, satellites are thought to 
leave a signature of their ``impact'' upon the host.  Whether these 
signatures are major, minor, long- or short-lived, remains an intense 
topic of debate; what is clear though is that the interplay between such 
merger/interaction-driven processes and later, internal, secular 
evolution, all combine to shape the galaxies we see now 
\citep[e.g.][]{1980ApJ...236..351D, 2000AJ....120..189D, 
2002A&A...393..389M, 2003MNRAS.346.1189L, 2006AJ....131.2004K, 
2008MNRAS.386L..82M}.

While idealised galaxy simulations are key to understanding secular 
processes \citep[e.g.][]{2005MNRAS.358.1477A, 2006ApJ...645..209D, 
2008ApJ...684L..79R, 2012MNRAS.426.2089R, 2009MNRAS.400.1706A, 
2009MNRAS.394...67A, 2010MNRAS.407.1433A}, simulations run in a 
cosmological context allow us to study the environmental effects of 
evolution, including the impact of satellite-host interactions through 
to the quiescent, settled, disc-phase 
\citep[e.g.][]{1991ApJ...377..365K, 1991ApJ...380..320N, 
1992ApJ...399L.109K, 1994MNRAS.267..401N, 1995MNRAS.276..549S, 
2002NewA....7..155S, 2003ApJ...597...21A}. These simulations are a 
critical tool for understanding the effect of accretion and interactions 
in the build-up of disc galaxies \citep[e.g.][]{2003ApJ...597...21A, 
2004ApJ...607..688G, 2008ApJ...688..254K}.

However, early cosmological simulations failed at replicating late-type 
massive spiral galaxies because of the so-called `angular momentum 
problem' \citep[e.g.][]{2002NewA....7..155S}. This problem consists of 
the over-production of the spheroid component of simulated galaxies due 
to enhanced star formation at early epochs. Nowadays, by modifying 
numerical and physical parameters in the simulations, more realistic 
galaxies have been reproduced \citep[see][for a study about the effect 
on the angular momentum of numerical parameters such as supernova 
feedback, spatial resolution, and star formation efficiency, amongst 
many others]{2012ApJ...749..140H}. State-of-the-art numerical 
simulations have overcome this `angular momentum problem' and are able 
to reproduce the scaling relations and galaxy properties found in nature 
\citep[e.g.][amongst others]{2004ApJ...606...32R, 2007MNRAS.374.1479G, 
2009MNRAS.396..696S, 2009MNRAS.398..591S, 2010MNRAS.408..812S, 
2010MNRAS.401.1826R, 2011ApJ...728...51B, 2012MNRAS.424.1275B, 
2012MNRAS.425.1270S, 2012MNRAS.427.1401C, 2013A&A...554A..47G, 
2014MNRAS.442.1794O, 2014MNRAS.444.3845F}.

Most of the successful cosmological simulations have been focused on 
reproducing Milky Way-type galaxies \citep[e.g.][]{2011MNRAS.415.2652H, 
2012ApJ...744L...9M, 2012MNRAS.427.1401C, 2013A&A...554A..47G, 
2014MNRAS.441..525W, 2014MNRAS.437.1750M}. Our privileged position 
inside the Milky Way allows us to take high quality spatial and 
spectroscopic data against which to confront simulations. A number of 
contemporary observational studies are exploring the solar neighbourhood 
with unprecedented quality, providing astronomers with a vast amount of 
information, including {\tt Gaia} \citep[][]{2001A&A...369..339P}, {\tt 
RAVE} \citep[][]{2006AJ....132.1645S, 2008AJ....136..421Z, 
2011AJ....141..187S, 2013AJ....146..134K}, and {\tt SEGUE} 
\citep[][]{2009AJ....137.4377Y}. Simulations play a pivotal role in 
interpreting the physical mechanisms which shape these datasets.

One of the most striking observations that needs a theoretical 
counterpart to be understood is the Age-Metallicity Relation (AMR) in 
the solar neighbourhood \citep[e.g.][]{1972MNRAS.155..483P, 
1972MmRAS..77...55H, 1980ApJ...242..242T}. Stars are born from the 
collapse of molecular clouds that imprint their chemical composition 
upon newly formed stars. Those stars undergo nuclear fusion reactions in 
their cores thereby changing their chemical composition. According to 
stellar evolution and galactic chemical evolution models 
\citep[e.g.][]{1995A&A...302...69P, 2003PASA...20..189F, 
2005A&A...430..491R}, once these stars reach the end of their lives they 
form new elements and expel part of their mass to the interstellar 
medium, increasing the latter's metal content. As a direct consequence, 
the subsequent generation of stars is more metal-rich than the 
former,\footnote{Note though that certain star formation and infall rate 
scenarios can lead, in some cases, to a subsequent \it decline \rm in 
metalicity \citep[e.g.][]{1995A&A...304...11M}.} resulting in a tightly 
correlated AMR, with younger stars being more metal-rich and older stars 
being more metal-poor. However, studies of the AMR in the solar 
neighbourhood have found an almost flat AMR for thin disc stars, with significant dispersion 
in metallicity at a given age \citep[e.g.][]{1985ApJ...294..674C, 
1993A&A...275..101E, 2001A&A...377..911F, 2004A&A...418..989N, 
2014A&A...565A..89B}, while thick disc stars seem to 
show a steep AMR \citep[][]{2013A&A...560A.109H}.

Different theoretical works have been able to reproduce aspects of the 
observed AMR of the solar neighbourhood, bulge, and nearby dwarf discs, 
by allowing for stellar radial motions in their models that contaminate 
the expected relation for stars born in the solar neighbourhood 
\citep[e.g.][]{2008ApJ...684L..79R, 2009MNRAS.398..591S, 
2009MNRAS.396..203S, 2012MNRAS.425..969P}. Radial motions (especially of 
older stars) can mix stars with different enrichment histories and 
different chemical composition (i.e. formed from different molecular 
clouds at different galaxy evolutionary stages). This effect can lead to 
flattening and greater dispersion in the AMR. Several authors have tried 
to explain what stimulates stars to move radially. Some of the proposed 
causes are i) the exchange of angular momentum at the corotation 
resonance of transient spiral arms \citep[e.g.][]{2002MNRAS.336..785S, 
2008ApJ...675L..65R}; ii) non-linear coupling between the bar and the 
spiral waves \citep[e.g.][]{2010ApJ...722..112M, 2012A&A...548A.126M, 
2012A&A...548A.127M}; and iii) the influence of satellites 
\citep[e.g.][]{2007ApJ...670..269Y, 2009MNRAS.397.1599Q, 2012MNRAS.420..913B}. These 
theoretical works have eased the debate about the observed AMR in the 
solar neighbourhood. However, there are some observational works still 
claiming that, although some changes to the AMR can be attributed to 
stellar migration, a large part of the observed scatter is intrinsic to 
the star formation processes or related to contamination of the solar 
neighbourhood by stars on apo- and pericentres 
\citep[][]{2001A&A...377..911F, 2013A&A...560A.109H}.

Despite the high quality, spatial resolution, and details of the data 
that we can acquire for stars in the Milky Way, the study of our Galaxy 
presents a number of significant disadvantages: i) the Milky Way is just 
one galaxy amongst the myriad different types of galaxies in the 
Universe; ii) our inside-view restricts our observations to the solar 
neighbourhood because the observable region is obscured by dust, 
preventing a global view of the Milky Way. To achieve a wider 
understanding of galaxies in the Universe, we need to study external 
systems \citep[e.g.][]{2009MNRAS.395...28M, 2012ApJ...752...97Y, 
2011A&A...529A..64P, 2011MNRAS.415..709S, 2012MNRAS.419.2031P, 
2013MNRAS.436...34C, 2014ApJS..211...17A, 2014A&A...570A...6S}. Those 
studies provide us with statistical information about integrated 
properties such as age and metallicity gradients (mass- and 
light-weighted), Star Formation Histories (SFHs), light and mass 
distributions, or age-metallicity relations. The new full spectrum 
fitting techniques used on these unresolved systems have been proved 
reliable when compared with the analysis of resolved stellar populations 
\citep[][]{2015A&A...583A..60R}.\footnote{Conversely, comparing resolved 
stellar populations with only a restricted sub-set of line indices can 
prove problematic \citep[e.g.][]{1999AJ....118.1268G}.}

The joint study of observations and realistic simulations can expand our 
knowledge about galaxy formation and evolution 
\citep[e.g.][]{2008ApJ...683L.103B, 2009MNRAS.398..591S, 
2009ApJ...705L.133M, 2012A&A...540A..56P, 2012ApJ...752...97Y}. 
Simulations allow one to isolate the primary impact of 
satellite merging in different environments, including for example,
the formation of 
thick discs \citep[e.g.][]{1993ApJ...403...74Q, 2004ApJ...612..894B,
2008MNRAS.391.1806V}, 
inflows of external material \citep[e.g.][]{2005MNRAS.363....2K, 
2006MNRAS.368....2D}, or radial mixing 
\citep[e.g.][]{2009MNRAS.397.1599Q, 2012MNRAS.420..913B}. The
establishment of testable predictions through simulations can be a
powerful tool in designing future observational campaigns.

In this paper, we make use of the cosmological hydrodynamical 
simulations of Milky Way-mass galaxies presented in 
\citet[][]{2012A&A...547A..63F} ({\tt RaDES}: Ramses Disc Environment 
Study). We study those simulated galaxies to search for signatures of 
their assembly processes in the age-metallicity-radius relationship. The 
study of assembly of the outer parts of the disc is especially important 
as, although it represents a small fraction of a galaxy's total mass, it 
is very sensitive to satellite accretion or external mass perturbations, 
providing a guide to galaxy formation and evolution.

In Sect. \ref{sample}, we present the sample of galaxies, the 
simulations, and their assembly histories. Sect. \ref{discs} presents the 
main observational characteristics of each simulation. The primary 
results are shown in Sect. \ref{chedy_imprints} where we discuss the 
impact of the assembly history on both the AMR and spatially-resolved 
SFH of the disc. The discussion and main conclusions are provided in 
Sect. \ref{discussion}.


\section{Sample of galaxies}
\label{sample}

In this work, we use the {\tt RaDES} \citep[][]{2012A&A...547A..63F} galaxies. The {\tt RaDES} set of 
galaxies was created to study differences between galaxies in loose 
group and field environments through cosmological simulations. {\tt 
RaDES} comprises 19 galaxies with masses similar to the Milky Way and 
disc characteristics similar to other observed disc galaxies 
\citep[][]{2012A&A...547A..63F, 2012A&A...540A..56P}.  While each system 
has similar properties in relation to their matter content (total, dark, 
stellar, baryonic, and gaseous mass), they also show important 
differences concerning their kinematic heating profiles \citep[see 
Figure 9 of][]{2012A&A...547A..63F}, star formation histories \citep[see 
Figure A.1 of][]{2012A&A...547A..63F}, disc fractions, and assembly 
histories.

For brevity, in this work we focus on the analysis of three discs from 
the {\tt RaDES} sample. These three span the range of assembly histories 
of the full sample, from intensive to quiescent; our conclusions are 
robust against the specific choice of systems.  

\subsection{Simulations}
\label{simulations}

The {\tt RaDES} galaxies are simulated with the adaptive mesh refinement 
code {\textsc{ramses}} \citep[][]{2002A&A...385..337T}. The simulations 
track dark matter, stars and gas on cosmological scales. The 
hydrodynamical evolution of gas uses a refining grid such that the 
resolution of the grid evolves to follow over-densities reaching a peak 
resolution of 436~pc (16 levels of refinement). {\textsc{ramses}} 
includes gas cooling/heating and a polytropic equation of state is used 
for dense gas to prevent numerical fragmentation.

Star formation occurs in gas that is more dense than $\rho_0 = $
0.1~n$_\mathrm{H}$~cm$^{-3}$ at a rate of $\dot{\rho} = 
-\rho/t_\mathrm{\star}$, where $t_\mathrm{\star} = t_0(\rho/\rho_0$)$^{-1/2}$, with 
$t_0=8$~Gyr. Stellar feedback is delayed by $10^7$ years and imparts 
kinetic energy, mass, and metals to the gas within a 2-cell radius 
sphere. The mass fraction of stellar particles that explode as SN is 
10\% and each SN provides 10$^{51}$~erg; 10\% of non-metals are 
converted to metals by each star particle.

The galaxies were simulated with cosmological parameters as follows: 
$H_0$=70~km s$^{-1}$Mpc$^{-1}$, $\Omega_{\mathrm{m}}$=0.28, 
$\Omega_{\mathrm{\Lambda}}$=0.72, $\Omega_{\mathrm{b}}$=0.045, and 
$\sigma_8$=0.8.\footnote{Here, $H_0$ is the Hubble constant, 
$\Omega_{\mathrm{m}}$ the fraction of total matter, 
$\Omega_{\mathrm{\Lambda}}$ the fraction of the dark energy, and 
$\sigma_8$ the strength of the primordial density fluctuations.} Two 
volumes are used with a size of 20 $h^{-1}$ Mpc and 24 $h^{-1}$ Mpc. The 
mass resolution of dark matter particles is either 
5.5$\times$10$^{6}$~M$_\odot$ or 9.5$\times$10$^{6}$~M$_\odot$ depending 
on which of two volumes the galaxy is drawn from. Details of the halo 
selection process and the simulation parameters may be found in 
\citet[][]{2012A&A...547A..63F} and in Sect. \ref{treeplots}, below.

\subsection{RaDES assembly histories}
\label{treeplots}

The assembly histories of the {\tt RaDES} galaxies have been obtained by 
a careful study of the merger trees discussed in 
\citet[][]{2012A&A...547A..63F}. In short, we use the {\it adaptahop} 
algorithm \citep[][]{2004MNRAS.352..376A} to create at each time step a 
catalogue of haloes and sub-haloes in the simulation. We are able to 
identify, not only each halo, but also their progenitors, descendants, 
sub-haloes, etc. By linking all these catalogues, we can create
merger trees for the different haloes following the ``most massive 
substructure method'' \citep[][]{2009A&A...506..647T}. The analysis of 
those merger trees shows that all {\tt RaDES} galaxies display, 
generally speaking, similar disc assembly histories in terms of phases 
or stages.

\begin{figure*}
\includegraphics[width=0.93\textwidth]{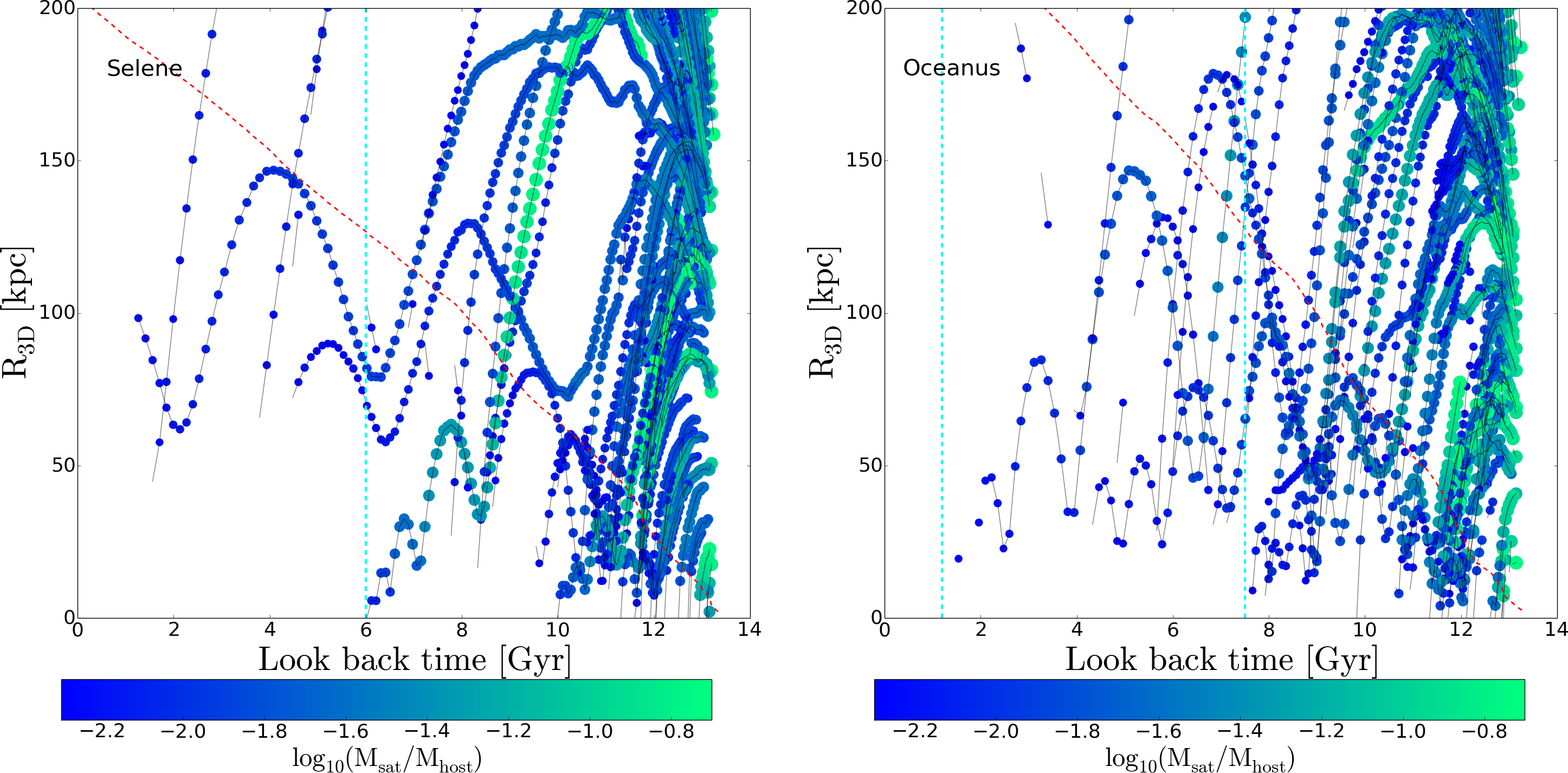} \\
\caption{Schematic representation of the satellite orbits for Selene 
(left) and Oceanus (right). On the y-axis we plot the three-dimensional 
distance (R$_\mathrm{3D}$) from the satellite centre to the host galaxy centre. 
On the x-axis we plot the lookback time (13.5~Gyr being the beginning of 
the simulation). Each point represents one satellite at a given snapshot 
with points corresponding to the same satellite at different timesteps 
linked by a solid black line. In some cases, the halo tracking between 
snapshots fails leading to missing connecting lines. We colour-code the 
points according to log$_\mathrm{10}(M_\mathrm{sat}/M_\mathrm{host})$ (the size of each point 
is also proportional to that value) as an indicator of the magnitude of 
the mergers. We have applied a cut in log$_\mathrm{10}(M_\mathrm{sat}/M_\mathrm{host})$ in 
order to clean the plot of the least massive satellites; satellites 
with $(M_\mathrm{sat}/M_\mathrm{host})_\mathrm{t}$ below 0.005 are ruled out in the plots. The 
dashed red line represents the time evolution 
of the host galaxy virial radius. Last merger times ($t_\mathrm{jump}$) are 
denoted via dashed cyan vertical lines. For Selene (left) note that the 
last merger happened 6.0~Gyr ago with $M_\mathrm{sat}/M_\mathrm{host}$ $\sim$0.089 
when it enters the virial radius and $\sim$0.0126 when the merger 
happens. Note the two merging events associated with the cessation of 
the two assembly phases in Oceanus (right) with $(M_\mathrm{sat}/M_\mathrm{host})_\mathrm{v}$ 
being $\sim$0.056 (0.05) and $(M_\mathrm{sat}/M_\mathrm{host})_\mathrm{t}$ being $\sim$0.006 
(0.005) for the first (second) merging event finishing each epoch.}
\label{tree_plot}
\end{figure*}

In Fig.~\ref{tree_plot}, we plot the distance between the centre of the 
host galaxy and its satellites for two {\tt RaDES} galaxies as a function of lookback time (Selene, 
left panel; Oceanus, right panel). Each point in Fig.~\ref{tree_plot} 
represents one satellite at a given timestep with the points 
colour-coded in logarithmic scale according to the virial mass of the 
satellite divided by the virial mass of the host halo 
($M_\mathrm{sat}/M_\mathrm{host}$). All {\tt RaDES} galaxies underwent three 
well-defined assembly `phases': i) an initial phase of merger-dominated 
evolution lasting $\sim$3.2$\pm$0.4~Gyr on average, while mergers of 
systems with similar masses occur, $M_\mathrm{sat}/M_\mathrm{host}$ (when the distance to the host galaxy is lower than 5~kpc) ranging from 0.1 
(the majority) to 3 (several of them). This first period is followed by 
ii) a quieter phase where the number of mergers is minimised by a factor 
of two with respect to the previous stage. During this phase (lasting 
$\sim$4.4$\pm$2.0~Gyr) several major satellites with $M_\mathrm{sat}/M_\mathrm{host}$ 
= 0.01--0.3,  
merge with the host galaxy. This epoch ends with a last merging 
event, occurring at different times depending on the galaxy in 
question. We refer to this stage in the galaxy's evolution as $t_\mathrm{jump}$ 
(see Table \ref{satellites_tab}), defined as the time at which this last 
merger dissolves in the host galaxy. Finally, iii) the disc settles and 
evolves ``secularly'', with just minor satellite mergers 
($M_\mathrm{sat}/M_\mathrm{host}$ < 0.01) disturbing the host disc (lasting 
$\sim$5.5$\pm$2.0~Gyr).

This last accreted satellite is characterised by a value of 
$M_\mathrm{sat}/M_\mathrm{host}$ $\sim$0.12 $\pm$ 0.09 at the time it enters the 
virial radius ($(M_\mathrm{sat}/M_\mathrm{host})_\mathrm{v}$) and $M_\mathrm{sat}/M_\mathrm{host}$ 
$\sim$0.013 $\pm$ 0.006 (averaged values) when it merges at $t_\mathrm{jump}$ 
($(M_\mathrm{sat}/M_\mathrm{host})_\mathrm{t}$). All the characteristics of these latest 
mergers are shown in Table \ref{satellites_tab}.

While most of the galaxies exhibit just one merger epoch (phases i and 
ii) followed by a quiescent phase (phase iii), some others show two 
different merging events (two phases ii) with the above outlined characteristics. After 
the first merger event ($t_\mathrm{jump-a}$), phase ii does not end, 
but continues until a second event takes place ($t_\mathrm{jump-b}$). Artemis 
and Oceanus are examples of such behaviour (see right panel of 
Fig.~\ref{tree_plot}).

Fig.~\ref{tree_plot} shows the satellites affecting the discs of Selene 
and Oceanus as two examples of the above outlined behaviour 
(one-merging-epoch and two-merging-epochs galaxies, respectively). For 
resolution/aesthetic purposes, we have applied a cut in 
Fig.~\ref{tree_plot} to remove the more insignificant satellites, i.e. $(M_\mathrm{sat}/M_\mathrm{host})_\mathrm{t}$ 
below 0.005. We have checked that those 
satellites not fulfilling our criterion have low masses, and thus, 
little influence in the disc evolution. For Selene, during the first 
3.5\,Gyr the evolution is characterised by an unsettled phase dominated 
by satellite accretion (phases i and ii). The lack of important mergers 
is the main characteristic during the rest of the simulation, although 
some low mass satellites do orbit the host. The last important merger 
that Selene underwent took place $\sim$6.0\,Gyr ago ($t_\mathrm{jump}$) with 
$(M_\mathrm{sat}/M_\mathrm{host})_\mathrm{v}$ $\sim$0.089 and $(M_\mathrm{sat}/M_\mathrm{host})_\mathrm{t}$ 
$\sim$0.0126. In the case of Oceanus (two-merging-epochs), the unsettled 
phase lasts for almost 6~Gyr, finishing with a satellite with 
$(M_\mathrm{sat}/M_\mathrm{host})_\mathrm{v}$ $\sim$0.056 and $(M_\mathrm{sat}/M_\mathrm{host})_\mathrm{t}$ 
$\sim$~0.006 (at $t_\mathrm{jump-a}$ $\sim$~7.5\,Gyr). After that, a 6~Gyr 
quiescent-phase follows with several low-mass satellites remaining 
isolated from the host disc, followed by another merging episode 
$\sim$1.2~Gyr ago with $(M_\mathrm{sat}/M_\mathrm{host})_\mathrm{v}$ $\sim$ 0.05 and 
$(M_\mathrm{sat}/M_\mathrm{host})_\mathrm{t}$ $\sim$ 0.005 ($t_\mathrm{jump-b}$ $\sim$1.2~Gyr).

In the next Sects. \ref{discs} and \ref{chedy_imprints}, we will 
concentrate on the effect that such assembly histories produce on the 
{\tt RaDES} galaxies.

\begin{table}
{\tiny
\centering
\begin{tabular}{llll}
\hline\hline
$Galaxy$ & $t_\mathrm{jump}$ & $(M_\mathrm{sat}/M_\mathrm{host})_\mathrm{v}$ & $(M_\mathrm{sat}/M_\mathrm{host})_\mathrm{t}$ \\ 
 & (Gyr) &  &  \\ 
$(1)$ & $(2)$ & $(3)$ & $(4)$ \\ \hline
 {\bf Apollo }      & {\bf 3.5 }  &  {\bf 0.112 }  &   {\bf 0.0125 }        \\  
 Artemis (a) & 7.5  &  0.063  &   0.01         \\
 \hspace{9mm} (b) & 1.75 &  0.126  &   0.01           \\
 Atlas       & 5.0  &  0.05   &   0.0126        \\
 Ben         & 6.0  &  0.223  &   0.031          \\
 Castor      & 8.0 &  0.004  &   0.0015          \\
 Daphne      &  2.0 &  0.1    &   0.0126      \\
 Eos         &  7.5 &  0.063  &   0.011       \\
 Helios      &  8.5 &  0.1    &   0.01       \\
 Hyperion    &  7.5 &  0.050  &   0.01         \\
 Krios       &  4.5 &  0.063  &   0.0126       \\
 Leia        &  7.5 &  0.125  &   0.019        \\
 Leto        &  1.2 &  0.316  &   0.0178        \\
 Luke        &  2.0 &  0.125  &   0.01          \\
 {\bf Oceanus (a) } &  {\bf 7.5 } &  {\bf 0.056 }  &   {\bf 0.006 }         \\
 {\bf \hspace{10.5mm} (b) } &  {\bf 1.2 } &  {\bf 0.05 }   &   {\bf 0.005 }         \\
 Pollux      &  5.5 &  0.281  &   0.019        \\
 {\bf Selene }      &  {\bf 6.0 } &  {\bf 0.089 }  &   {\bf 0.0126 }       \\  
 Tethys      &  6.5 &  0.25   &   0.0251      \\
 Tyndareus   &  4.0 &  0.316  &   0.012         \\
 Zeus        &  4.0 &  0.039  &   0.011     \\ \hline
\end{tabular}
\caption{Characteristics of the last accretion episode. (1) Name of the 
galaxy; (2) time of the last accretion (look-back time); (3) ratio 
between the mass of the satellite that merges and the host galaxy in the 
moment the satellite enters the virial radius; (4) ratio between the 
mass of the satellite that merges and the host galaxy in the moment of 
the merge ($t_\mathrm{jump}$). In the case of Artemis and Oceanus the two 
events are labelled as {\it a} and {\it b}.}
\label{satellites_tab}
}
\end{table}

\section{Discs characteristics}
\label{discs}

\subsection{Disc decomposition}
\label{Jcut}

One of the classic problems plaguing the simulation of disc galaxies 
in a cosmological context is the overproduction of stars associated
with the spheroid (bulge and/or halo). Significant advances have been 
made over the past decade in ameliorating this problem, although
this manifestation of the classical over-cooling and angular momentum
catastrophe problems has not yet been fully eliminated. Instead, 
as is standard practice in the field, we post-process the simulations
to mitigate the contamination of the disc by spheroid stars.

To separate the disc from the spheroid component, we apply a kinematic 
selection criterion based upon the circularity ($J_{z}$/$J_\mathrm{circ}$) 
distribution \citep[e.g.][]{2009MNRAS.396..696S, 2010MNRAS.408..812S}. 
We have labelled disc stars as those with circularities ranging from 
$0.9$ to $1.1$; in this way, we ensure we are considering particles on 
circular orbits in the plane of the disc. Although different selection 
criteria can be found in the literature 
\citep[e.g.][]{2009MNRAS.398..591S, 2012A&A...547A..63F, 
2012A&A...540A..56P} ending with subtly different sub-sets of `disc' 
particles, we have checked that our results are robust to the specific 
selection criteria. While our chosen circularity criterion might appear 
restrictive, the characteristics of the discs so defined (mass, age, and 
metallicity profiles, etc.) are consistent with those of observed galaxy 
discs.

\begin{figure}
\includegraphics[width=0.45\textwidth]{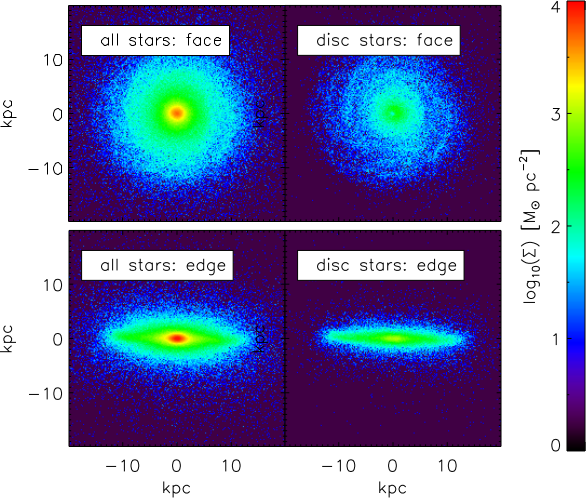} \\
\caption{Face-on (top panels) and edge-on (bottom panels) surface
  density maps of Selene considering every particle in the simulation (left panels) or just disc 
particles (right panels).}
\label{disc_selec_plot}
\end{figure}

Figure~\ref{disc_selec_plot} shows the spatial distribution of the stellar particles 
for Selene when our kinematic criterion is applied (right) compared to the distribution 
when all the particles are represented (left). A careful inspection of the edge-on views allows 
us to conclude that our criterion properly clean our discs from the spheroid component. The thick disc
is not strictly spatially resolved in these simulations and our disc selection is an aggregate thick and thin
disc. The face-on view of this particular galaxy highlights 
some disc characteristics such as the spiral structure and we conclude that our
strict selection of disc particles allows us to perform our analysis without contamination by halo stars.

To spatially sample each disc's characteristics from the inner to the 
outer disc, we have scaled each one to the radius corresponding to a 
stellar mass surface density of $10^7$ $\rm M_{\odot}$ $\rm kpc^{-2}$ 
(hereafter $R_7$); this `normalisation' corresponds to the typical 
`break radius' in the surface density profile of disc galaxies 
\citep[e.g.][]{2008ApJ...675L..65R, 2009MNRAS.398..591S}, and has been 
chosen as useful scaling, but does not impact on the conclusions 
whatsoever. We study four radial regions (see Table \ref{regions} and 
Sects. \ref{AVR}, \ref{AMR_sect}, and \ref{chedy_imprints}), defined as 
i) inner, this region accounts for the particles in the inner disc scale 
length ($h_{\rm in}$) obtained analysing the SDSS r-band mock images from {\tt 
SUNRISE} \citep[][]{2006MNRAS.372....2J}; the images generated by {\tt 
SUNRISE} can be seen in \citet[][]{2012A&A...547A..63F} (see Appendix 
\ref{app_SB} for further information); ii) middle disc, this is a 
$h_{\rm in}$-width region around $R_7/2$; iii) outer disc, $h_{\rm in}$-width 
region around $R_7$; and iv) outskirts, 3$h_{\rm in}$-width region around 
$R_7$ + 2 $\times$ $h_{\rm in}$. We apply these spatial cuts for every galaxy, avoiding 
overlap between regions. We have followed the same approach for every 
{\tt RaDES} galaxy, with no assumptions regarding the surface brightness 
profile of the discs. Again, we tested various ways of radially scaling 
our discs and the results are not contingent upon this convenient 
normalisation.

\begin{table*}
\centering
\begin{tabular}{ccccccc}
\hline\hline
    Regions & \multicolumn{2}{c}{General} & \multicolumn{2}{c}{Selene} & \multicolumn{2}{c}{Oceanus} \\
    & Lower limit & Upper limit & Lower limit & Upper limit & Lower limit & Upper limit \\
    &  &  & \multicolumn{2}{c}{[kpc]} & \multicolumn{2}{c}{[kpc]} \\
 (1) & (2) & (3) & (4) & (5) & (6) & (7) \\ \hline
1 & 0.0 & $h_{\rm in}$ & 0.00 & 5.67 & 0.0 & 8.07 \\
2 & $R_7$/2 - $h_{\rm in}$/2 & $R_7$/2 + $h_{\rm in}$/2 & 5.67 & 8.59 & 8.07 & 11.09 \\
3 & $R_7$ - $h_{\rm in}$/2 & $R_7$ + $h_{\rm in}$/2 & 8.67 & 14.34 & 11.09 & 18.14 \\
4 & $R_7$ + 0.75$\times h_{\rm in}$ & $R_7$ + 3.5$\times h_{\rm in}$ & 15.75 & 31.35 & 20.15 & 42.35 \\ \hline
\end{tabular}
\caption{Definition of the four radial regions studied in this analysis. 
(1) Number of the region; (2) General definition of the lower limit in 
units of inner disc scale-length ($h_{\rm in}$, 5.67 and 8.07~kpc for Selene and 
Oceanus) and the radius corresponding to a stellar mass surface density 
of $10^7$ $\rm M_{\odot}$ $\rm kpc^{-2}$ ($R_7$, 11.5 and 14.1~kpc for 
Selene and Oceanus); (3) General definition of the upper limit in units 
of $h_{\rm in}$ and $R_7$; (4) Lower limit for Selene (kpc); (5) Upper limit 
for Selene (kpc); (6) Lower limit for Oceanus (kpc); (7) Upper limit for 
Oceanus (kpc). $h_{\rm in}$ was obtained analysing the SDSS r-band mock 
images from {\tt SUNRISE} \citep[][]{2006MNRAS.372....2J}, those images 
can be seen in \citet[][]{2012A&A...547A..63F}. The values of $h_{\rm in}$ for 
all the simulated discs can be found in Table~\ref{SB_properties} summarising 
the results from our surface brightness analysis. Several of the values in 
columns (4), (5), (6), and (7) coincide as a consequence of applying the 
general definition avoiding overlapping between regions.} 
\label{regions}
\end{table*}

\begin{figure}
\includegraphics[width=0.45\textwidth]{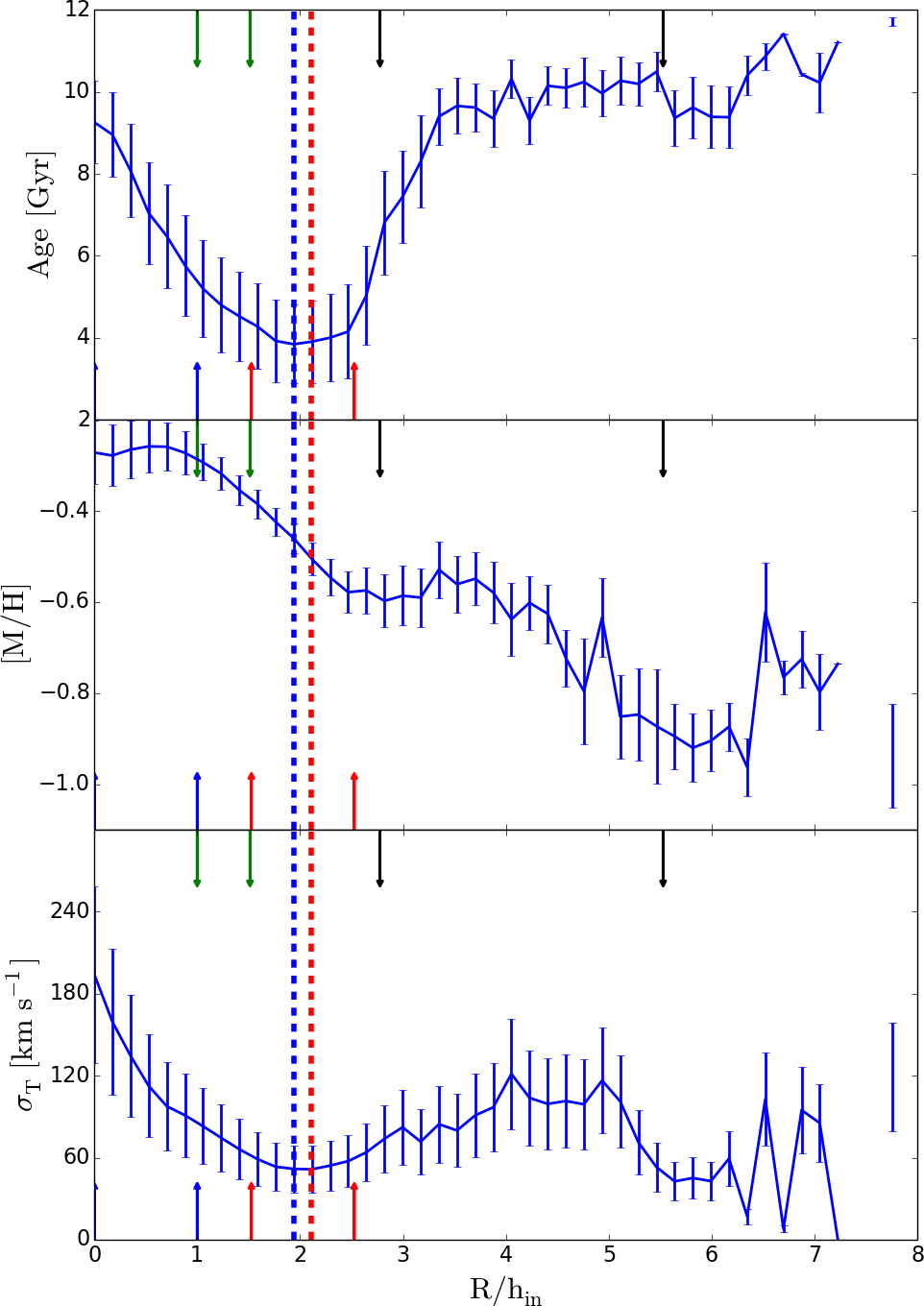} \\
\caption{Upper panel: Selene stellar disc age profile. Middle panel: 
Selene stellar disc metallicity profile. Bottom panel: Selene total 
velocity dispersion profile (disc stars). Red (blue) vertical dashed lines are located at the 
surface brightness break (minimum age) radius. The errorbars indicate 
the standard deviation of the values of every particle within each 
radial bin. $h_{\rm in}$ is the inner disc scale-length in SDSS r-band mock images 
from {\tt SUNRISE} \citep[][]{2006MNRAS.372....2J}. These mock images 
can be seen in \citet[][]{2012A&A...547A..63F}. Vertical arrows are 
located at the beginning and the end of the four regions defined in 
Table~\ref{regions}. Blue, region 1; green, region 2; red, region 3; 
black, region 4.}
\label{age_met_prof}
\end{figure}

\subsection{Stellar age and metallicity distribution of disc stars}
\label{ageandz}

Different groups have tried to study, both observationally and 
theoretically, the properties of the stellar populations as a function 
of radius in spiral galaxies. In \citet{2008ApJ...684L..79R}, using 
N-body + SPH simulations of an isolated and idealised disc, the authors 
generated age profiles with a characteristic ``U-shape'' attributed to 
radial migration induced by transient spiral arms. On the other hand, 
\citet{2009MNRAS.398..591S}, using fully-cosmological hydrodynamical 
simulations, also found such an age profile in their simulations, but 
proposed the combination of two different processes to explain it: i) a 
radial change in the star formation history linked to a drop in the gas 
density (itself, due to a warp) as the main cause and ii) radial 
migration of stars towards larger radii; it is important to note though 
that said U-shaped age gradient was found even in the absence of radial 
migration. Observationally, \citet[][]{2008ApJ...683L.103B} stacked the 
SDSS data from \citet[][]{2006A&A...454..759P}, obtaining a clear 
U-shaped $g-r$ colour profile for type II galaxies (those 
showing a down-bending surface brightness profile). However, this work 
has two main disadvantages: i) it is based on stacked profiles and a 
careful study of individual galaxies show that not all of them display 
this inversion in the age profile, nor the minimum in age (colour) 
located at the break radius \citep[][]{2012ApJ...758...41R}; and ii) as 
they use colour profiles, it is highly affected by the age-metallicity 
degeneracy, so the $g-r$ colour cannot be interpreted either as the age 
or the metallicity of the stellar populations. The difficulty of 
obtaining good spectroscopic data in the outer parts of the spiral 
galaxies to minimise the age-metallicity degeneracy 
\citep[e.g.][]{2009MNRAS.395.1669G} has hampered the study of reliable 
stellar age profiles reaching beyond the break radius. 
\citet[][]{2012ApJ...752...97Y} studied the stellar content of 12 spiral 
galaxies using Mitchell Spectrograph IFS data 
\citep[][]{2008SPIE.7014E..70H}. They obtained U-shaped age profiles in 
three out of the six galaxies they were able to study beyond their break 
radius. Currently, we are analysing the stellar content focused on the 
outer parts of the CALIFA galaxies \citep[][]{2012A&A...538A...8S} 
obtaining no clear relation between the occurrence of U-shape age 
profiles and their surface brightness profiles \citep[][]{2016MNRAS.456L..35R}.

A recent study has carefully analysed the stellar content in the outer 
parts of M31 by means of 14 resolved fields 
\citep[][]{2015MNRAS.446.2789B}. They find that the outermost fields 
show that a significant fraction of their mass had already formed by z 
$\sim$1, while fields a bit closer to the centre are on average younger. 
They suggest this is the consequence of a complicated galaxy evolution 
due to mergers. The study of age and metallicity profiles in simulated 
galaxies as well as more observational effort will help us to better 
understand the behaviour of the stellar age in the outer parts of spiral 
galaxies.

With such purpose, we have studied the age and metallicity distribution 
of the {\tt RaDES} galaxies' discs by means of one-dimensional profiles. 
We obtain those profiles by applying a mass-weighted average to all disc 
particles within 0.5\,kpc-wide radial bins. Fig.~\ref{age_met_prof} 
shows the age, metallicity, and velocity dispersion profiles 
characterising the disc of one such galaxy, Selene. Each of the 19 
analysed galaxies show similar profiles regardless their morphology and 
characteristics: i.e. negative mass-weighted 
metallicity gradients, and U-shaped mass-weighted age profiles. 
For all of them, the radius where the minimum 
in age is reached is roughly located in region 3. The outermost parts of 
every galaxy show an extended, old plateau (older than 10~Gyr, beyond 
$\sim$~6.4 h$_r$). The metallicity profiles show a flattening in the 
centre, mainly caused by the low metallicity of the older stars 
populating the inner region of our discs (see Figs.~\ref{AMR} 
and~\ref{AMR_oceanus}). These profiles will be extensively analysed in 
Sect.~\ref{U_back}.

In addition to the age and metallicity profiles, we have obtained the 
velocity dispersion profiles (bottom panel of Fig.~\ref{age_met_prof}) 
for each of the above defined {\tt RaDES} discs. They also show a U-shaped form. 
In fact, there is a correlation in the shapes of the age 
and velocity dispersion profiles, with the minimum of both distributions 
located roughly at the same radial position. This is related to the 
age-velocity dispersion relation expected for galactic discs (discussed 
in the following section) where, in general, younger stars have lower 
velocity dispersions. As stars age they are heated by the effect of the 
galaxy dynamics and/or satellites (see Fig.\,\ref{sigma_age}). This 
aspect might deepen the U-shape age profile as the stars that are being 
born at the minimum age radius have less time to move away 
and make other regions younger, on average. Similar U-shapes in the 
velocity dispersion profiles have been found in the 
\citet[][]{2008MNRAS.391.1806V} simulations, as the consequence of the 
accretion of a discy satellite with an initial orbital inclination for 
the satellite with respect to the mid-plane of the host disc of 
30$^{\circ}$ (they obtain similar results using spherical satellites and 
an orbital inclination of 60$^{\circ}$).

\subsection{Age-velocity dispersion relation}
\label{AVR}

\begin{figure}
\includegraphics[width=0.45\textwidth]{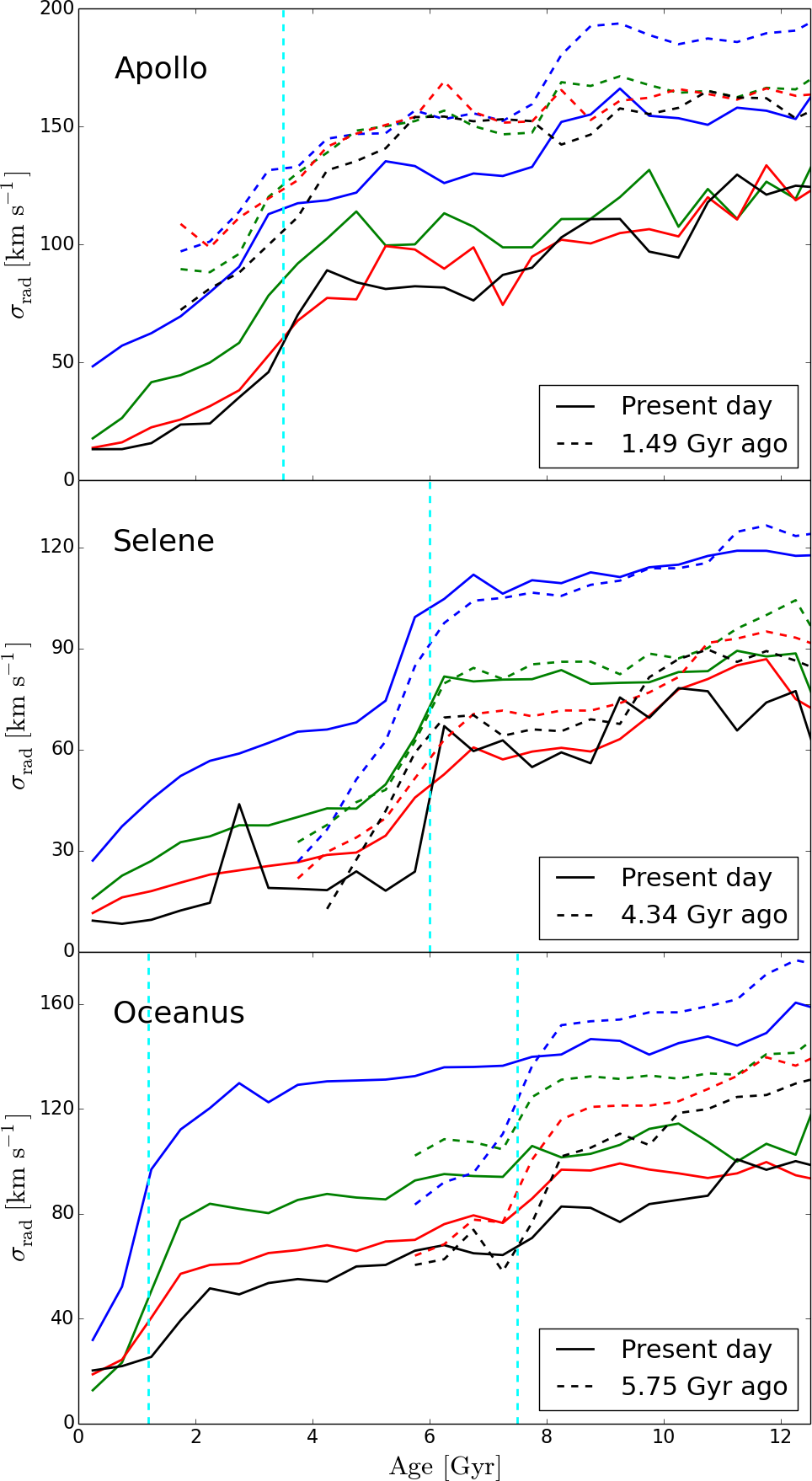} \\
\caption{Radial velocity dispersion vs. age for Apollo (upper panel), 
Selene (middle panel), and Oceanus (bottom panel) disc stars, colour-coded according to their radial 
positions. Blue: Region 1: inner region (one disc scale-length width, 
$h_{\rm in}$ using the SDSS r-band mock images from {\tt SUNRISE} 
\citep[][]{2006MNRAS.372....2J}, these mock images can be seen in 
\citet[][]{2012A&A...547A..63F}). Red: Region 2: a 1-$h_{\rm in}$ width 
annulus between the centre and the $10^7$ $\rm M_{\odot}$ $\rm kpc^{-2}$ 
mass surface density. Green: Region 3: a 1-$h_{\rm in}$ width annulus centred 
at the $10^7$ $\rm M_{\odot}$ $\rm kpc^{-2}$ mass surface density. Black: 
a 3.5-$h_{\rm in}$ width annulus further out (see text, Table \ref{regions}, 
and Fig.~\ref{sigma_age} for further details). $h_{\rm in}$ is 2.34, 5.67, 
and 8.07~kpc for Apollo, Selene, and Oceanus respectively. $t_\mathrm{jump}$ is 
denoted for each galaxy by dashed cyan vertical lines (the moment when 
the last main satellite is accreted). This plot is aimed at illustrating 
the different epochs for different galaxies when the last major 
accretion event happens. Dashed lines show the AVR at intermediate simulation stages 
between the formation of the galaxy and the present day to address the effect of the last 
merger episode in the AVR (1.49~Gyr ago for Apollo, 4.34 for Selene, and 5.75~Gyr for Oceanus).}
\label{sigma_age}%
\end{figure}

In the solar neighbourhood of the Milky Way, the stellar velocity 
dispersion increases with the stellar age, however the exact 
shape/behaviour of the relationship remains a matter of debate 
\citep[e.g.][]{1977A&A....60..263W, 2000MNRAS.318..658B, 
2007MNRAS.380.1348S}. \citet{2004A&A...418..989N} made use of the 
Geneva–Copenhagen Survey to study the Age-Velocity dispersion Relation 
(AVR) in the solar neighbourhood. They studied each velocity component 
separately, finding that power laws can fit the relation found in every 
direction. They interpreted those results as evidence for a continuous 
heating of the disc. Conversely, several works have found some 
saturation in the V and W velocity dispersion components 
\citep[][]{2008A&A...480...91S}. \citet[][]{2009MNRAS.397.1286A}, using 
{\it Hipparcos} and Geneva–Copenhagen data, were not able to rule out 
such saturation for stars older than 4~Gyr, combined with an abrupt 
increase in velocity dispersion for the oldest stars. 
\citet[][]{2014MNRAS.443.2452M}, based on seven simulated galaxies, 
proposed continuous heating takes place to explain the slope of the AVR 
for stars younger than 8-9~Gyr, but also found a step in the AVR for the 
oldest stellar population, related to an early merger phase. The authors 
showed that the maximum in $\sigma_z$ is strongly decreased when age 
errors of 30\% are implemented, suggesting that observations can easily 
miss such a jump with the current accuracy of age measurements. Although 
hard to distinguish among the different heating agents (e.g., bars, 
spirals, mergers, and stars born hot at high redshift), Martig et al. 
showed that radial migration does not heat the disc, in agreement with 
\citet[][]{2012A&A...548A.126M} and \citet[][]{2014ApJ...794..173V}.

To confront our simulations with both the existing empirical and model 
AVRs, in order to identify the heating mechanisms underpinning \tt 
RaDES\rm, in Fig.~\ref{sigma_age} we show how the stellar radial velocity 
dispersion varies as a function of stellar age for three of the {\tt 
RaDES} galaxies (Apollo, Selene, and Oceanus). We study this relation in 
the four different regions outlined earlier (Sect. \ref{Jcut}). The 
behaviour for the individual velocity components (R, $\phi$, and $z$) 
parallels that of Figure 9 of \citet[][]{2012A&A...547A..63F}. The shape 
of the AVR is quite similar across the four regions suggesting that heating 
is consistent over the entire disc, albeit, there are qualitative
differences in the strength of this heating as a function of radius.

The most remarkable feature we can find in these age-velocity diagrams 
is the presence of an increase in the velocity dispersion values around 
a given age ($\sim$~$t_\mathrm{jump}$), a feature that is apparent throughout 
the entire disc (in the four radial regions). Although for some galaxies 
this jump in the AVR is smooth (as is the case for Apollo), other 
galaxies show a clear sudden jump (e.g. Selene and Oceanus, see 
Fig.~\ref{sigma_age}). Stars older than $t_\mathrm{jump}$ have larger velocity 
dispersion than the stars with ages younger than $t_\mathrm{jump}$ for all {\tt 
RaDES} galaxies: this jump is clearly linked to their last important 
merger (happening at $t_\mathrm{jump}$ with $(M_\mathrm{sat}/M_\mathrm{host})_\mathrm{t}$ 
$\sim$0.0126, see Sect. \ref{treeplots}). This jump is more evident for 
the outer parts and fades towards the centre; mergers affect the outer 
disc more strongly because of its lower surface density. 

Dashed lines in Fig.~\ref{sigma_age} represents the AVR at different evolutionary 
stages in order to appreciate how our galaxies have evolved since the last merger. For Apollo 
and Selene we show the AVR from 1.49 and 4.34~Gyr ago, respectively. We observe that  
the AVR taken in the present day and just after the mergers have
similar shapes across the 4 regions, i.e. the disc kinematics respond
qualitatively the same at all radii. Some differences do exist in how
the 4 regions evolve however: in the case of Apollo (the galaxy with
the smoothest jump of the three) the AVR was also smooth 1.49~Gyr ago, yet since that time,
regions 2--4 have relaxed more than region 1. For Selene, the AVR 4.34~Gyr ago and
nowadays are similar in that the step in velocity dispersion exists
immediately after the merger and persists to the present day. The differences 
between the AVR at intermediate evolutionary epochs and the one displayed at 
present time suggest that, apart from the effect caused by the mergers, secular
evolution following the merger clearly affects the final AVR shape (a similar behaviour is 
displayed by Oceanus).

In the case of Oceanus (see Fig.~\ref{sigma_age}), one of the galaxies where two merger 
episodes can be identified by its assembly history (see Fig.~\ref{tree_plot}), we can also 
distinguish two jumps in the AVR. The more recent event presents a more prominent jump, while 
the earlier one is more subtle because it is `swamped' by the 
combination effect of the latest merger and secular heating. 
This swamping of the earlier jump might be caused by a differential 
heating between cold and hot stars minimising the extant jump, affecting cold stars more than 
hot ones. To check our hypothesis, in Fig.~\ref{sigma_age} we show the AVR displayed by Oceanus disc stars 
5.75~Gyr ago, i.e. after the first merger event and prior to the second one; while at the 
end of the simulation, the ``jump'' at 7.5~Gyr is almost negligible, the 
AVR 5.75~Gyr ago shows a more prominent ``jump'' for stars older than 
7.5~Gyr being more evident for the inner region. This result favours our idea that the jump 
experienced by stars older than 7.5~Gyr was caused by a merger event 7.5~Gyr ago but has been 
partially swamped by the last merger and the subsequent secular heating (especially true for region 1).

We note that major mergers are not the only source of kinematic
heating affecting our discs: internal secular evolution processes and
other minor mergers can also shape the AVR. These plots do,
however, prove that satellite merging is an important heating mechanism
affecting the kinematic state of the stars experiencing such events
and therefore the AVR.

Another striking point displayed by the AVRs that is worth discussing is 
the values found for the velocity dispersion. Our velocity dispersions are higher than 
those observed in the solar vicinity \citep[e.g.][]{2009A&A...501..941H}. Old stars 
display values as high as twice the value for Milky Way old stars and young stars 
show those values presented by solar-neighbourhood old stars. It is expected that 
higher velocity dispersions are found within our simulated discs than for the Milky Way as high velocity 
dispersions are endemic to almost all of the cosmological simulations of disc galaxies 
\citep[see ][]{2011MNRAS.415.2652H}. High velocity dispersions may also be due to contamination
of the disc by spheroid stars, however we have minimised that effect by applying our disc particle criterion.
Furthermore, none of our galaxies are a perfect match to the Milky Way, so we do not expect the same behaviour as displayed 
by the Milky Way which has a different assembly 
histories than the simulated discs underwent.

\subsection{Age-metallicity relation}
\label{AMR_sect}

\begin{figure*}
\includegraphics[width=0.93\textwidth]{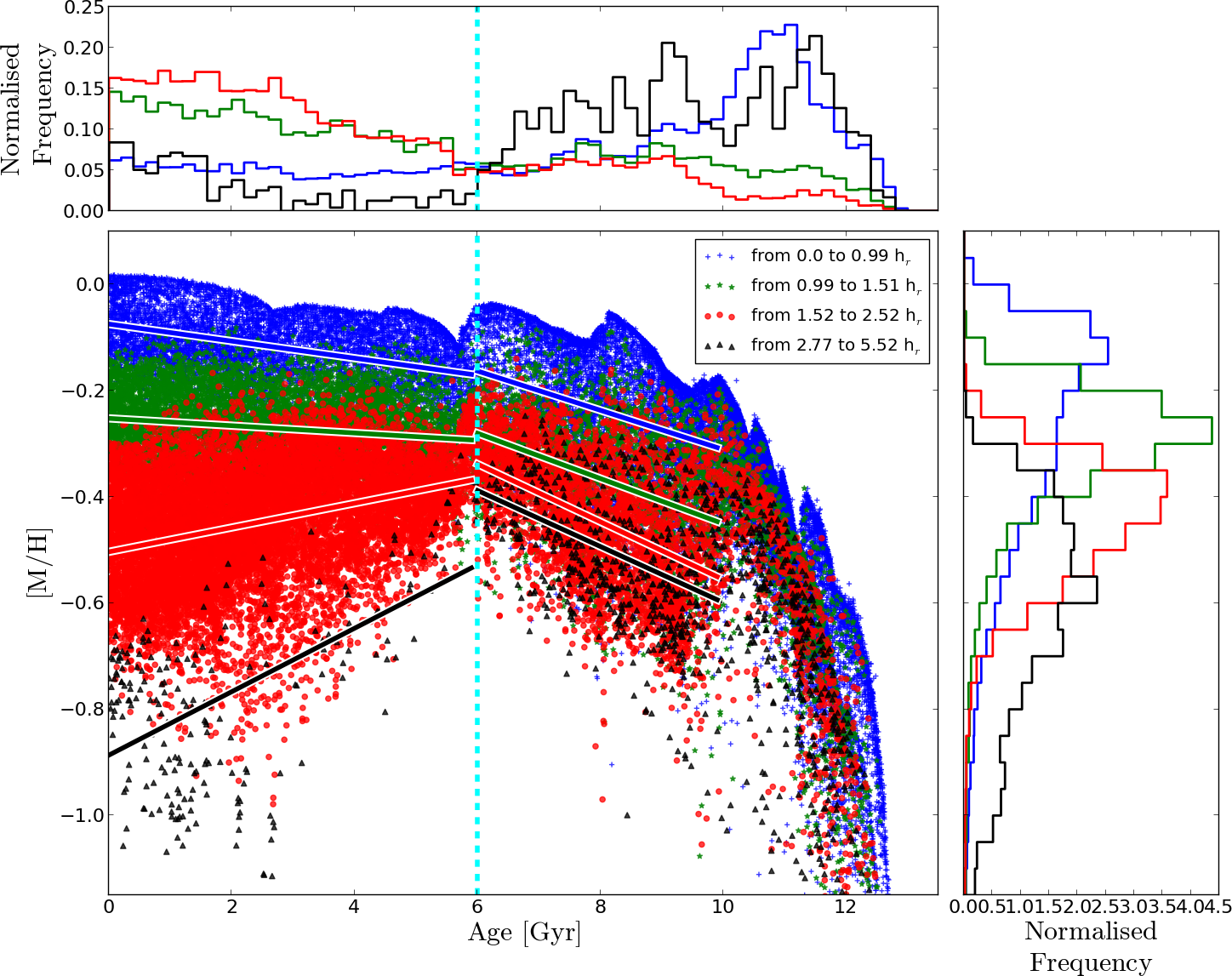} \\
\caption{Age-metallicity relation for Selene disc stars, colour coded 
according to their current radial positions (see text, 
Table~\ref{regions}, and Fig.~\ref{sigma_age} for more information). 
Blue crosses represent stars currently located in region 1. Green 
asterisks represent stars currently in region 2. Red dots represent 
stars currently in region 3. Black triangles represent stars currently 
in region 4. The two histograms on top of the AMR and on the right 
represent the age and metallicity distribution function, respectively 
(colour-coded in the same way). $t_\mathrm{jump}$ is denoted by the dashed cyan 
vertical line. Solid lines correspond to the linear fits performed to 
the AMR of the stars in the four radial regions for epoch ii (from $t_\mathrm{jump}$ 
till 10 Gyr ago) and epoch iii (from the present day till $t_\mathrm{jump}$).}
\label{AMR}%
\end{figure*}

\begin{figure}
\includegraphics[width=0.45\textwidth]{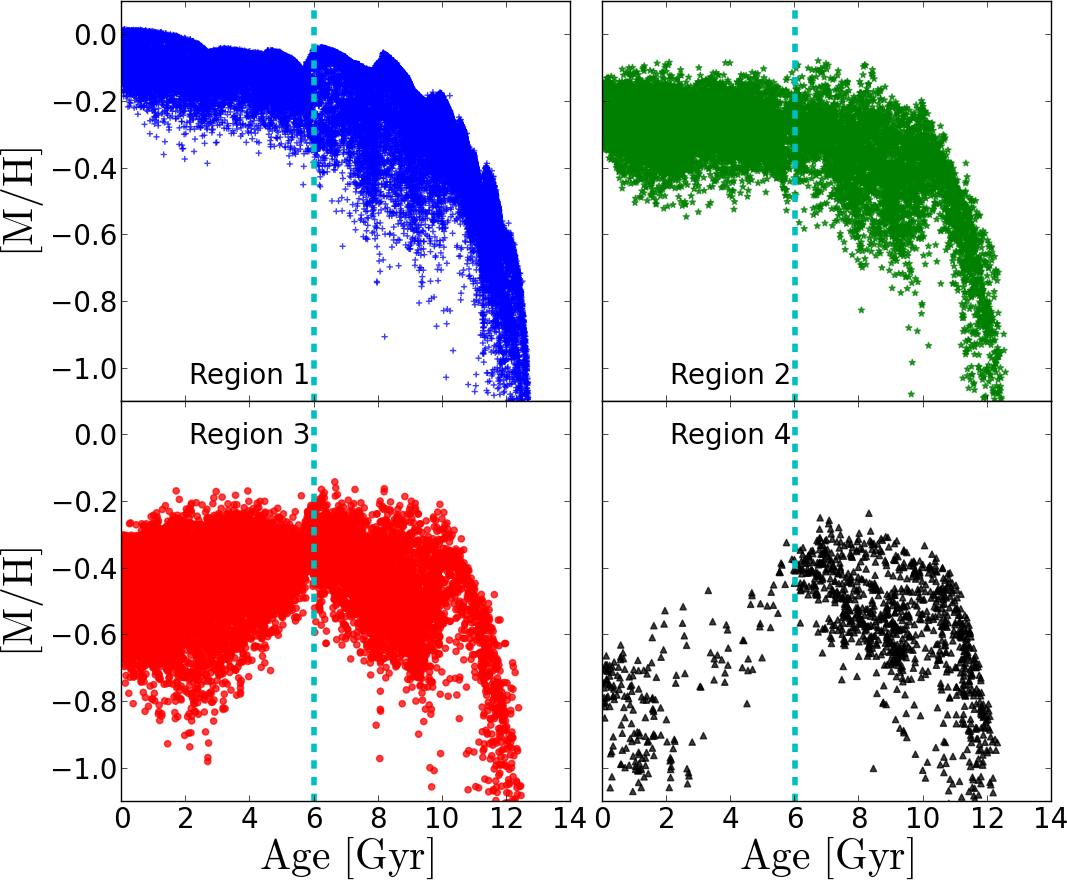} \\
\caption{Age-metallicity relation for Selene disc stars, equivalent to Fig.~\ref{AMR} but 
split into 4 different panels; one per region.}
\label{four_panels}%
\end{figure}

\begin{figure*}
\includegraphics[width=0.93\textwidth]{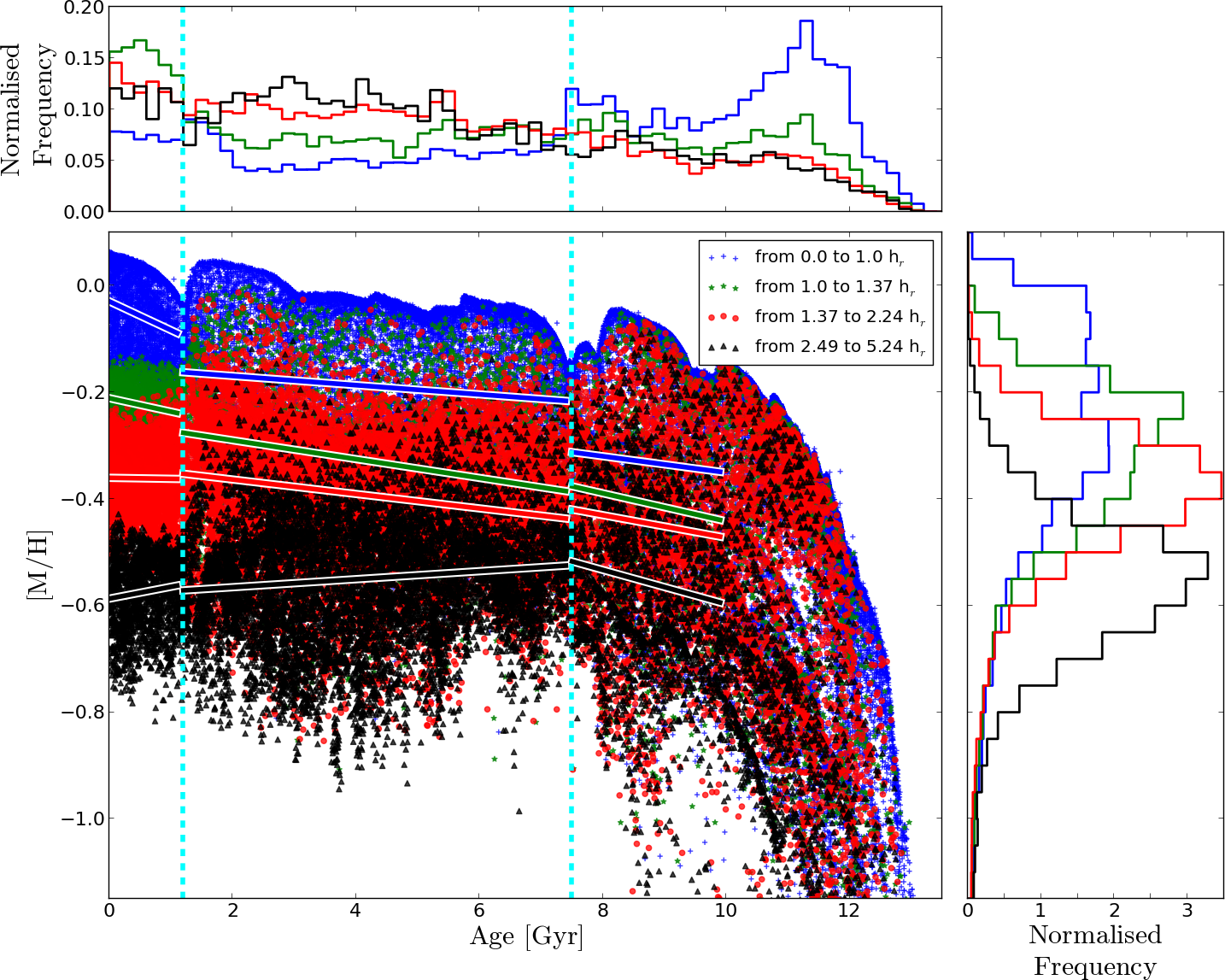} \\
\caption{Age-metallicity relation for Oceanus disc stars; see 
Fig.~\ref{AMR} for more information. $t_\mathrm{jump-a}$ and $t_\mathrm{jump-b}$ are 
denoted by the dashed cyan vertical lines.}
\label{AMR_oceanus}%
\end{figure*}

We have analysed the Age-Metallicity Relation (AMR) for the discs of the 
{\tt RaDES} galaxies. In Figs.~\ref{AMR} and \ref{AMR_oceanus} we show, 
as two representative examples, the AMR for the Selene and Oceanus disc 
stars, colour-coded according to their radial location (see Table 
\ref{regions} for the definition of the extent of each region). In these 
figures we also show the age and metallicity distribution functions for 
each region (top and right-hand panels respectively). To properly 
appreciate the details of Fig.~\ref{AMR} we have split such figure in 4 
different panels corresponding to the 4 radial regions under study (see 
Fig.~\ref{four_panels}).

Common to all the analysed galaxies, the chemical 
properties of the stars show a different behaviour depending on the 
assembly history stage they were born in. Figure~\ref{four_panels} clearly shows 
well-differentiated epochs according to the dispersion of the AMR displayed by the stars 
corresponding to the three-phase assembly history described in Sect.~\ref{treeplots}: 
stars formed during epoch i (older than 10~Gyr) show very small dispersion while stars born at 
epochs ii and iii display a larger dispersion at all ages. This tight AMR for 
old stars is a direct consequence of the origin of old stars, accreted
from satellites or formed in the early host galaxy: these environments
are spatially small and have comparatively homogeneous metal
abundances and enrichment histories. Afterwards, when quiescent phases 
dominate the mass assembly, the diversity of satellite metallicity
increases and the galaxy develops a metallicity gradient that enhances
the dispersion of the AMR. The metallicity distribution is also
broadened as late-forming satellites with lower metallicities are
accreted by the enriched host galaxy. Radial stellar motions also increase 
the dispersion of each region as stars move from the metal enhanced
center of the galaxy and the relatively metal-poor outskirts to other
regions. These radial motions tend to affect older populations more strongly (as
we will show in Sect.~\ref{SFH_rad}), yet will only broaden the dispersion of
the AMR if stars exist with distinct metallicities in other regions of
the galaxy. It is for this reason that radial motions broaden the
AMR dispersion for a given region for the epoch ii stars (because a metallicity
gradient exists at this time) but does not for epoch i (when the
metallicity is quite homogeneous) or epoch iii stars (when less radial
migration has taken place, see Sect.~\ref{SFH_rad}).

\begin{figure*}
\includegraphics[width=0.31\textwidth]{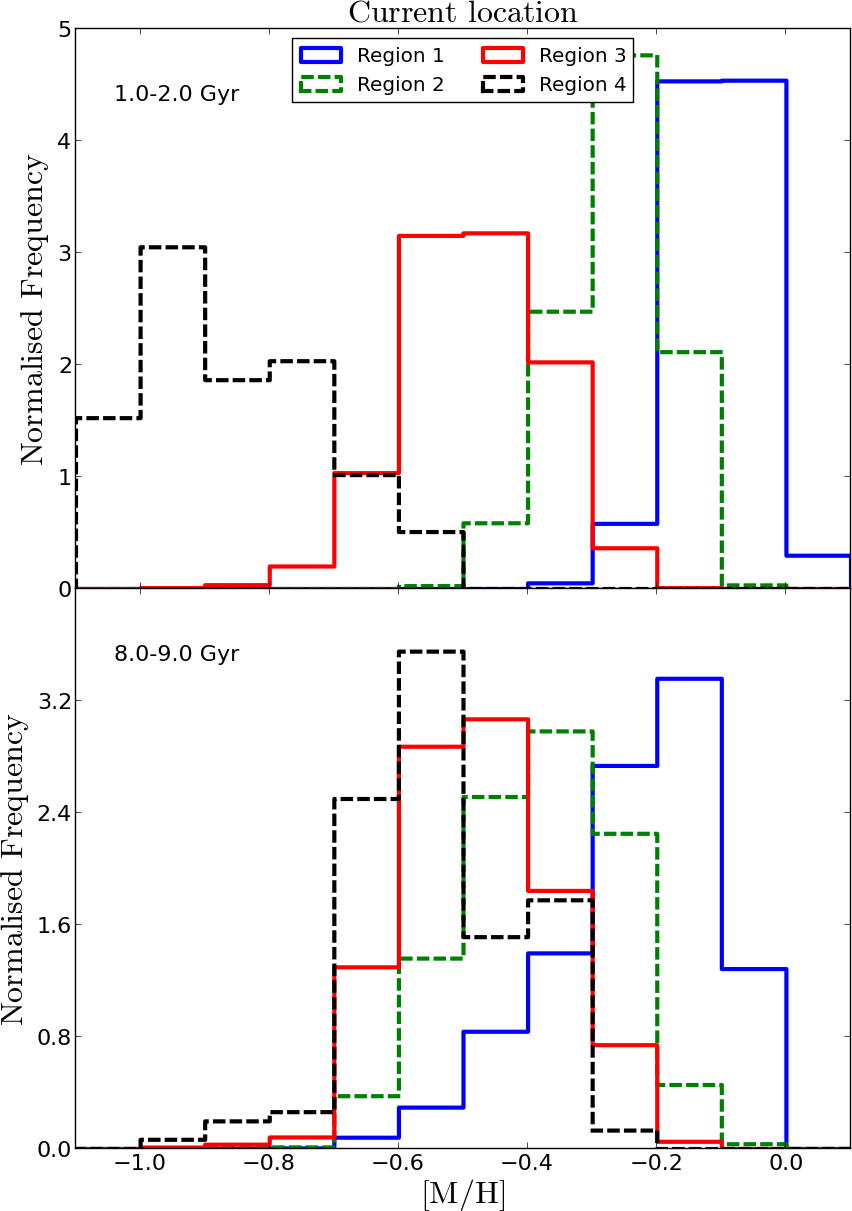} ~
\includegraphics[width=0.303\textwidth]{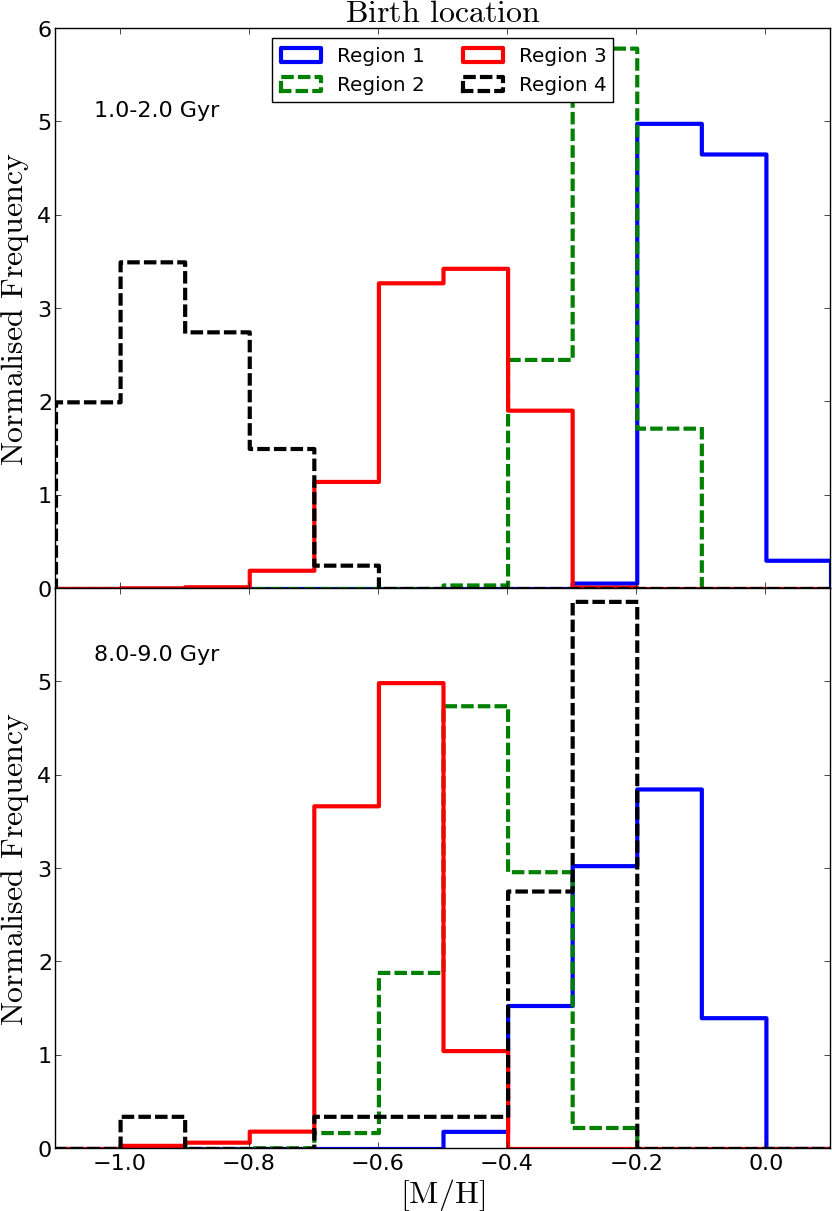} ~
\includegraphics[width=0.31\textwidth]{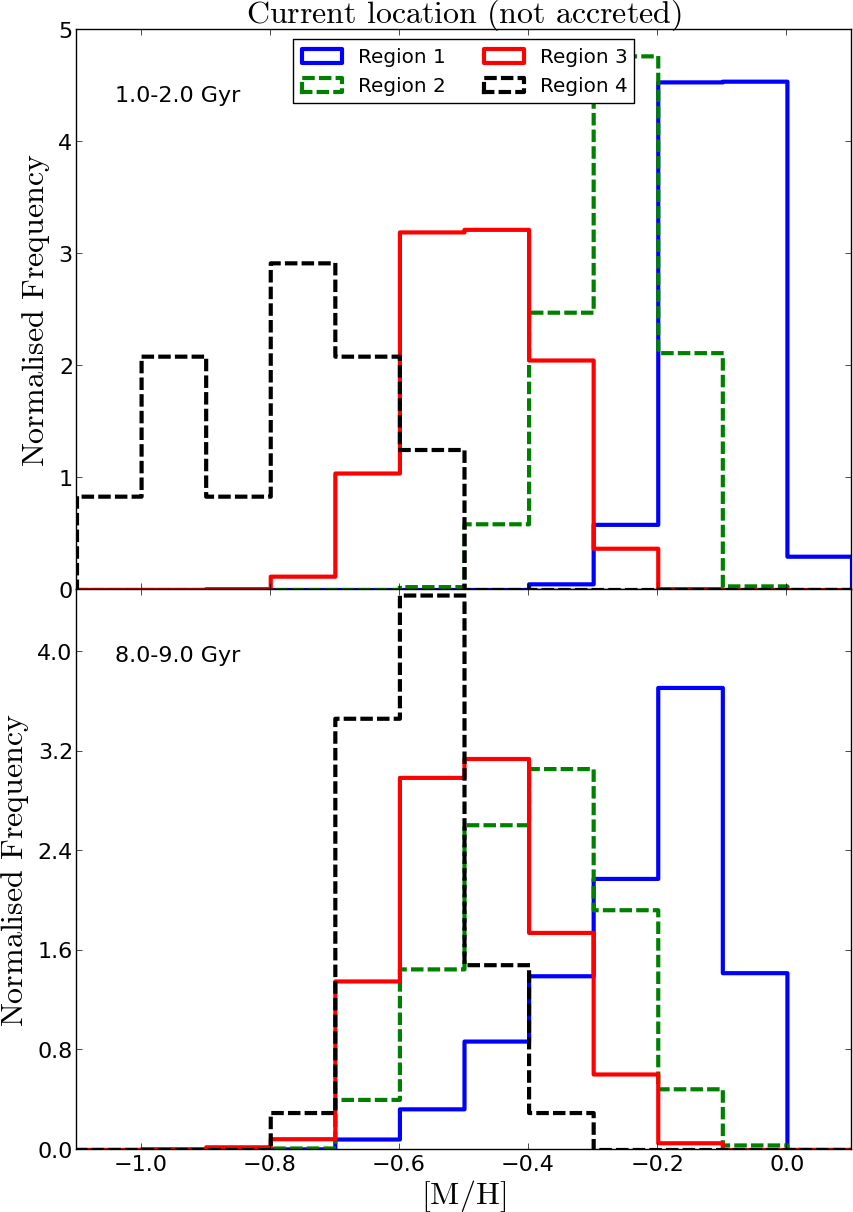} \\
\caption{Metallicity distribution in the four different radial regions 
distinguishing between young (1-2~Gyr, upper panel) and old stars 
(8-9~Gyr, bottom panel) for Selene (left, current locations; middle, birth locations; 
right, current location not considering accreted stars). Blue, green,
red, and black histograms represent particles 
in regions 1, 2, 3, and 4 respectively. Note the different distributions for the young stars compared to the 
similar distributions in the case of the old component.}
\label{met_distrs}
\end{figure*}

In addition, we have carefully analysed other aspects of the AMR such as: 
i) the metallicity distribution, suggesting different degrees of 
chemical mixing according to the Age-Metallicity plots; and ii) the 
slopes in the AMR, we should note that we compute these slopes in the 
AMR for epochs ii and iii, avoiding the more turbulent epoch i.

i) We have analysed the MDF (see Fig.~\ref{met_distrs}) for stars in regions 1 
to 4, distinguishing between young (younger than $t_\mathrm{jump}$, 1 to 2~Gyr) 
and old (older than $t_\mathrm{jump}$, 8 to 9~Gyr) stars (see left panel of 
Fig.~\ref{met_distrs} for Selene). For every galaxy, the metallicity 
dispersion is larger in the outer parts than in the inner parts 
(irrespective of age) ranging from 0.04 vs. 0.10~dex in region 1 to 0.07 
 vs. 0.13~dex in region 4 for the young vs. old components respectively (averaged values 
for all 19 {\tt RaDES} galaxies). This is consistent with 
\citet[][]{2014A&A...572A..92M}, where the increase in the AMR scatter 
with radius was related to the expected increase in contamination from 
migration and heating with radius. The metallicity dispersion for old 
stars is also larger than for young stars. If we define the change in 
the metallicity dispersion between old and young component as 
$100*(\sigma_\mathrm{met, old} - \sigma_\mathrm{met, young})/\sigma_\mathrm{met, old}$, its 
radial evolution is 58.7\% in region 1, 37.9\% in region 2, 26.4\% in 
region 3, and 21.6\% in region 4, i.e. the differences are
  minimised at greater radii.

If we do not focus on the radial evolution of the MDF for old or young 
stars, but in the discrepancies between old and young populations we can find 
interesting differences. Young stars show very different MDFs for the 
different radial bins while more similar distributions are displayed by 
old stars in regions 1 to 4. To quantify such a statement we have 
analysed, using Kolmogorov-Smirnov (KS) statistical tests, the 
differences between the metallicity distributions of different radial 
regions for young and old stars (at their current positions). The KS test 
effectively determines 
whether two sets of data are drawn from the same statistical 
distribution and how different those distributions are. For our 
purposes, larger KS values when comparing different radial regions would 
imply a greater degree of chemical mixing. According to these tests, 
young stars show distinct metallicity distributions between the 
different radial regions, while old stars in each region show similar 
metallicity distributions to one another. The average of the KS 
statistic values for all the galaxies comparing all the possibilities of 
pair or regions (i.e. 1-2, 1-3, 1-4, 2-3, 2-4, and 3-4) is 0.85 for 
young stars and 0.45 for old stars. These KS tests allow us to conclude 
that the degree of chemical mixing is greater for old stars than for 
young stars.

To properly assess the cause of the different mixing degrees 
for old and young stars, Fig.~\ref{met_distrs} not only shows the MDF 
for young and old stars at the current position (left panel) 
but also at their birth location (middle panel) for Selene. 
If the well-mixed MDFs displayed by old stars were due to the
merger event, we should observe that old stars, taken at their birth
location, exhibit distinct MDFs. However, Fig.~\ref{met_distrs} 
(left and middle panels for old stars) 
reveals that while there is some degree of mixing of the MDFs after
the formation of the stars, there is no clear separation as found in
the younger populations, and the majority of the similarity in the
MDFs of old stars in different regions is imprinted at birth. This statement 
is backed by the mean and dispersion values displayed by these distributions. 
Although slightly lower dispersion values and larger mean metallicity ranges are found for 
stars at their birth position than at the current position, this is not enough as to suggest 
that the mixing was imprinted by merger events\footnote{The mean metallicities ([M/H]) for old 
stars at their current position are -0.23, -0.38, -0.48, and -0.56 from region 1 to 4, 
with metallicity dispersion of 0.13, 0.12, 0.12, and 0.11 respectively. However, at their 
birth location those values change to -0.20, -0.44, -0.58, and -0.31 (mean metallicities) 
and 0.09, 0.07, 0.07, and 0.16 (metallicity dispersions) for the particular case of Selene.}.

To further quantify the importance of mixing induced by accretion (in contrast to mixing imprinted at birth, 
the right panel of Fig.~\ref{met_distrs} shows the MDF of the Selene stars of star particles at regions 1 to 4 (current 
position) excluding accreted stars (i.e. $R_\mathrm{birth} > 20$ kpc and $|z| > 3$ kpc). The mean metallicity values of those MDFs are -0.23, 
-0.39, -0.48, and -0.56 while their dispersions are 0.13, 0.12, 
0.11, and 0.07 for old stars from region 1 to 4. If we compare these 
values with those computed for the MDFs for old stars at their 
current positions (mean metallicities of -0.23, -0.38, -0.48, and -0.56 and dispersions of 0.13, 0.12, 0.12, and 0.11 for regions 1--4 
respectively) and comparing the MDF shapes, we can conclude that the 
mixing is particularly important for region 4.

\begin{figure}
\includegraphics[width=0.45\textwidth]{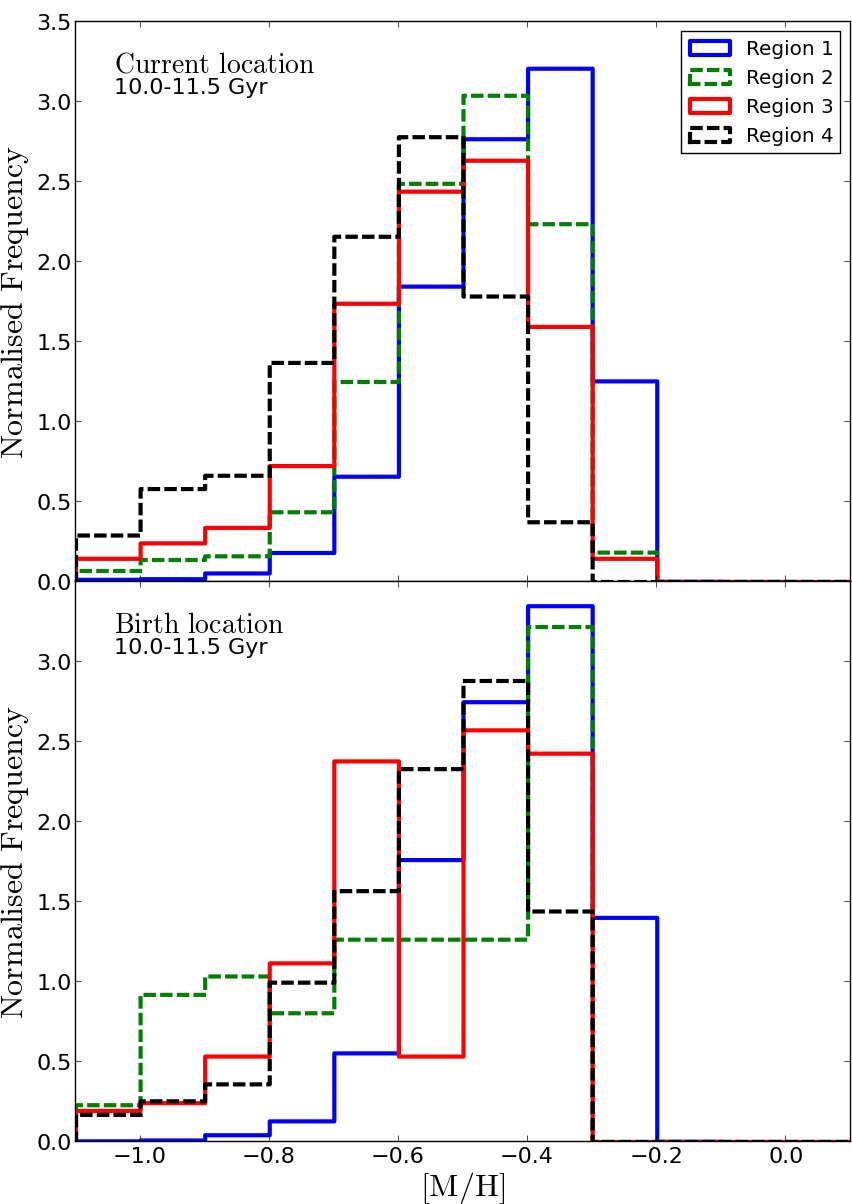} \\
\caption{Metallicity distribution in the four different radial regions 
for very old stars (10.0-11.5~Gyr) for Selene. Coloured histograms follow 
the colour code explained in Fig.~\ref{met_distrs}.}
\label{met_distrs_old_old}
\end{figure}

To determine if the MDFs of epoch i stars (10 to 11.5~Gyr old)
have changed during the evolution of the galaxy we show, in
Fig.~\ref{met_distrs_old_old}, the MDFs of these stars at their
current location (top panel) and at their birth location (bottom
panel). The time evolution of these old stars is much weaker than
for other age ranges because the MDFs of our four regions are so
well mixed when the stars are born\footnote{The displayed metallicity 
dispersions are 0.13, 0.15, 0.16, and 0.18 (0.12, 0.24, 0.20, and 0.18) 
from regions 1 to 4 at their current (birth) locations and their mean 
values are -0.44, -0.51, -0.55, and -0.65 (-0.42, -0.61, -0.58, and 
-0.57).}. The MDFs of regions do change
over time but there is no clear trend present in the MDFs taken at
birth locations. Taken together with Fig.~\ref{met_distrs} this
further demonstrates that the clearly separated MDFs seen in the
youngest stars was never present in older populations which form
with more homogeneous MDFs.

\begin{figure}
\includegraphics[width=0.45\textwidth]{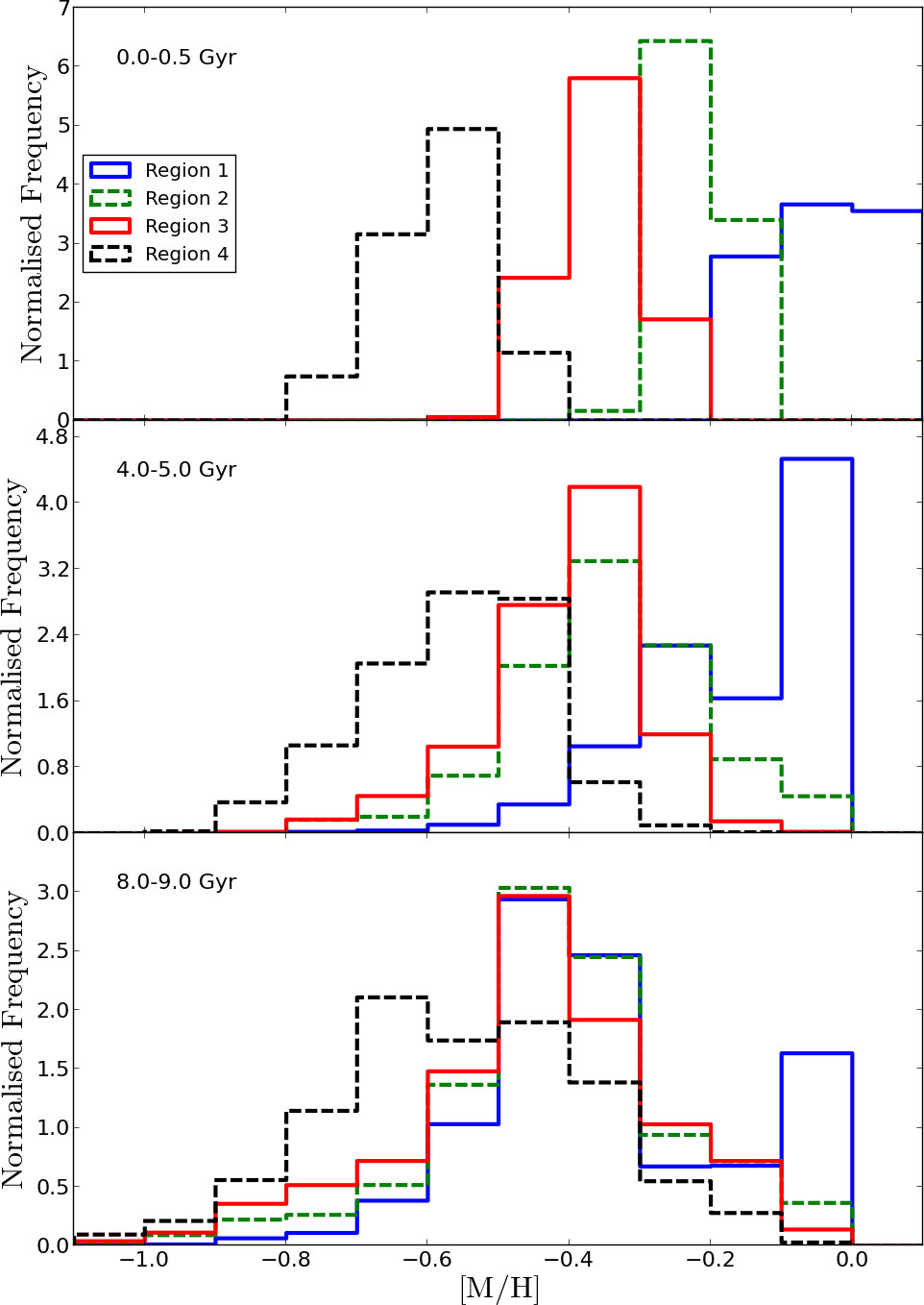} \\
\caption{Metallicity distribution in the four different radial regions 
distinguishing between young (0-0.5~Gyr, upper panel), intermediate 
(4-5~Gyr, middle panel) and old stars (8-9~Gyr, bottom panel) for 
Oceanus (current locations). Coloured histograms follow the colour code explained in 
Fig.~\ref{met_distrs}. Note the different 
distributions for the young stars compared to the similar distributions 
in the case of the old component. Intermediate age stars show an 
intermediate behaviour.}
\label{met_distrs_oceanus}%
\end{figure}

In summary, the well-mixed MDFs displayed by stars born during phases 
i and ii is not directly due to a single satellite accretion event (although it also 
has an effect) but because of the collective activity in phases i and
ii that homogenise the chemical characteristics of the gas from which stars are born.

ii) The slopes of the AMR for the four regions are always negative for 
the old stars (for every galaxy with just one $t_\mathrm{jump}$). These slopes 
are computed by means of a linear fit to the points in the AMR in the 
plane $[M/H]$-Age (thus in units of dex$/$Gyr). These old-star AMR 
slopes are roughly similar, as expected from stars born at a more
chemically homogeneous time in the assembly history. The relative slope differences 
(100$\times(slope_{i}-slope_{j})/slope_{j}$, with $i$ and $j$ ranging 
from 1 to 4 accounting for all the possibilities) are under 40\% for the 
entire sample of the {\tt RaDES} galaxies (40\% is reached just for the 
most extreme cases, the median is around 20\%). When considering the 
younger part of the AMR (stars younger than $t_\mathrm{jump}$), the slopes 
range from negative (regions 1 and 2) to positive (region 4) in most of 
the galaxies. Region 3 is the region with the shallowest AMR slope. The 
average values considering the 19 {\tt RaDES} galaxies are -0.031, 
-0.027, -0.004, and 0.008~dex$/$Gyr for regions 1 to 4 respectively. In 
the particular case of Selene the values are -0.016, -0.007, 0.002, and 
0.06~dex$/$Gyr from region 1 to 4.

The case of Oceanus (as well as Artemis, not shown here) is a special 
one that deserves to be analysed in further detail. As we have already commented, 
Oceanus shows two merger episodes during the simulation with similar 
characteristics to the ones described in Sect. \ref{treeplots}. Both 
episodes imprint a jump in the AVR (see Sect. \ref{AVR}, with the jump 
at 7.5~Gyr being later erased by the effect of the last merger event) 
and leave a signature in the AMR. In Fig.~\ref{AMR_oceanus} 
we can clearly distinguish three main age regions, delimited by 
$t_\mathrm{jump-a}$ $\sim$7.5 and $t_\mathrm{jump-b}$ $\sim$1.2~Gyr. 
Fig.~\ref{met_distrs_oceanus} shows the metallicity distributions for 
three different age bins (young, intermediate, and old) for Oceanus. In 
this case the radial evolution of the MDF is weaker for the older age bins. The 
young component shows metallicity dispersions of 0.071, 0.036, 0.056, 
and 0.071~dex (regions 1 to 4). For the intermediate (old) component, we 
find a radial evolution of 0.132, 0.133, 0.111, and 0.125~dex (0.176, 
0.170, 0.182, and 0.188~dex). However, it is still true (even for this 
two-event galaxy) that the metallicity dispersion is higher for older 
stars (those with ages in the range 7.5--10~Gyr), the metallicity dispersion has intermediate values for 
intermediate-age stars, and lower values for young stars. As already 
outlined in the case of Selene, these well-mixed MDFs displayed by stars born 
during epochs i and ii are mainly a consequence of gas mixing that is
imprinted on the stars at birth. The radial 
evolution of the slope in the AMR from negative to positive is also found 
for the youngest stars ($-0.054$, $-0.025$, $-0.0018$, and $0.022$~dex$/$Gyr), 
while negative AMR slopes are obtained for the intermediate and old 
components (except for the slope of intermediate aged stars in region 4 which
is consistent with the increasingly positive slope as a function of radius).
To assess the degree of chemical mixing we decided, as 
described above, to make use of KS tests. We can draw similar 
conclusions as before, i.e. the youngest component of the AMR has 4 
different metallicity distributions (KS statistic of 0.96), the 
intermediate component show a higher degree of mixing (KS statistic of 
0.60), and the old component show the most similar metallicity 
distributions (KS statistic of 0.24). 

The negative-to-positive change (with decreasing age) in the slopes of
the AMRs deserves a further explanation. Negative slopes in the AMR
are a direct consequence of chemical enrichment due to stellar
evolution. Positive slopes are difficult to produce (and indeed are
not frequently found in nature) due to chemical evolution alone,
requiring efficient inflows of un-enriched gas into the galaxy.
Inversion of the AMR is more easily achieved through the 
dynamical effects of satellite accretion (metal-rich stars being formed
in the core of the satellites while passing through region 4),
accretion of metal-poor gas (to form young, metal-poor stars), and
radial migration. Further details can be found in Sect.~\ref{discussion}.

\begin{figure*}
\includegraphics[width=0.45\textwidth]{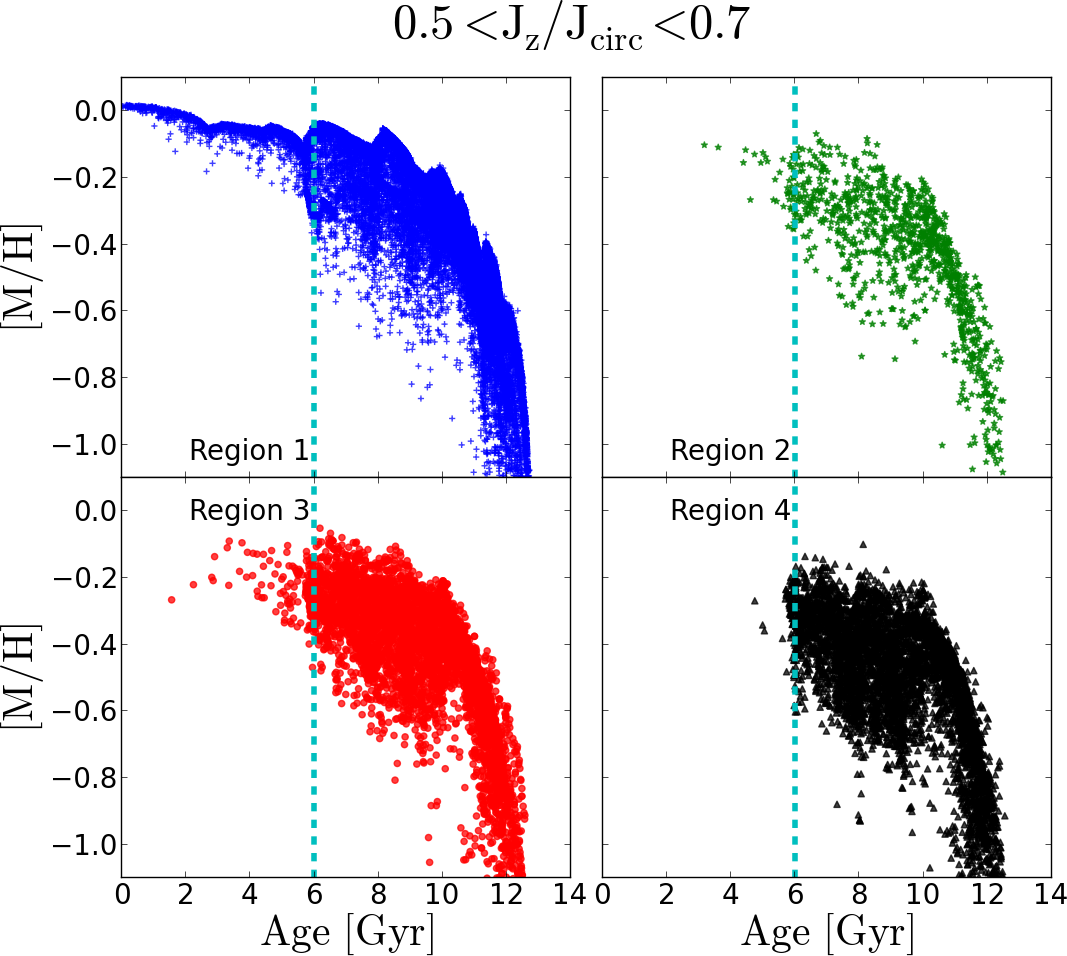} ~
\includegraphics[width=0.45\textwidth]{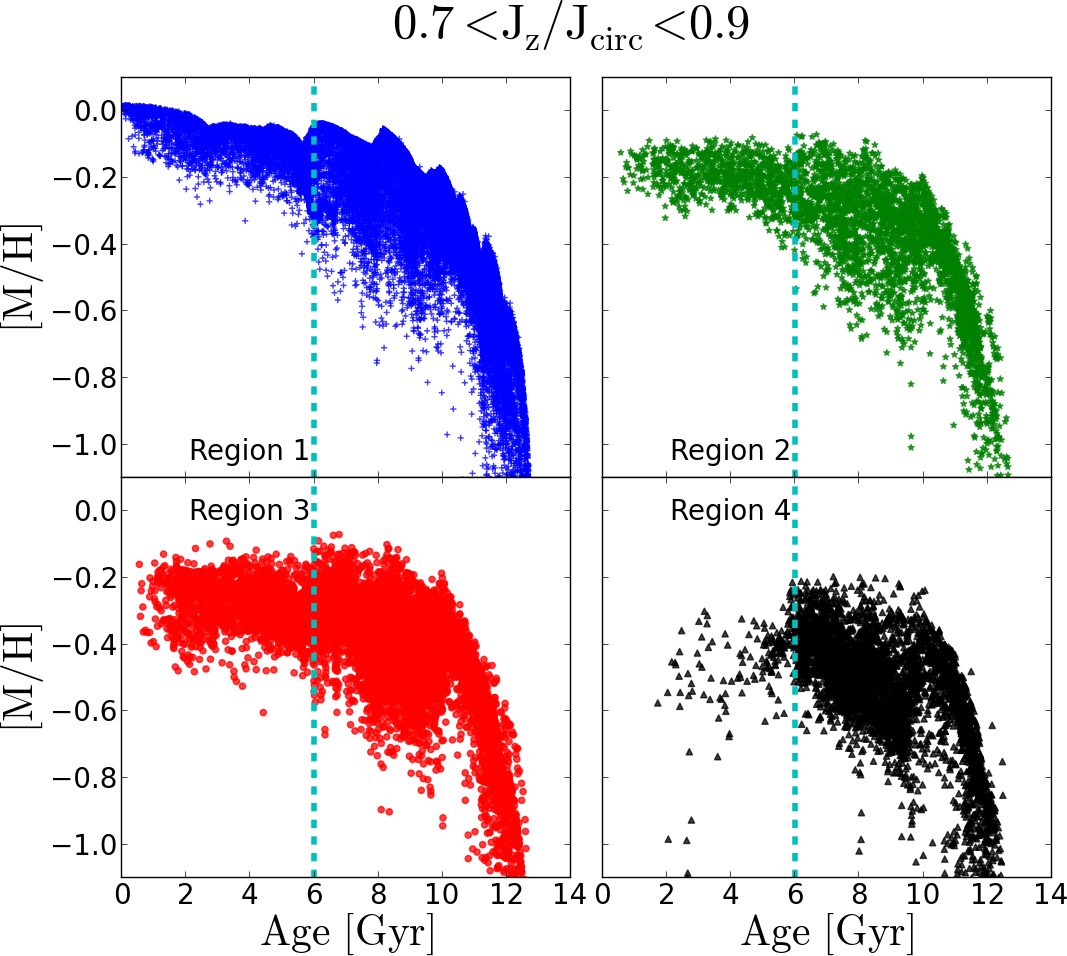} \\
\includegraphics[width=0.45\textwidth]{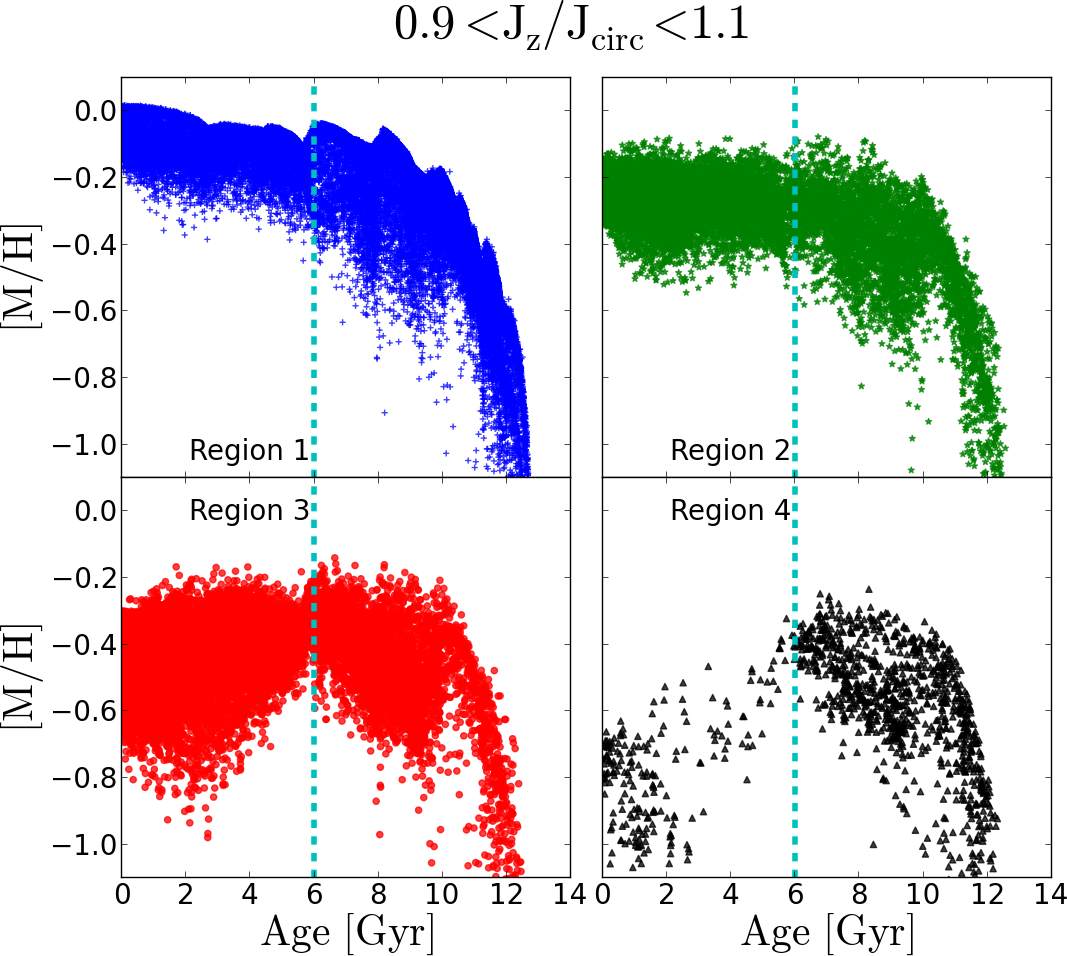} ~
\includegraphics[width=0.45\textwidth]{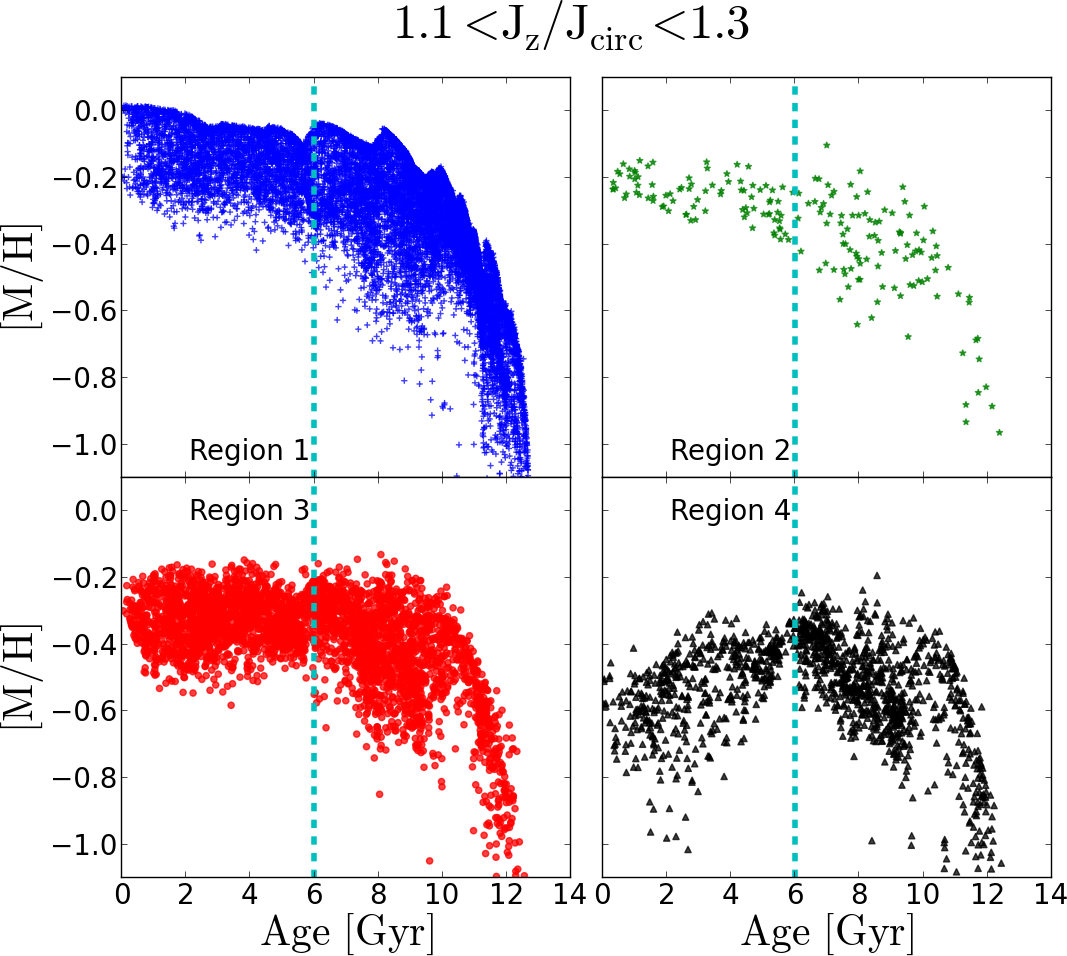} \\
\caption{Age-metallicity relation for Selene applying different cuts in the $J_{z}/J_\mathrm{circ}$ ratio. Upper left, particles with $J_{z}/J_\mathrm{circ}$ between 0.5 and 0.7; upper right, particles with $J_{z}/J_\mathrm{circ}$ between 0.7 and 0.9; lower left, particles with $J_{z}/J_\mathrm{circ}$ between 0.9 and 1.1; lower right particles with $J_{z}/J_\mathrm{circ}$ between 1.1 and 1.3}
\label{AMR_J_z_cut}%
\end{figure*}

In order to check the influence of our kinematic selection
criterion for disc/non-disc particles on the AMR we have repeated
our analysis of the AMR for Selene considering four different
$J_{z}/J_\mathrm{circ}$ ranges (0.5--0.7, 0.7--0.9, 0.9--1.1, and 1.1--1.3, see
Fig.~\ref{AMR_J_z_cut}). The main results outlined
in this section remain true for each of the $J_{z}/J_\mathrm{circ}$
selections, albeit with slight differences: there tend to be fewer
stars forming post-$t_\mathrm{jump}$ in the lower $J_{z}/J_\mathrm{circ}$ cuts
(particularly at larger radii) making the inverted portion of the
AMR non-existent. Our disc star criterion
(0.9$<$$J_{z}/J_\mathrm{circ}$$<$1.1) strengthens our results as this choice
is  the most representative of a disc population, though it should
be clear from Fig.~\ref{AMR_J_z_cut} that the inversion will be
present if stars with $J_{z}/J_\mathrm{circ}$ in the range 0.5--1.3 were
all included.

\vspace{6mm}

\section{Chemo-dynamical imprints of satellite accretion: radial stellar motions}
\label{chedy_imprints}

Sections \ref{AVR} and \ref{AMR_sect} clearly point towards the effect 
of satellite accretion on observables such as the AVR or the AMR. 
Considering the richness of cosmological simulations we can however 
study galaxy properties other than those apparent at the present day, 
such as time evolution, stellar motions, and angular momentum. A careful 
inspection of these simulations can help us to understand the mechanisms 
responsible for the present-day observed properties.
 
In this section, we further investigate the effect of satellite 
accretion in the stellar age distribution at different radii (see Sect. 
\ref{SFH_rad}) and the signatures that these events leave in the AMR 
(see Sect. \ref{AMR_sect_colour}).

\subsection{Hints from the stellar age distribution}
\label{SFH_rad}

\begin{figure*}
\includegraphics[width=0.85\textwidth]{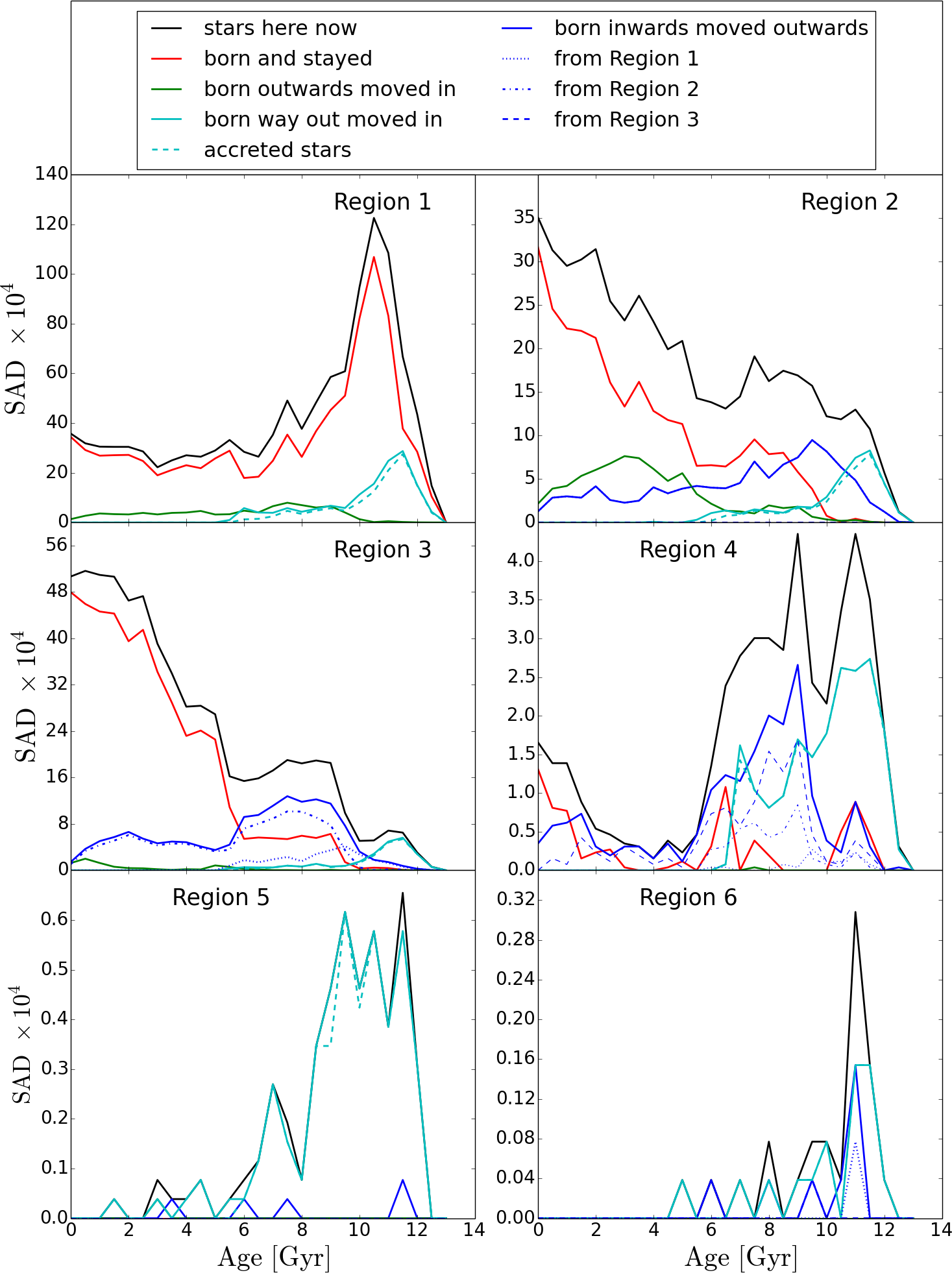} \\
\caption{Radially resolved, stellar age distribution for the Selene disc 
normalised to the total stellar mass. The different panels show the 
stellar age distribution of the stars currently located in regions 1 
(upper-left), 2 (upper-right), 3 (middle-left), and 4 (middle-right), 
see Table~\ref{regions} for more information. We have added two more 
regions in the outer most part of the galaxy. Region 5 goes from 4.4$h_{\rm in}$ to 
6.2$h_{\rm in}$ and region 6 goes from 6.2$h_{\rm in}$ to 7.9$h_{\rm in}$.  These two regions 
have been added to study the causes of the old plateau in the age 
distribution seen in the outermost disc. Black solid lines represent 
particles currently in that region; solid red line, particles born and 
stayed there; solid blue line, particles born inwards and moved outwards 
currently in that region; solid green line, particles born outwards and 
moved in; solid cyan line, particles born way out and moved in; dashed 
cyan line, accreted particles (i.e. $R_\mathrm{birth} > 20$ kpc and $|z| > 3$ kpc); dotted 
blue line, particles born in region 1 that end up there; dotted-dashed 
blue line, particles born in region 2 that end up there; dashed blue 
line, particles born in region 3 that end up there. The definitions to 
the lines are described in full in the main text.} 
\label{SFH}%
\end{figure*} 

The study of stellar properties such as age or metallicity of the stars 
currently located at specific regions, as well as their original 
positions, can provide us with keys to understanding galaxy evolution. 
Fig.~\ref{SFH} shows the current Stellar Age Distribution (SAD, 
normalised to the mass of the whole galaxy) for Selene's disc stars, in the 
four scaled regions (solid black line). At this stage we add 
two more regions from the outermost parts corresponding to the old 
plateau: region 5 at 4.4$h_{\rm in}$--6.2$h_{\rm in}$ and region 6 at 6.2$h_{\rm in}$--7.9$h_{\rm in}$.
We separate the components of this 
current SAD according to i) particles born in each region that have 
stayed there ({\it in-situ} stars, solid red line); ii) born outwards 
then moved in (solid green line less than 5.0~kpc, solid cyan line more 
than 5.0~kpc); and iii) stars born inwards and moved out (solid blue 
line). In this third case, we divide them as well depending on the birth 
region: iiia) born in region 1 (dotted blue line); iiib) born in region 
2 (dotted-dashed blue line); and iiic) born in region 3 (dashed blue 
line). Our birth radius is very close to the instantaneous radius 
at the time of the birth of the stars. It is defined as 
the location of the star in the first snapshot.

As a general trend, radial region 1 is mainly populated by old stars 
(older than 10~Gyr whose distribution shows a peak at 12~Gyr) that were 
born and stayed in that region (red line, accounting for $\sim$23\% of 
the mass in region 1), with a non-negligible fraction of stars coming 
from the outer parts ($\sim$20\% of the mass in region 1). The oldest 
stars (older than 10~Gyr) in radial region 2 were formed inside 
($\sim$5\% of the mass in region 2) or far outside ($\sim$7\% of the 
mass in region 2), while the majority of the young stars (younger than 
4~Gyr) formed {\it in-situ} ($\sim$33\% of the mass in region 2) with 
also an important contribution of stars born inside and outside that 
region ($\sim$6\% of the mass in region 2). The same dichotomy is found 
in regions 3 and 4, older stars were born inside or much further out 
while young stars mainly formed there or moved outward from the inner 
regions with fewer {\it in-situ} stars, especially for region 4, which 
is mainly dominated by migrated stars with a small percentage of in-situ 
stars ($\sim$16\% of the mass in region 4). Studying outward travellers 
in region 4, we can see that the older the star is the more it has 
travelled outwards (from regions 1, 2, or 3). The radial evolution of 
the red lines reflects a clear inside-out growth of the disc, and we can 
analyse when star formation starts in each radial region: Stars were formed 
in region 1 since the beginning of the simulation; region 2 started forming 
stars 10~Gyr ago; region 3, 6~Gyr ago; and region 4, 2~Gyr ago. We note 
that the vast majority ($\sim$92\%) of stars migrating more than 5~kpc (solid cyan lines
in Fig.~\ref{SFH}) are accreted from outside the galaxy,
i.e. $R_\mathrm{birth} > $20~kpc and $|z_\mathrm{birth}| > $3~kpc. The age
distribution of these accreted stars is shown in Fig.~\ref{SFH} as a
dashed cyan line.

Regions 5 and 6 are mainly populated by old stars on circular orbits 
coming from well outside the host galaxy (cyan lines in 
Fig.\,\ref{SFH}). Some stars from the inner parts are also found, 
however, the main cause of the old plateau observed in all the {\tt 
RaDES} galaxy discs is the accretion of old stars from early satellites. 
This is in contrast to the interpretation by 
\citet[][]{2008ApJ...684L..79R}, who suggested that the oldest stars in 
the outskirts have migrated from the inner disc. We find no stars that 
were born in the outer parts, that are still in that region, and that 
meet our criterion for disc stars.

\subsection{Hints from the age-metallicity relation}
\label{AMR_sect_colour}

\begin{figure}
\includegraphics[width=0.45\textwidth]{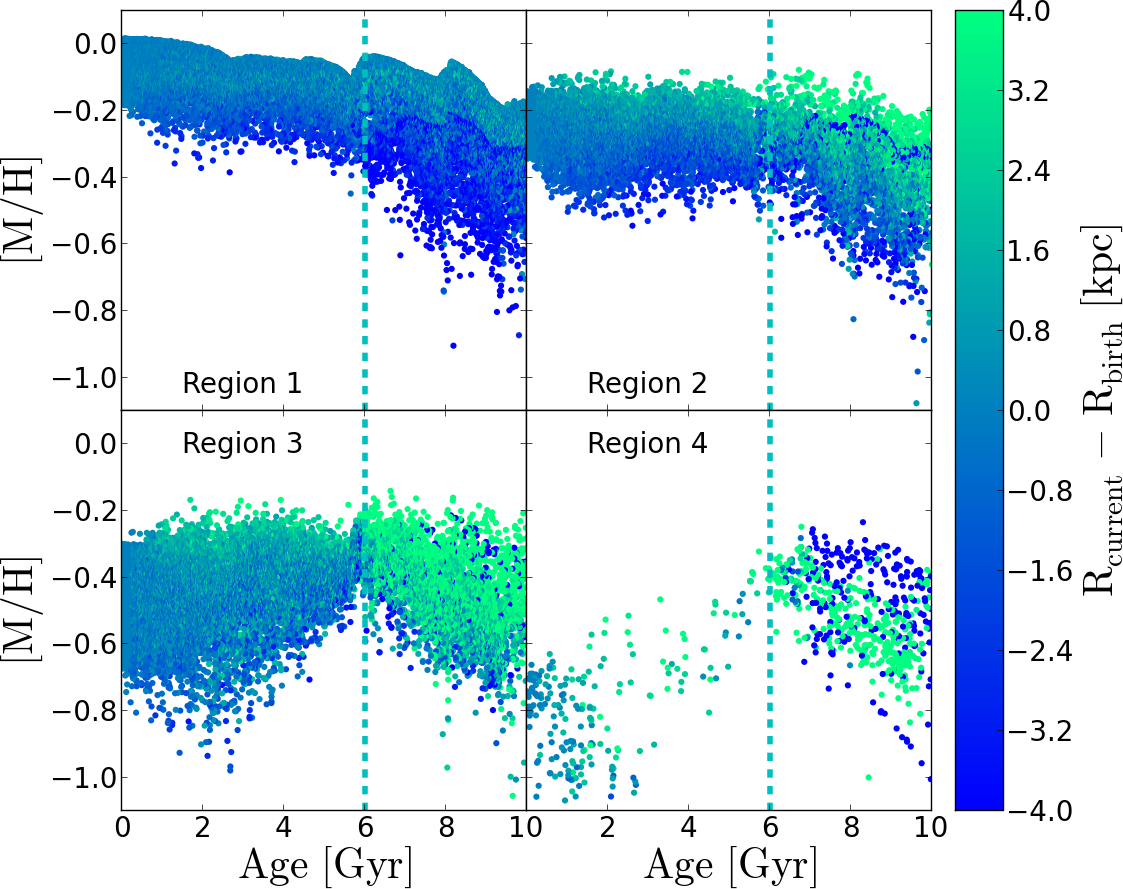} \\
\caption{Age-metallicity relation for Selene disc particles in 
four radial bins (current positions), every point represents a stellar 
particle colour-coded according to its $R_\mathrm{current}$ - $R_\mathrm{birth}$. 
Distribution of plots as in Fig.~\ref{SFH} (regions 1 to 4). Dashed cyan 
vertical lines are located at $t_\mathrm{jump}$. We must highlight that, 
although the colour-coded range goes from -4 to 4~kpc, we are plotting 
all the disc particles in that region, i.e. light-green (dark-blue) 
really means $R_\mathrm{current} - R_\mathrm{birth} \geq 4.0$ ($R_\mathrm{current} - 
R_\mathrm{birth} \leq -4.0$).}
\label{AMR_radius}%
\end{figure}

\begin{figure}
\includegraphics[width=0.45\textwidth]{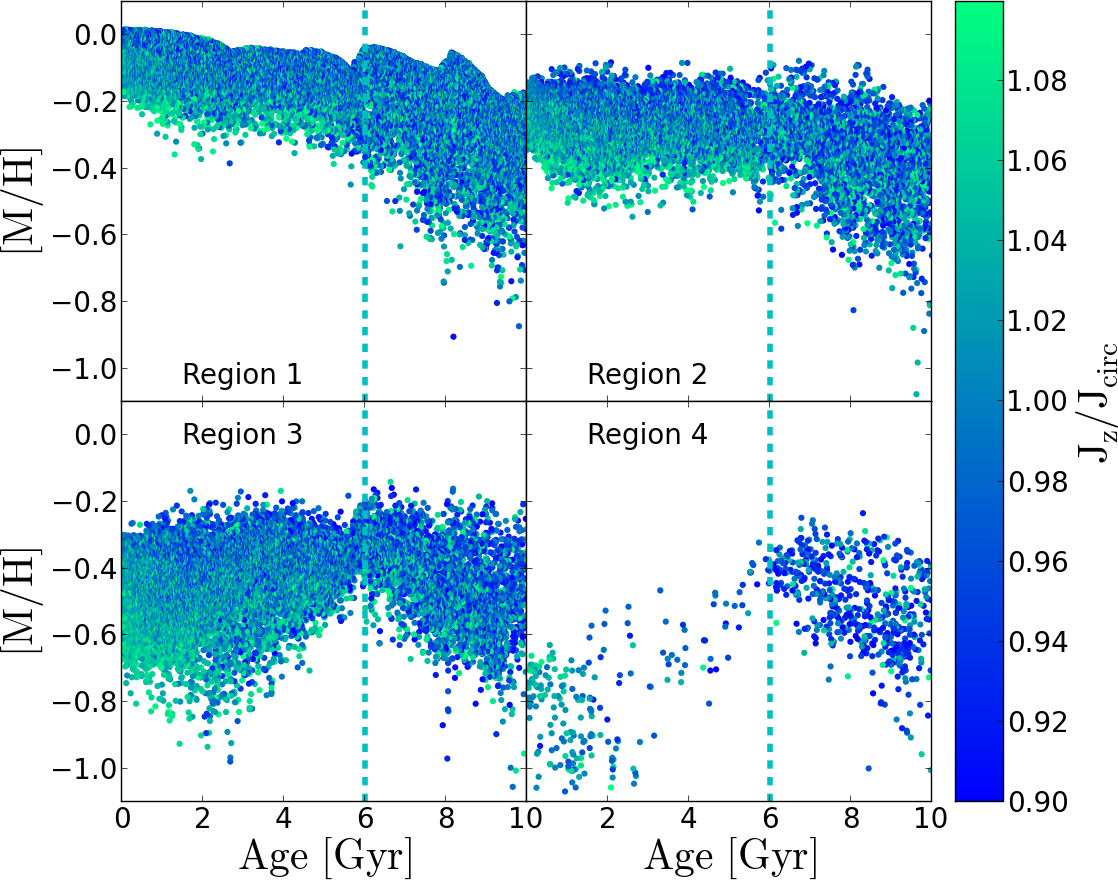} \\
\caption{Age-Metallicity Relation for Selene disc particles in 
four radial bins (current positions), every point represents a stellar 
particle colour-coded according to its $J_{z}/J_\mathrm{circ}$. Distribution of 
plots as in Fig.~\ref{SFH} (regions 1 to 4). Dashed cyan vertical lines 
are located at $t_\mathrm{jump}$.}
\label{AMR_jz_jcirc}%
\end{figure}

In Sect. \ref{AMR_sect} we showed the radial evolution of the AMR in our 
disc galaxies and outlined the first signatures left by satellite 
accretion in such relation.  Here, we study in more detail the AMR and 
expand the imprints of satellite accretion on the chemistry and the 
dynamics of our galaxies.

Fig.~\ref{AMR_radius} shows the same AMR displayed in Fig.~\ref{AMR} but 
this time split into four different sub-panels (one for each radial 
region) and colour-coded according to the $R_\mathrm{current}$ - $R_\mathrm{birth}$ 
for each particle. We can draw similar conclusions to those outlined in 
the previous section: region 1 is mainly populated by {\it in-situ} 
stars at all ages with a significant number of stars that have moved 
inwards, particularly from the last satellite (cyan line, stars born 
before $t_\mathrm{jump}$, see Fig.~\ref{SFH}). As we compare the inner with the 
outer regions, more and more stars moving outwards appear (especially 
amongst older populations) and fewer stars move inwards, except for 
those stars moving inwards that were born at the furthest distances in 
satellites (dark blue points in Fig.~\ref{AMR_radius}). The extreme case 
is found in region 4 where the majority of particles have been formed at 
smaller radii, some old stars come from satellites, and just a few have 
been born in-situ (stars younger than 2~Gyr, when the inside-out growth 
of the disc reached region 4). Every old particle now located in regions 
3 or 4 is either coming from the inner parts or from satellites 
(consistent with Fig.~\ref{SFH}).

The AMR of the last accreted satellite can be discerned from the older 
and dark-blue points in regions 1 and 2 in Fig.~\ref{AMR_radius} (due to 
aesthetic purposes we restrict the colour bar range from -4 to 4~kpc, 
i.e. dark-blue points mean $R_\mathrm{current} - R_\mathrm{birth} \leq -4.0$). The 
shape of this AMR mimics that of the host galaxy. We note a correlation 
between {\it ``dips''} in the AMR envelope of the host disc stars and of 
the satellite. We can explain such correlation as the consequence of 
repeated flybys that the satellite has experienced during its approach.

In \citet[][]{2012A&A...540A..56P}, we studied the time evolution of one 
of the {\tt RaDES} galaxy's (Apollo) `instantaneous' or gas-phase 
metallicity gradient, finding a shallower profile nowadays than at early 
simulation stages, i.e. the metallicity gradient of the galaxy is 
flattening with time. According to that work, we would expect the mean 
metallicity of each region to be similar for young stars and more 
different for older stars. However, Fig.~\ref{AMR} shows the opposite 
behaviour, i.e. the difference in the mean metallicity at a given age 
between two adjacent regions is increasing as we are moving towards 
younger ages. However, this neglects the effect of stellar migration 
which may flatten the metallicity profile. Radial motion causes this 
effect as stars from the metal-rich center of the galaxy move to the 
outskirts. This pushes up the mean metallicity for intermediate-aged 
stars found at larger radii and inverts the expected trend of the AMR 
because the youngest stars are born from gas with a metallicity that is 
lower than the center had in the past.

A striking segregation can be seen in the metallicity of stars depending upon
how they have migrated. Stars that have moved 
outwards are systematically more metal-rich than particles that have 
stayed in a region or moved inwards (see Fig.~\ref{AMR_radius}). This is 
a direct consequence of the negative metallicity gradients and stellar 
radial motions and consistent with \citet[][]{2013A&A...558A...9M}. In 
fact, if we infer the metallicity gradients from such plots 
($\sim$-0.071~dex/kpc) we obtain similar results to those presented in 
\citet[][]{2012A&A...547A..63F} (Table~2, $\sim$-0.061~dex/kpc).

Satellite accretion not only produces a mixture in the spatial 
distribution of stars and an exchange of stars moving inwards/outwards, 
it also leaves a signature in the dynamics of the stars. In 
Fig.~\ref{AMR_jz_jcirc} we show the AMR split in four sub-panels, as in 
Fig.~\ref{AMR_radius}, but now colour-coding each stellar particle 
according to its circularity ($J_{z}$/$J_\mathrm{circ}$). Despite the narrow 
range of $J_{z}$/$J_\mathrm{circ}$ we are considering in our discs (see Sect. 
\ref{Jcut}) we find another interesting trend.  Specifically, in every 
analysed region there is little or no correlation between the stellar 
metallicity and $J_{z}/J_\mathrm{circ}$ prior to $t_\mathrm{jump}$. This might suggest 
that during the epoch of satellite accretion (phases i and ii in 
Sect.~\ref{treeplots}), the disc was hot and well mixed, and those stars 
still preserve that state (as already shown and quantified in 
Sect.~\ref{AMR_sect}). If we consider the subsequent quiescent 
phase after satellite accretion, the stars born after $t_\mathrm{jump}$ display 
a new segregation with the more metal-rich (which tend to be migrating 
outwards) stars having the least circular orbits and vice versa.

We should emphasise that Fig.~\ref{tree_plot} shows (by means of 
$t_\mathrm{jump}$, dashed cyan vertical lines) the moment when the last 
satellite merges with the disc of the host galaxy. After this time the 
satellites do not dynamically affect the host disc any longer. We can 
conclude from the above outlined analysis that satellites affect the 
disc properties in a continuous way until the last one totally merges 
with the host galaxy (not when it enters the virial radius), after that, 
our discs evolve mostly quiescently.

\section{Discussion and conclusions}
\label{discussion}

We have analysed the stellar content of the {\tt RaDES} 
galaxies \citep[][]{2012A&A...547A..63F} in order to study the effect of 
satellite accretion in shaping the chemo-dynamical properties of their 
discs. In this work, we concentrate on the characterisation of the 
assembly history of these galaxies and the influence of such assembly on 
stellar disc properties such as the age and metallicity profiles (Sect. 
\ref{ageandz}); the Age-Velocity dispersion Relation (AVR, see Sect. \ref{AVR}); 
the Age-Metallicity Relation (AMR, see Sect. \ref{AMR_sect}); and other 
dynamic and kinematic properties (see Sect. \ref{chedy_imprints}). We 
study the radial evolution of several of those properties by dividing our 
discs into 4 radial regions (see Table \ref{regions}).

The assembly histories of the {\tt RaDES} galaxies are characterised by 
three main epochs: i) A first merger-dominated epoch, when most of the 
mass of the galaxy is formed via accretion of massive satellites 
($M_\mathrm{sat}/M_\mathrm{host}$ ranging from 0.1 to 3). ii) During the second phase 
the host galaxy still suffers several mergers but decreasing in number 
and mass ratio ($M_\mathrm{sat}/M_\mathrm{host}$ ranging from 0.01 to 0.3, at a 
distance smaller than 5~kpc). iii) This second phase is followed by a 
quiescent period (with some sporadic satellite accretion with 
$M_\mathrm{sat}/M_\mathrm{host}$ $\sim$0.001) in which the disc has settled. We 
define $t_\mathrm{jump}$ as the time boundary between phases ii and iii, at 
which time the last satellite merges with the host galaxy. This latest 
accretion event is characterised by its satellite having similar 
characteristics for all the {\tt RaDES} galaxies ($M_\mathrm{sat}/M_\mathrm{host}$ 
$\sim$0.12 $\pm$ 0.09 when it enters the virial radius and 
$M_\mathrm{sat}/M_\mathrm{host}$ $\sim$0.013 $\pm$ 0.006 when the merger happens, 
averaged values, see Table \ref{satellites_tab}). This event marks the 
end of the active merging period and the beginning of quiescent 
evolution.

This assembly history leaves an imprint on the chemical and dynamical 
properties of the stellar discs of our galaxies. The AVR shows a clear 
jump (sudden for some galaxies, smooth for others) at around $t_\mathrm{jump}$ 
(see Fig.~\ref{sigma_age}). This feature is a direct consequence of the 
heating produced by the satellite accretion. Those stars that underwent 
phases i and ii (older than $t_\mathrm{jump}$) are in a hotter state now (high 
velocity dispersion) than young stars that were born in the quiescent 
phase iii. The smoothness of the jump depends on the ``merging time 
scale'' of the latest accretion event and subsequent internal 
evolution.

These differences in the velocity dispersion between old and young stars 
lead to differential chemical mixing in our discs. While different 
radial regions show different AMRs (with very different metallicity 
distributions) for young stars, 
Kolmogorov-Smirnov statistical tests show that the metallicity 
distributions of old stars display similar characteristics at different 
radii (see Figs.~\ref{AMR}, \ref{AMR_oceanus}, \ref{met_distrs}, and 
\ref{met_distrs_oceanus}). Although the radial redistribution of
stars due to satellite accretion and secular activity does take place, we find that the
radial distribution of metals in the gas from which old stars were formed 
was homogeneous during phases i and ii
compared with the phase iii and this is the main cause of the similar MDFs 
displayed by old stars. Thus, radial redistribution of stars does little to 
change the MDFs of these older stars that are born with a flat metallicity gradient.

Most of the analysed galaxies show the above outlined assembly scenario 
leaving the same imprints in the AMR and the MDFs. However, two of them show two 
different accretion events after the disc is settled (Artemis and 
Oceanus, see Figs.~\ref{sigma_age} and \ref{AMR_oceanus}). Similar 
results and conclusions as the ones outlined before for the rest of galaxies 
apply to these two cases. The bottom 
panel of Fig.~\ref{sigma_age} shows how this second event is able to 
heat the stars of the entire disc, regardless of their ages. The fact that the 
jump in the AVR at 7.5~Gyr is smoother than the one at 1.2~Gyr might be 
a consequence of such heating produced by this latest accretion event 
(at 1.2~Gyr). To test this hypothesis, we have analysed the AVR 5.75~Gyr 
ago (dashed lines in the bottom panel of Fig.~\ref{sigma_age}). We can 
observe that the ``jump'' in velocity dispersion at 7.5~Gyr was more 
important in the past than it is nowadays. This confirms the suggestion 
that the merger event is able to heat the entire disc of the host 
galaxy, swamping previous jumps in the AVR, especially in the inner regions. 
Thus, old steps in the AVR 
can be erased by later mergers as heating produced by mergers is more 
efficient on kinematically cold stars than on hot stars. In the 
particular case of Oceanus, the latest merger event blurs the ``jump'' 
observed at 7.5~Gyr. KS tests comparing the radial metallicity 
distribution of three age bins according to these two merger events also show 
that old stars display the higher mixing degree while intermediate-age and 
young stars show a lower one (with the youngest stars displaying the 
lowest mixing, see Fig.~\ref{met_distrs_oceanus}). Although part of the 
observed mixing might be induced by the merging of structures, the greater 
mixing degree displayed by old stars was imprinted at birth (as for the case 
of galaxies with just one merger event). The activity in phases i and ii homogenises 
the chemical characteristics of the gas along the entire galaxy and, as a 
consequence, stars formed from such homogenised gas display similar MDFs 
regardless their location.

\begin{figure*}
\includegraphics[width=0.93\textwidth]{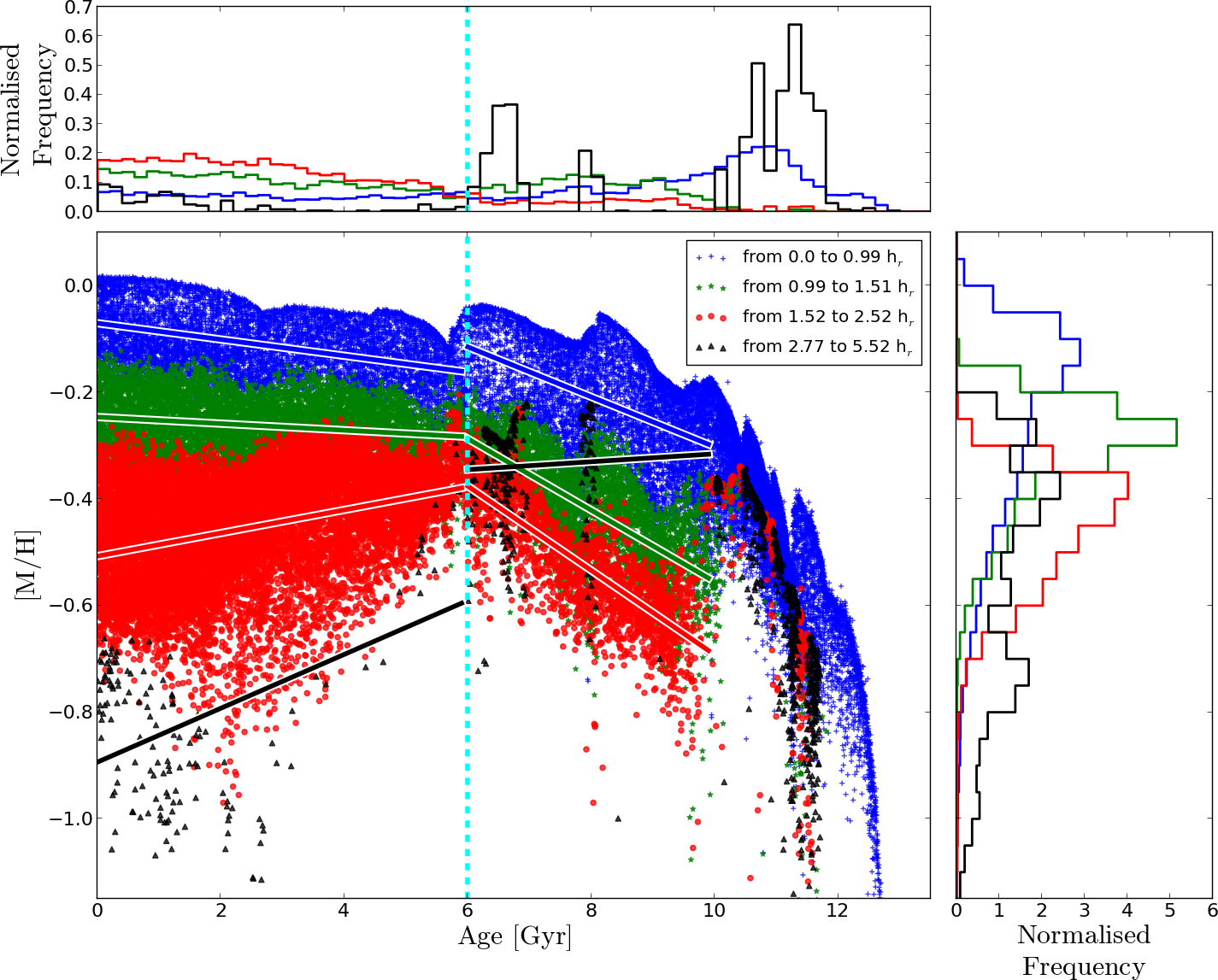} \\
\caption{Age-metallicity relation for Selene disc stars, 
colour-coded according to their birth radius. See Fig.~\ref{AMR} for 
further information.}
\label{AMR_birth_position}%
\end{figure*}

Figures~\ref{AMR}, \ref{AMR_oceanus}, and~\ref{AMR_birth_position} show 
{\it ``dips''} in the maximum [M/H] of the central region (region 1) 
located at $t_\mathrm{jump}$ as well as at other time steps (to a lesser 
degree). Those shallower {\it dips} are related to the galaxy accretion 
history as they correspond to close encounters between the host galaxy 
and minor satellites. The {\it dips} are caused by temporary dilution of 
the gas in the central region by the merger either bringing metal-poor 
gas from the halo of the galaxy or simply adding the metal-poor gas from 
the surroundings of the satellite. In both cases this gas has a lower 
metallicity than the gas in the centre of the galaxy.

We observe two interesting segregations in our AMR plots (see 
Figs.~\ref{AMR_radius} and \ref{AMR_jz_jcirc}). The innermost regions 
(regions 1 and 2) of our galaxies show that metal-rich stars have moved 
outwards, while metal-poorer stars have moved inwards. This behaviour 
can be easily explained by the observed negative metallicity profile (in 
fact, their gradients can be inferred from such plots, see Sect.~
\ref{AMR_sect_colour}). The outermost parts of the galaxies (regions 3 
and 4) show such behaviour for the young stars (younger than 
$t_\mathrm{jump}$), while the old stars populating those regions have migrated 
from the inner parts or come from satellites, i.e. formed outside of the 
galaxy itself.

The segregation in circularity ($J_{z}/J_\mathrm{circ}$) deserves a further 
explanation. Fig.~\ref{AMR_jz_jcirc} shows that old stars display well 
mixed values of their circularity regardless of their metallicity 
(a consequence of the heating mechanisms that they have undergone). 
However, in the case of the young stars (younger than $t_\mathrm{jump}$), the 
more metal-rich the star is, the lower the circularity is. We can 
explain this by the expectation that stars with higher $J_{z}/J_\mathrm{circ}$ 
and lower metallicities are born outside the studied annulus (as shown 
by Fig.~\ref{AMR_radius}). There are two ways in which these stars can 
appear in an annulus inward of their birth places: i) Their angular 
momenta decreased (i.e., inwards migration) to the value of the stars 
born locally, in which case $J_{z}/J_\mathrm{circ}$ will be close to 1, or ii) 
they have reached the final annulus on their pericenters, but with 
angular momenta still larger than the one appropriate for the annulus 
under consideration. It is the second case that causes the green outline 
at the lower metallicity range. Those stars have guiding radii outside 
the studied region. Similarly, lower angular momentum disc stars 
(dark-blue points) are accumulated at the upper metallicity edge.

In Sect. \ref{AMR_sect}, we describe and quantify the main 
characteristics of the AMR in different radial bins for Selene. Apart 
from the differences in the chemical mixing between old and young stars, 
one of the most interesting aspects outlined is the progressive 
change in slope from regions 1 to 4. A negative slope in the AMR is 
easily explained by the fact that young stars (in a simple scenario) are 
born from enriched gas, and thus, are expected to be more 
metal-rich than old stars populating the same region. The presence of 
radial motions (as those described in Fig.~\ref{AMR_radius}) combined 
with a negative metallicity profile can produce a higher dispersion in 
the AMR and a change in the AMR slope \citep[][]{2008ApJ...684L..79R, 
2009MNRAS.398..591S}. Extreme radial motions can even cause the 
inversion that we observe: old stars migrate from the metal-rich centre 
of the galaxy to the outer parts as well as stars from intermediate 
radii (less metal-rich) also move inwards. However, if we now recreate 
Fig.~\ref{AMR}, this time colour-coding the point according to the 
region in which the stars were born instead of their present-day 
position, we still have some inversion (see 
Fig.~\ref{AMR_birth_position}). A careful analysis of the time evolution 
of the AMR at birth location for region 4 can give us the explanation. 
Satellites orbiting around the host galaxy are forming stars in their 
inner parts with metallicities similar to those displayed by the centre 
of the host galaxy. Before merging with the host galaxy, the satellites 
pass through region 4 several times (the clusters of black triangles at 
$\sim$6.5 and 8.0~Gyr in the case of the satellite responsible for the 
$t_\mathrm{jump}$) forming stars within region 4, particles that after the 
merger took place acquire a circular orbit (fulfilling our disc 
criterion). These particles are metal-rich stars. In addition, we find a 
low number of particles that are born in region 4 from metal-poor gas 
accreted from the halo or other satellite haloes with ages between 3.0 
and 6.0~Gyr. As a consequence of the metal-poor gas accretion and the 
low star formation activity, gas in region 4 becomes progressively more 
metal-poor. Finally, we find a group of newly-born stars (younger than 2.0~Gyr) 
due to inside-out growth of the disc reaching region 4 (see red 
line in region 4, Fig.~\ref{SFH}). Once true disc star formation is 
established the normal negative gradient to the AMR reappears (the last 
2.0~Gyr). This scenario naturally leads to an inverted AMR in region 4 
even when radial motions are not considered. The same reasoning can be 
applied to region 3 (outer parts in general), with the difference that 
inside-out growth reaches this region much sooner (6 Gyr ago, see 
Fig.~\ref{SFH}).

Positive slopes in the AMR of the outer parts of nearby galaxies have 
already been observed. \citet[][]{2012MNRAS.420.2625B} studied the SFH 
and the AMR of two outer fields in M31 and another field in M33 using 
the data from \citet[][]{2011MNRAS.410..504B}, both based on HST/ACS 
Colour-Magnitude Diagrams (CMDs). The SFH and AMR found in both 
galaxies show enhanced SF at $\sim$~2\,Gyr and an inverted AMR for 
stars younger than $\sim$~3\,Gyr, similar to the one in the outer parts 
of Oceanus in this work. They suggest that this SF enhancement and the 
inverted AMR is caused by a close encounter between M31 and M33 2-3\,Gyr 
ago according to the N-body simulations by 
\citet[][]{2009Natur.461...66M}. In addition, 
\citet[][]{2012MNRAS.420.2625B} find that the analysed outer fields in 
M31 and M33 are older than the inner parts, suggesting an U-shaped age 
profile, in contrast to the purely negative profile found by 
\citet[][]{2009ApJ...695L..15W} in M33. To compare with our results we 
must keep in mind that, while they are observing the consequences of a 
close encounter between two massive spiral galaxies, we are considering 
the entire assembly history of the {\tt RaDES} galaxies. Our work, along 
with these observational and theoretical works confirm that satellite 
merging or galaxy encounters leave a signature in observable quantities 
such as the AMR or the age profile, especially in the outer parts.

In \citet[][]{2015MNRAS.446.2789B}, the team have expanded upon their 
previous analysis in M31 by analysing 14 outer fields located along the 
disc, the Giant Stellar Stream (GSS, caused by satellite accretion), and 
in regions in-between. They are able to confirm the inverted AMR already 
outlined in \citet[][]{2011MNRAS.410..504B} and 
\citet[][]{2012MNRAS.420.2625B} for all the analysed fields. In 
addition, they deduce that most of the mass in the disc regions was 
already formed by z $\sim$1, while stream-like regions are on average 
older, as expected since the GSS origin is linked to satellite 
accretion. Joining the above result with our findings we confirm that 
satellite merging or galaxy encounters leave a signature in observable 
quantities such as the AMR or the age profile, especially in the entire 
outer disc.

In \citet[][]{2008MNRAS.391.1806V} and \citet[][]{2009MNRAS.399..166V}, 
the authors study the effect that 1:5 
mass-ratio (at virial radius) satellite mergers have upon the thick disc 
properties using N-body simulations of a satellite being accreted by a 
pre-existing disc galaxy. They find that $\sim$2~Gyr after the merger 
event occurs the properties of the heated discs have settled and stop 
evolving further (roughly consistent with our findings). They claim that 
the kinematic impact on their host (thin) discs is almost negligible in 
the case of the vertical and azimuthal components while in the case of 
the radial velocity dispersions they usually increase their value by 
5-10~km s$^{-1}$ at all radii. We find increases from $\sim$50\% in the 
inner parts to $\sim$100\% in the velocity dispersion of the outer parts 
for Selene (latest satellite with a mass ratio of $\sim$1:10). This 
suggests strong disc flaring, an important aspect of the formation of 
thick discs in a cosmological context \citep[][]{2015ApJ...804L...9M}. 
However, although the effect we see in our simulations in the heating of 
the discs from satellite accretion is greater than in the case of 
\citet[][]{2008MNRAS.391.1806V} and \citet[][]{2009MNRAS.399..166V}, 
we must take into account that the jump in the AVR is a 
consequence of phases i and ii. While in the \citet[][]{2008MNRAS.391.1806V} 
and \citet[][]{2009MNRAS.399..166V} works they 
analyse controlled experiments in which one unique satellite merges with 
a host galaxy, our simulations are embedded in a cosmological context 
and thus, this greater heating is not only caused by the latest merging 
event, but it is a consequence of all the previous mergers.

\subsection{Revisiting the ``U-shape'' age profile}
\label{U_back}

\begin{figure}
\includegraphics[width=0.45\textwidth]{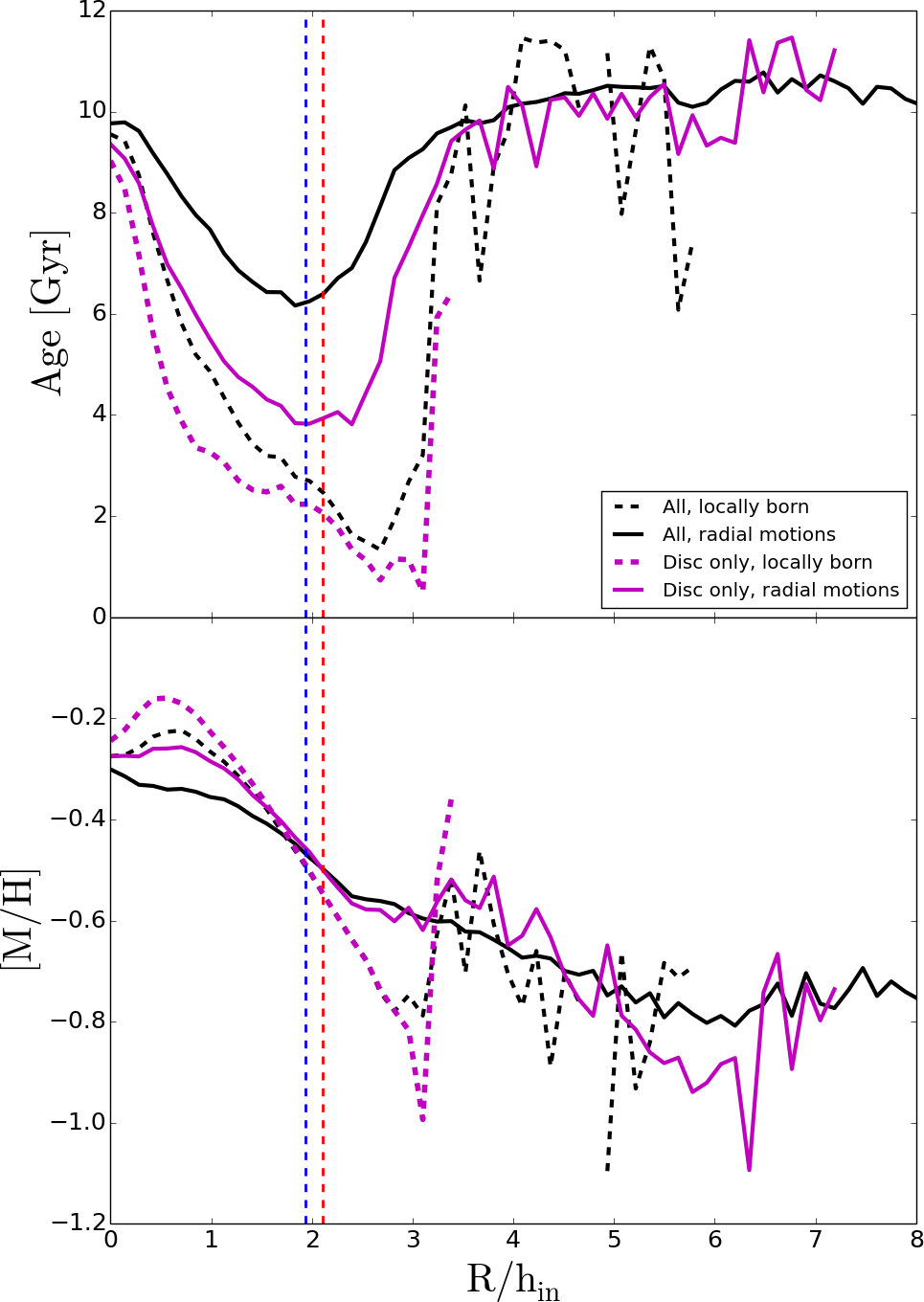} \\
\caption{Selene age and metallicity profiles for all particles and just 
for disc particles either including or removing radial motions. Solid 
lines: we allow stars to move radially to compute the age and 
metallicity profiles. Dashed lines: stars born locally, not allowing 
radial motions. Black lines: all the particles in the simulation 
(disc+spheroid). Magenta lines: particles fulfilling our disc criterion. 
This plot is focused on understanding the effect of radial motions in 
the age and metallicity profiles. Red (blue) vertical dashed lines are 
located at the break (minimum age) radius. $h_{\rm in}$ is the inner disc 
scale-length in SDSS r-band mock images from {\tt SUNRISE} 
\citep[][]{2006MNRAS.372....2J}, these mock images can be seen in 
\citet[][]{2012A&A...547A..63F}.}
\label{mig_effect}%
\end{figure}

\begin{figure}
\includegraphics[width=0.45\textwidth]{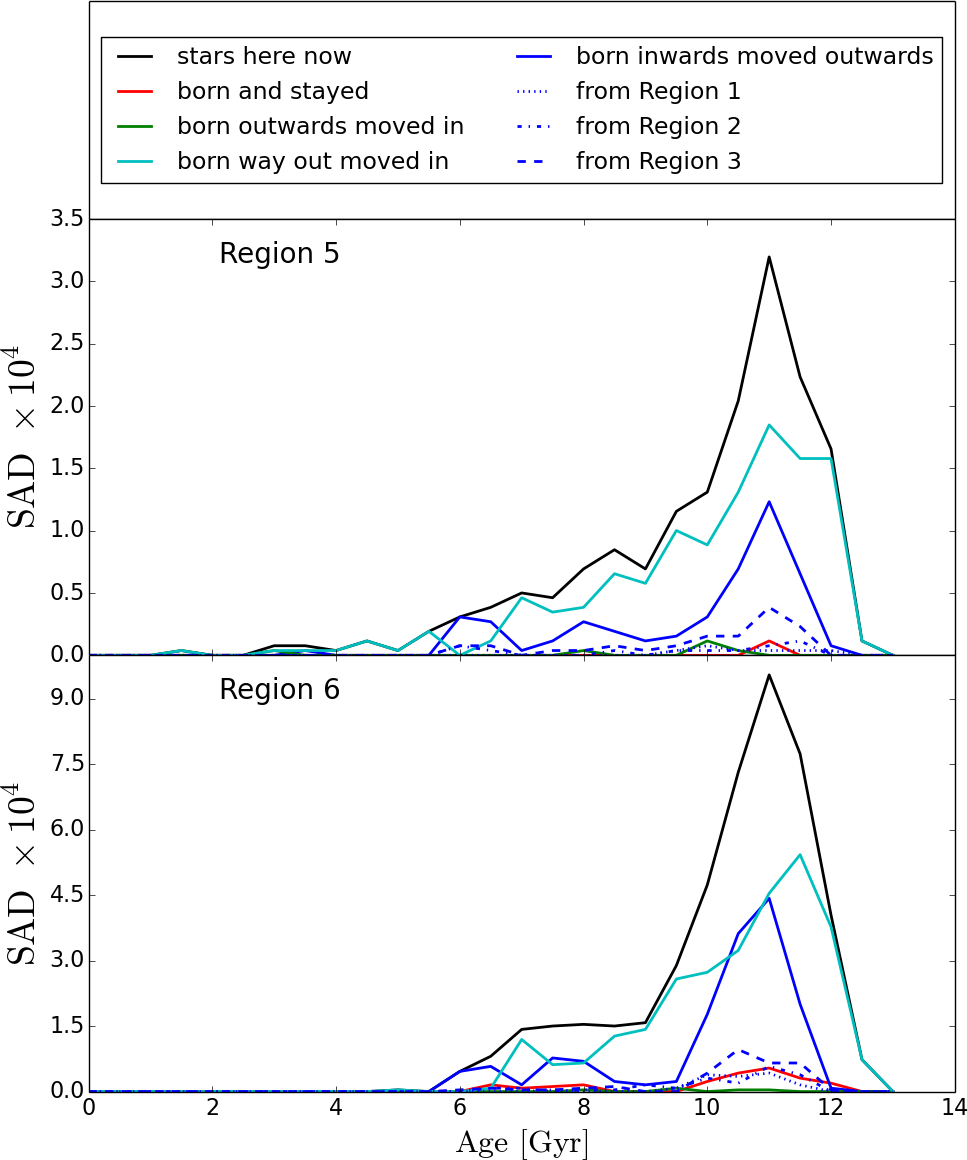} \\
\caption{Radially resolved, stellar age distribution for Selene, 
normalised to the total stellar mass. In this plot we study all the 
particles in the simulation (in Fig.~\ref{SFH} we show only disc 
particles). The upper panel shows the stellar age distribution of the 
stars currently located in region 5 (from 4.4$h_{\rm in}$ to 6.2$h_{\rm in}$) while the 
bottom panel shows stars currently located in region 6 (6.2$h_{\rm in}$ to
7.9$h_{\rm in}$). Lines and colours represent the same as in Fig.~\ref{SFH}. Note 
the presence of the low number of old, {\it in-situ} stars that are 
responsible for shaping the old outer plateau even in the absence of 
radial motions if we do not apply our disc criterion.}
\label{SFH_outer}%
\end{figure}

Every age profile for the {\tt RaDES} galaxies exhibits the same 
behaviour (see Sect. \ref{ageandz} and Fig.~\ref{age_met_prof}): A 
negative profile, followed by an upturn that end up in an old plateau at 
large galactocentric radius. Recent simulations find such U-shape 
profiles in the stellar age distribution of their simulated discs 
\citep[e.g.][]{2008ApJ...675L..65R, 2008ApJ...684L..79R, 
2009MNRAS.398..591S, 2009ApJ...705L.133M}. Different mechanisms have 
been proposed to explain such profiles, but there is still a lack of 
consensus (see Sect. \ref{ageandz}).

The findings presented in this paper point to three main aspects 
responsible for shaping the disc age profiles:

a) An inside-out growth of the disc: As discussed in Sect. \ref{SFH_rad} 
and shown by Fig.~\ref{SFH} (red lines), our discs display a clear 
inside-out growth of the disc. This mechanism can account for the inner 
part (negative gradient) of the age profile.

b) Radial motions (inwards and outwards) of stars belonging to the disc: 
Fig.~\ref{SFH} (blue lines) also show that the regions where the upturn 
in age appears are dominated by old stars that have move outwards 
(especially region 4) with the contribution of some young, {\it in-situ} 
stars.

c) Accreted, old stars from satellites: Phases i and ii populate our 
discs with old stars coming from the accretion of satellites. Such 
accretion is also responsible for the U-shape in the age profile as part 
of the old stars in regions 3 and 4 come from satellites (cyan lines in 
Fig.~\ref{SFH}). However, these old accreted stars are especially 
responsible for the old plateau observed at large galactocentric 
distances (see Fig.~\ref{age_met_prof}). Regions 5 and 6 are mainly 
populated by these old accreted stars.

However, although radial motions might play an important 
role in creating the U-shaped age profile \citep[][]{2008ApJ...675L..65R, 
2008ApJ...684L..79R}, they are not necessarily needed to build up such a 
profile \citep[][]{2009MNRAS.398..591S, 2015A&A...578A..58H}. In 
Fig.~\ref{mig_effect} we analyse the effect that radial motions and our 
disc-particle criterion (based on $J_{z}/J_\mathrm{circ}$) imprint in the shape 
of the age and metallicity profiles.

Radial motions have their greatest effect on the age minimum, i.e. a 
younger minimum in the age profile is found when suppressing radial 
motions. This is because the minimum age is in a prime location to be 
migrated to (it can receive stars from larger and smaller radii easily) 
and because migrating stars tend to be older. However, U-shaped age 
profiles are found whether we allow stars to migrate or not, i.e. radial 
motions affect the shape of the age profile but are not in fact the 
origin of the U-shape. 

To explain why the U-shaped age profile is still recovered after 
suppressing radial motions we need to analyse the stellar content in the 
outer parts of our simulated galaxies (regions~4, 5, and 6, see 
Fig.~\ref{SFH}). Region~4 shows some {\it in-situ} star formation around 
6, 7, and 11~Gyr ago, this star formation can easily explain the upturn 
in the age profile. However, the old plateau disappears if we suppress 
radial motions for the disc stars (while still visible if we analyse all 
the particles, see Fig.~\ref{mig_effect}). This old plateau disappears 
for disc particles because of the lack of {\it in-situ} disc stars at 
large radii. However, if we study the entire sample of stellar particles 
(spheroid and disc component), we can find that an old, non-disc 
population exists in this outermost regions of the {\tt RaDES} galaxies, 
suggesting early star formation in these outer parts (see 
Fig.~\ref{SFH_outer}). 

In recent years, there is a growing body of works 
claiming that the formation of the outer disc of the Milky Way may have begun 
as early as 10~Gyr ago \citep[e.g. see discussions in ][]{2013A&A...560A.109H, 2014ApJ...781L..31S, 
2015A&A...578A..58H}. This suggests that, apart from radial migration, other mechanisms might be
in place while shaping these outer regions. In particular, \citet[][]{2014ApJ...796...38N}, 
analysing {\tt APOGEE} data 
\citep[][]{2010IAUS..265..480M}, find a complex Milky Way formation 
from a well-mixed and turbulent interstellar medium with different SFHs for 
inner and outer disc. In addition, more studies are finding upturns 
in the age profiles in the outer parts of, not only spiral galaxies, but 
also dwarfs \citep[][]{2013ApJ...778..103H}.

Both radial motions and the application of less restrictive 
disc-particle criteria, have a flattening effect on the metallicity 
profiles. The reason for this is straightforward; particles that move 
radially inwards the greatest distances are old, metal-poor stars that 
strongly affect the regions where the average metallicity is higher 
(lowering the average metallicity). In addition, particles moving 
outwards were formed in high-metallicity gas and move towards regions 
dominated by low-metallicity stars, thus, this outwards motion increases 
the mean stellar metallicity in the outer parts. The impact of relaxing 
our disc-particle criterion has the same effect. Relaxing the 
criteria allows more spheroid particles to be considered in the 
analysis, particles that are characterised as being old and metal-poor 
and thus reducing the mean metallicity in the metal-rich regions more 
strongly while making little difference to the metal-poor regions. In 
the case of Selene the metallicity gradient evolves from 
-0.20, -0.19, -0.13, and -0.12~dex/$h_{\rm in}$ from locally-born disc stars, 
all locally-born stars, disc stars with radial motions, and all stars with 
radial motions, respectively.

The assembly history of disc galaxies leaves important fingerprints in 
their chemical and dynamical properties. Based on this assembly history, 
we have been able to explain previously observed features such as the 
dispersion in the AMR, inverted AMR gradients, and U-shaped age profiles. 
Satellites mainly affect the host disc when they completely merge with 
it, although some star formation and gas dilution can be observed during 
the different satellite flybys. These fingerprints should be easily 
observed in present and upcoming spectroscopic surveys such as {\tt 
Gaia} \citep[][]{2001A&A...369..339P}, {\tt RAVE} 
\citep[][]{2006AJ....132.1645S, 2008AJ....136..421Z, 
2011AJ....141..187S, 2013AJ....146..134K}, {\tt SEGUE} 
\citep[][]{2009AJ....137.4377Y}, {\tt APOGEE} 
\citep[][]{2010IAUS..265..480M}, or {\tt 4MOST} 
\citep[][]{2012SPIE.8446E..0TD} helping us to understand the physics
shaping the present-day observable properties of the Milky Way.

\begin{acknowledgements}
We thank the anonymous referee for very useful comments.
This research has been partly supported by the Spanish Ministry of 
Science and Innovation (MICINN) under grants AYA2011-24728 and 
AYA2014-53506-P, by the Junta de Andaluc\'ia (FQM-108), 
and the UK's Science \& Technology Facilities Council (ST/J001341/1: 
BKG, ST/F007701/1: CGF). The generous allocation of resources from 
STFC’s DiRAC Facility (COSMOS: Galactic Archaeology), the DEISA 
consortium, co-funded through EU FP6 project RI-031513 and the FP7 
project RI-222919 (through the DEISA Extreme Computing Initiative), and 
the PRACE-2IP Project (FP7 RI-283493), are gratefully acknowledged. TRL 
thanks the support of the Spanish Ministerio de Educaci\'on, Cultura y 
Deporte by means of the FPU fellowship. CGF acknowledges the support of 
the European Research Council for the FP7 ERC starting grant project 
LOCALSTAR. P.S-B acknowledges support from the Ramón y Cajal program 
and to the grant ATA2010-21322-C03-02 from the Spanish Ministry of Economy 
and Competitiveness (MINECO). This research made use of python (\url{http://www.python.org}), 
of Matplotlib \citep[][]{Hunter:2007}, a suite of open-source python modules that 
provides a framework for creating scientific plots, and Astropy, a community-developed 
core Python package for Astronomy \citep[][]{2013A&A...558A..33A}.
\end{acknowledgements}

\bibliographystyle{aa} 
\bibliography{bibliography} 

\appendix

\section{Surface Brightness profiles}
\label{app_SB}
 
\begin{figure*}
\centering
\includegraphics[width=0.45\textwidth]{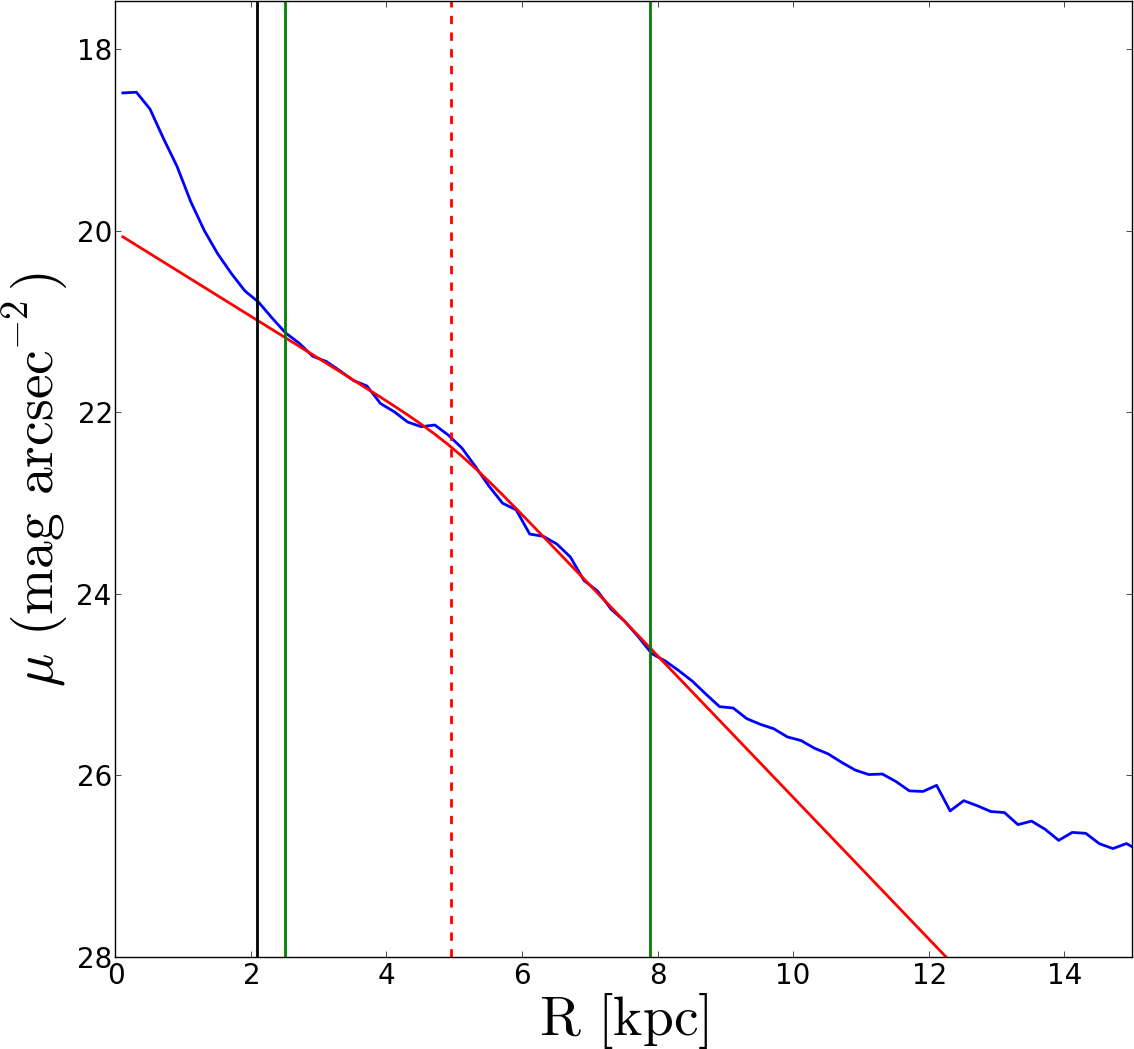} ~ 
\includegraphics[width=0.45\textwidth]{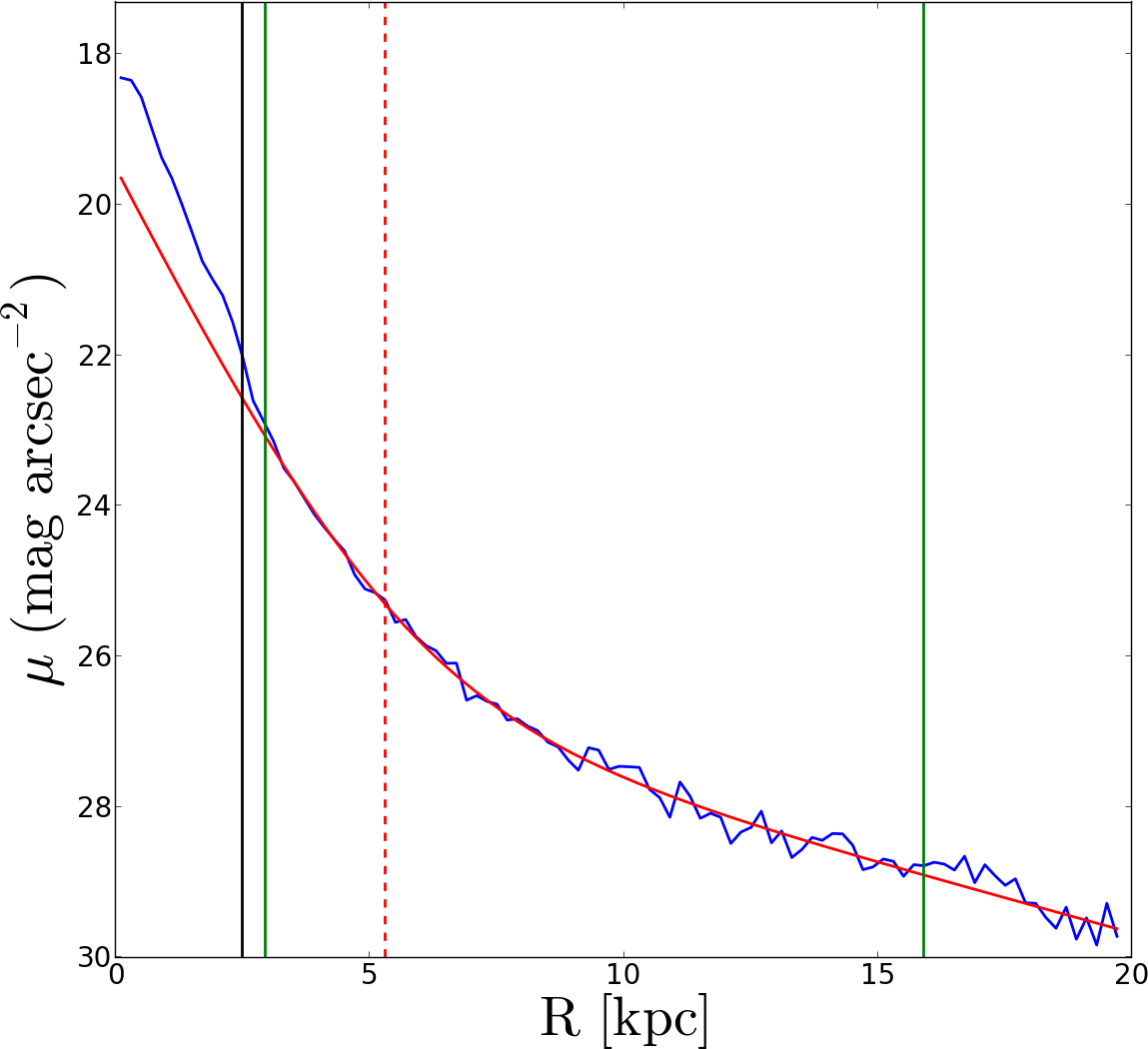} \\
\includegraphics[width=0.45\textwidth]{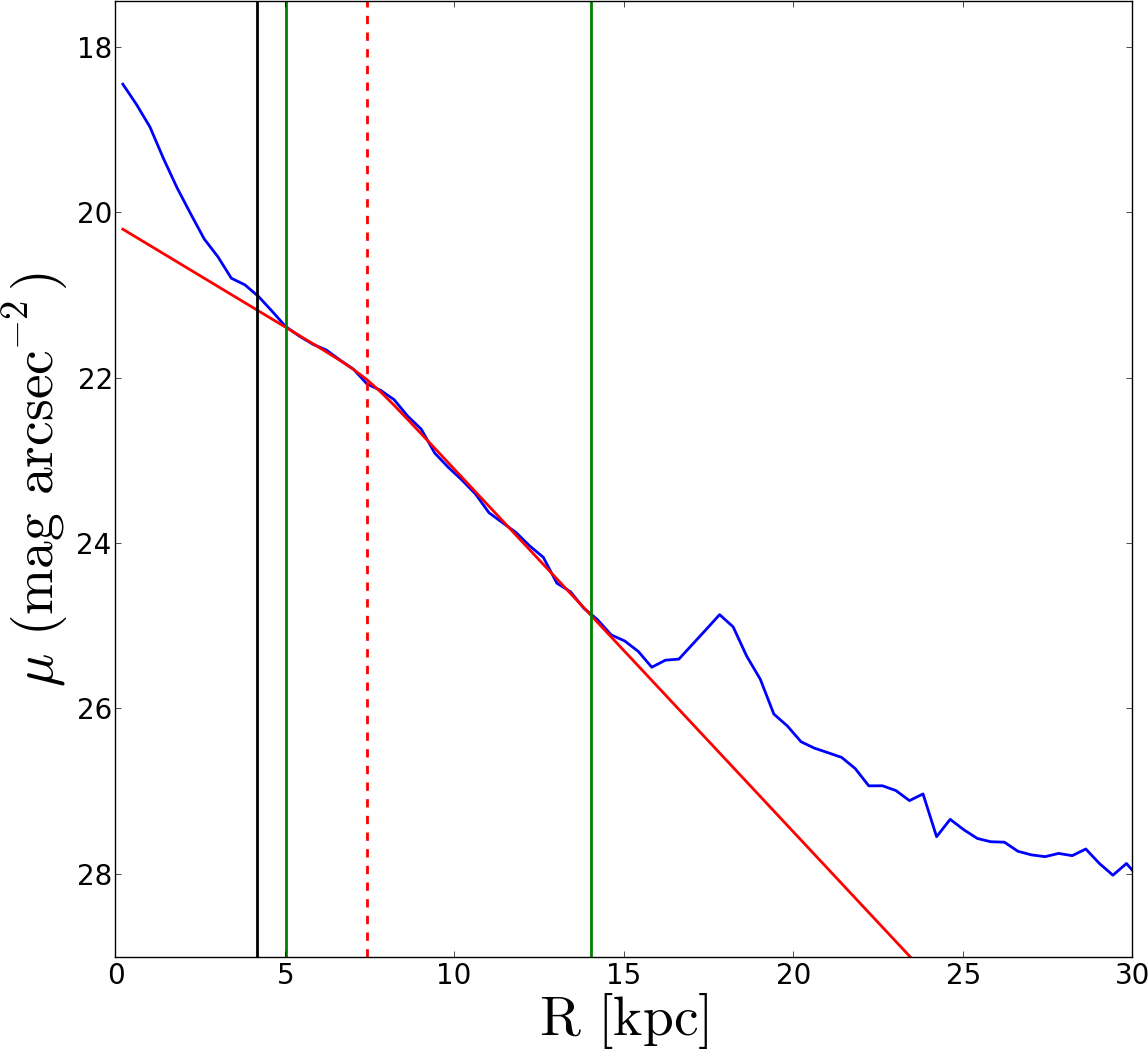} ~
\includegraphics[width=0.45\textwidth]{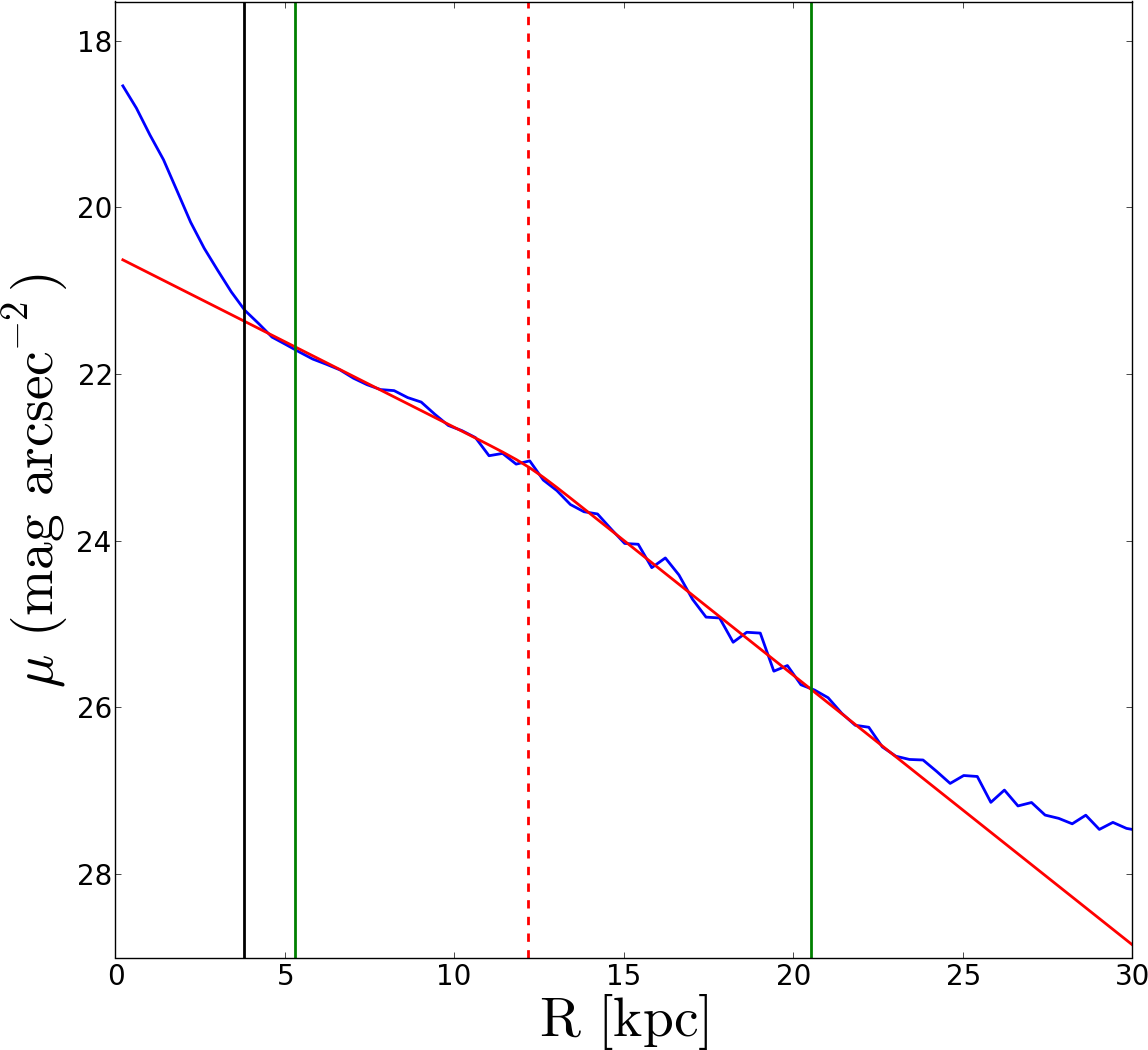} \\
\includegraphics[width=0.45\textwidth]{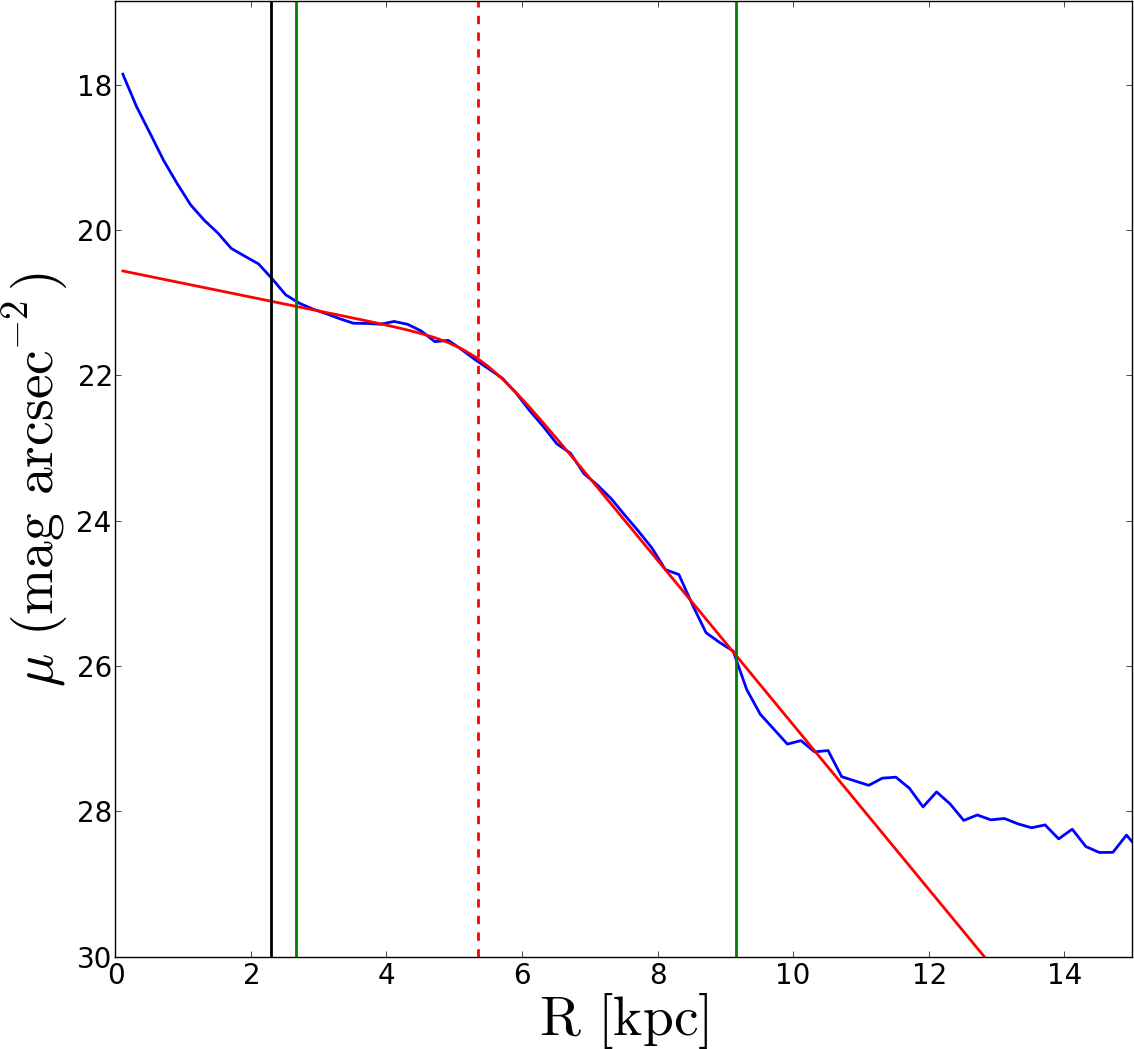} ~
\includegraphics[width=0.45\textwidth]{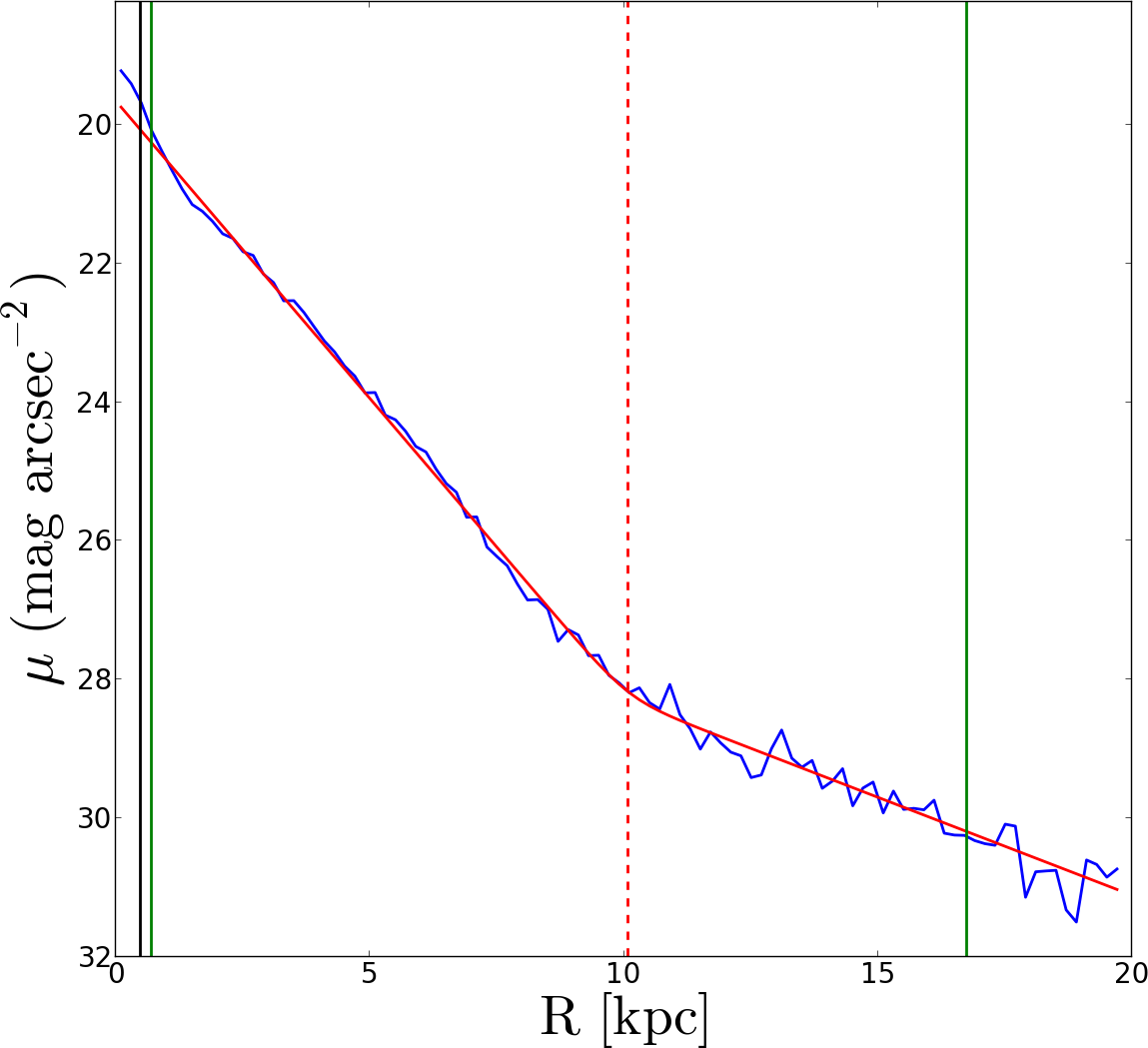} \\
\caption{Surface Brightness profiles in SDSS r band for Apollo (top left), Artemis (top right), Atlas (middle left), Ben (middle right), Castor (bottom left), and Daphne (bottom right). Blue solid line: ``observed'' SB profile using the {\tt SUNRISE} \citep[][]{2006MNRAS.372....2J} mock images, those images can be seen in \citet[][]{2012A&A...547A..63F}. Red solid line: single$/$double disc fit. Green, vertical, solid lines: these lines delimit the extent where the fit has been performed. Black vertical solid line: Boundary between the spheroid component and the disc, it is computed as the point where the distance between red and blue solid lines is less than 0.1 mag arcsec$^{-1}$. Red, vertical, dashed line: location of the break.}
\label{SB_profiles}
\end{figure*}

The study of the light distribution in observed spiral galaxies has led 
recently to a rapid evolution in our thinking regarding its structural 
characteristics. The wide variety of surface brightness (SB) profiles 
reported \citep[e.g.][amongst others]{2006A&A...454..759P, 
2008AJ....135...20E, 2011AJ....142..145G} intimates that such profiles 
may hold a key piece of the puzzle towards understanding galaxy 
formation and evolution; the plethora of profiles now found in the 
literature are a contrast to the canonical wisdom that discs were single 
exponentials \citep[][]{1970ApJ...160..811F}. For completeness, we show 
here the light distribution of the {\tt RaDES} sample, derived from 
one-dimensional SB profiles of the mock images presented in 
\citep[][]{2012A&A...547A..63F}; these images were produced using {\tt 
SUNRISE} \citep[][]{2006MNRAS.372....2J}. {\tt SUNRISE} uses the stellar 
and gaseous distributions, as well as Spectral Energy Distributions 
(SEDs) for each composite stellar particle, drawn from the {\it 
Starburst99} stellar population models \citep[][]{1999ApJS..123....3L}, 
in order to generate the bandpass-dependent mock images. One-dimensional 
profiles were then fitted with the function presented in 
\citep[][equations 5 and 6]{2008AJ....135...20E} with a broken 
exponential profile implemented. This way, we can distinguish between 
galaxies with exponential, up-bending, or down-bending profiles, as well 
as characterise the light distribution of the inner and outer discs. 
Fig.~\ref{SB_profiles} shows those profiles. Table \ref{SB_properties} 
summarises the light distribution information for the 19 simulated 
galaxies.

\begin{table*}
\centering
\begin{tabular}{rrrrrrrrrrr} 
\hline 
Galaxy & SB type & \multicolumn{3}{c}{$h_{\rm in}$ (kpc)} & \multicolumn{3}{c}{$h_\mathrm{out}$ (kpc)} & \multicolumn{3}{c}{$R_\mathrm{break}$ (kpc)} \\
 & & g & r & i & g & r & i & g & r & i \\\hline 
Apollo    &  II &  2.68 & 2.34 & 2.19 & 1.24 & 1.39 & 1.44 &  5.14 &  4.96 &  4.94 \\
Artemis   & III &  0.60 & 0.79 & 0.78 & 5.56 & 5.87 & 5.97 &  4.27 &  5.30 &  5.30 \\
Atlas     &  II &  5.08 & 4.39 & 3.43 & 2.31 & 2.47 & 2.53 &  7.63 &  7.43 &  7.91 \\
Ben       &  II &  5.88 & 5.28 & 5.03 & 3.15 & 3.36 & 3.36 & 12.33 & 12.18 & 12.42 \\
Castor    &  II & 10.06 & 5.70 & 5.32 & 0.95 & 0.96 & 0.96 &  5.37 &  5.35 &  5.32 \\
Daphne    & III &  1.23 & 1.26 & 1.27 & 3.76 & 3.84 & 3.81 & 10.05 & 10.08 & 10.08 \\
Eos       & III &  2.96 & 2.95 & 2.91 & 8.38 & 8.55 & 8.37 & 19.32 & 18.32 & 17.84 \\
Helios    & III &  1.92 & 1.93 & 2.01 & 7.73 & 7.76 & 7.82 & 11.54 & 11.49 & 11.74 \\
Hyperion  &  II &  4.54 & 4.31 & 4.24 & 2.51 & 2.77 & 2.78 & 14.94 & 14.97 & 15.15 \\
Krios     & III &  2.63 & 2.62 & 2.63 & 8.51 & 8.45 & 8.33 & 16.82 & 16.11 & 15.85 \\
Leia      & III &  3.65 & 3.81 & 3.90 & 7.37 & 7.48 & 7.58 & 20.95 & 21.21 & 21.62 \\
Leto      & III &  0.95 & 0.99 & 1.00 & 4.11 & 4.07 & 3.99 & 6.65  &  6.59 &  6.51 \\
Luke      &   I &  5.84 & 5.78 & 5.77 &  -   &  -   &  -   &  -    &  -    &  -    \\
Oceanus   &  II &  8.19 & 8.08 & 7.93 & 5.17 & 4.18 & 4.39 & 21.19 & 23.17 & 23.21 \\
Pollux    & III &  1.20 & 1.25 & 1.22 & 4.33 & 4.03 & 3.93 &  8.80 &  8.61 &  8.41 \\
Selene    &  II &  5.74 & 5.68 & 5.36 & 1.73 & 2.01 & 1.98 & 12.75 & 11.98 & 12.13 \\
Tethys    &  II &  5.23 & 4.43 & 4.20 & 1.97 & 2.12 & 2.16 &  9.92 &  9.82 &  9.85 \\
Tyndareus & III &  1.34 & 1.33 & 1.33 & 3.67 & 3.39 & 3.22 &  7.69 &  7.12 &  6.89 \\
Zeus      &   I &  0.90 & 0.92 & 0.92 &  -   &  -   &  -   &  -    &  -    &   -   \\
\end{tabular}
\caption{Main disc properties from the light distribution in the g, r and i SDSS bands for the {\tt RaDES} galaxies. First column: Galaxy name. Second column: Surface Brightness type according to the \citet[][]{2006A&A...454..759P} classification. Third to fifth columns: Inner disc scalelength in kpc (g, r and i SDSS bands). Sixth to eighth columns: Outer disc scalelength in kpc (g, r and i SDSS bands). Ninth to eleventh columns: Break radius in kpc (g, r and i SDSS bands).}
\label{SB_properties}
\end{table*}

\newpage
\begin{figure*}
\includegraphics[width=0.45\textwidth]{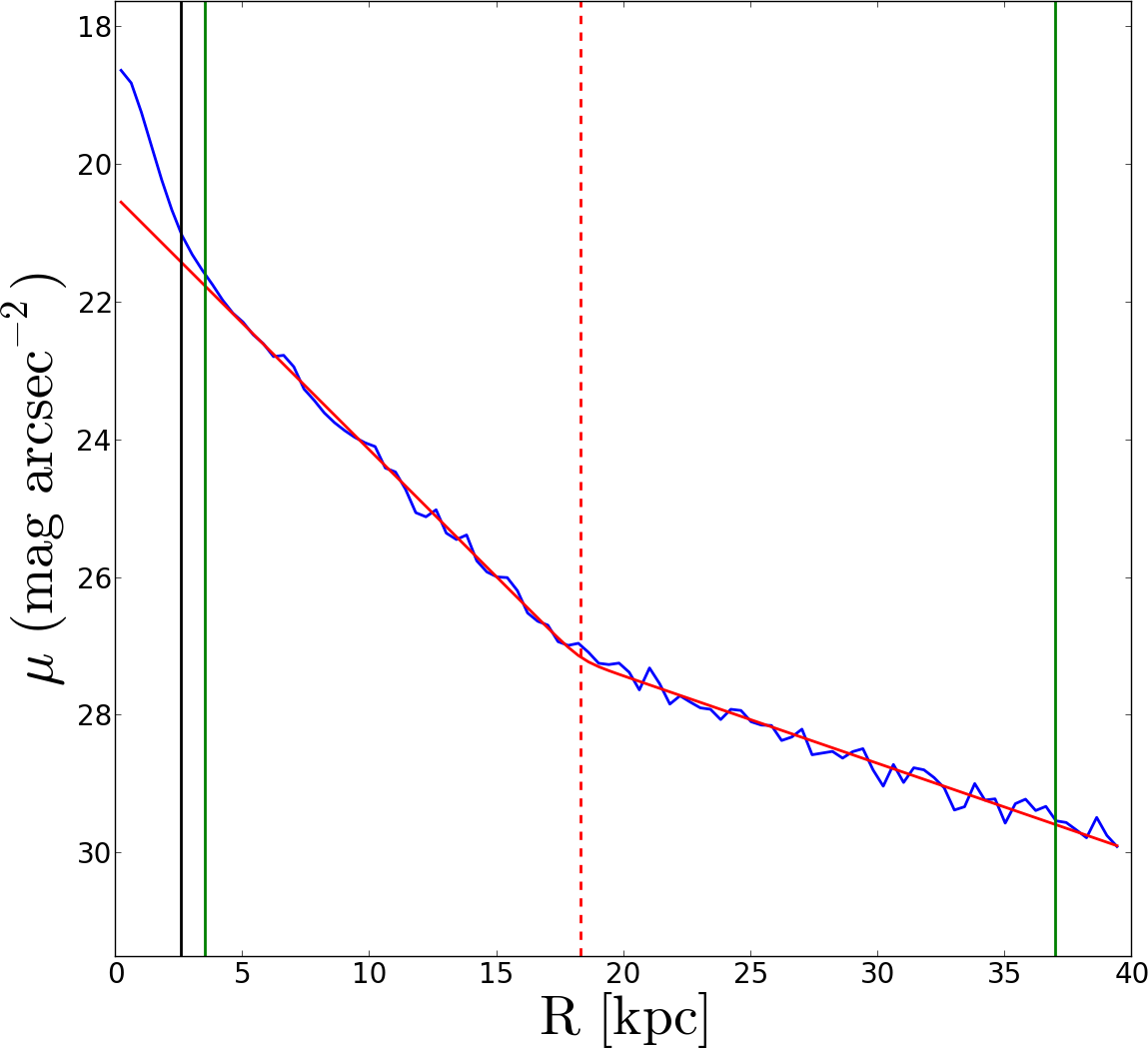} ~
\includegraphics[width=0.45\textwidth]{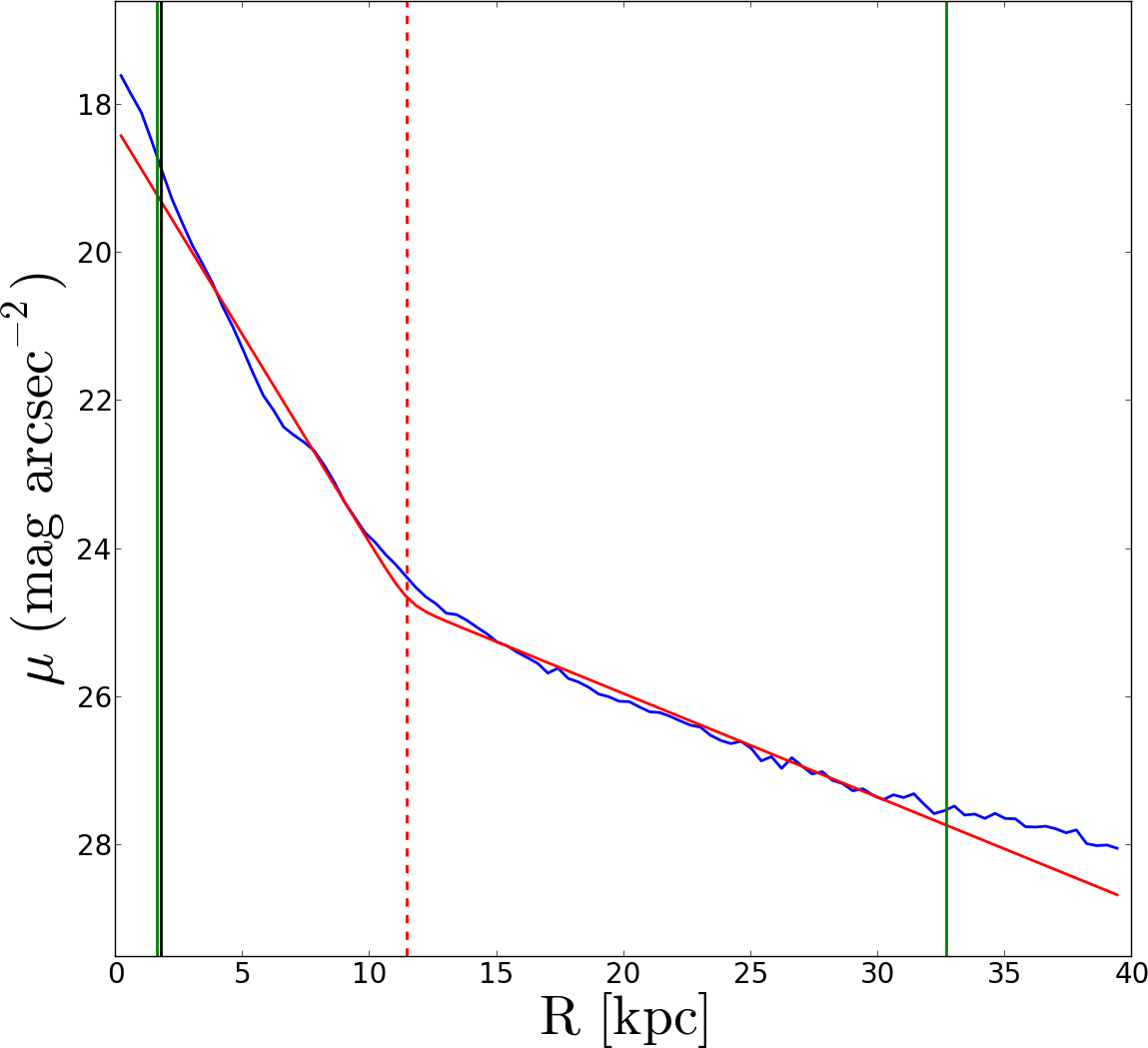} \\
\includegraphics[width=0.45\textwidth]{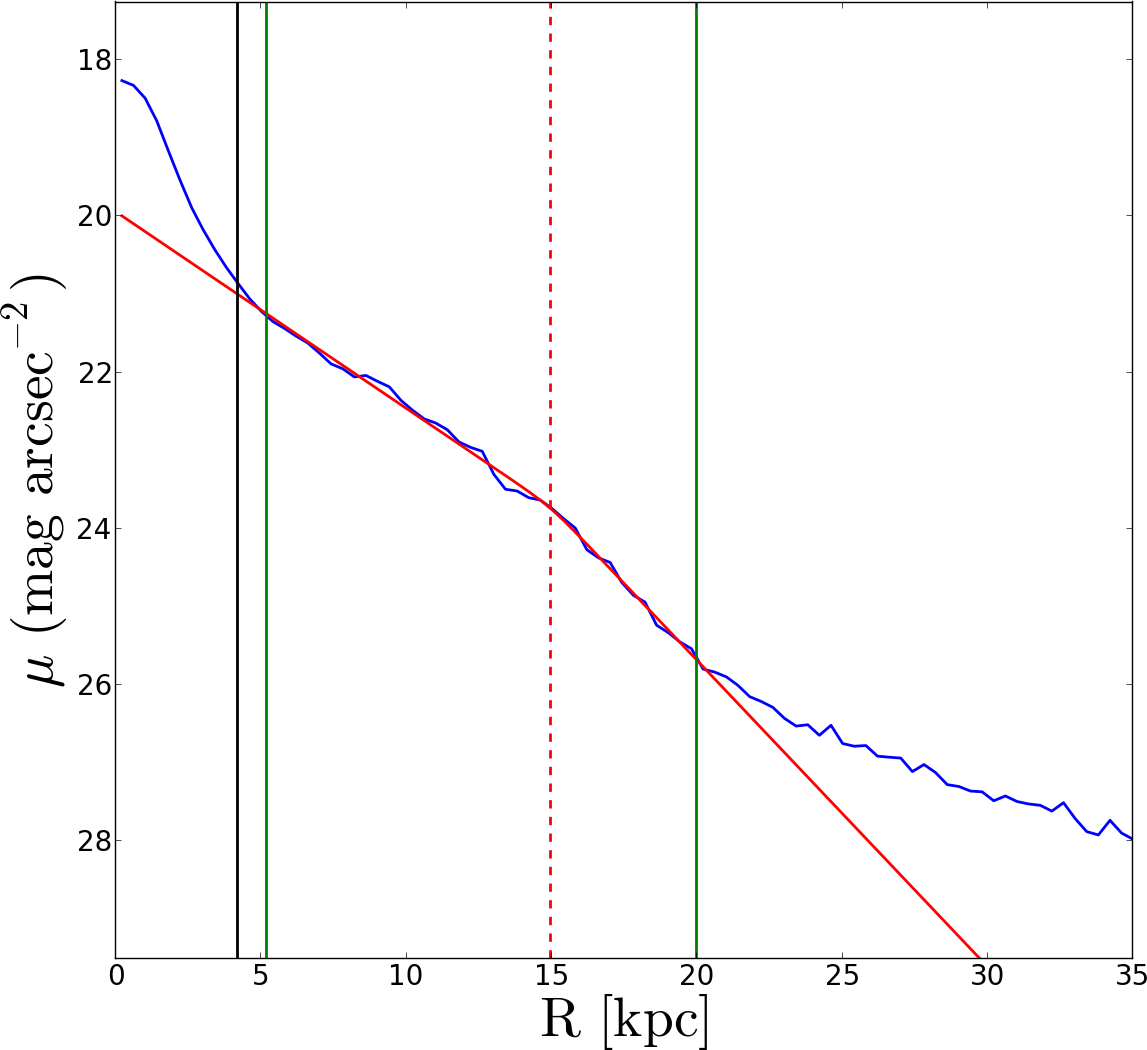} ~ 
\includegraphics[width=0.45\textwidth]{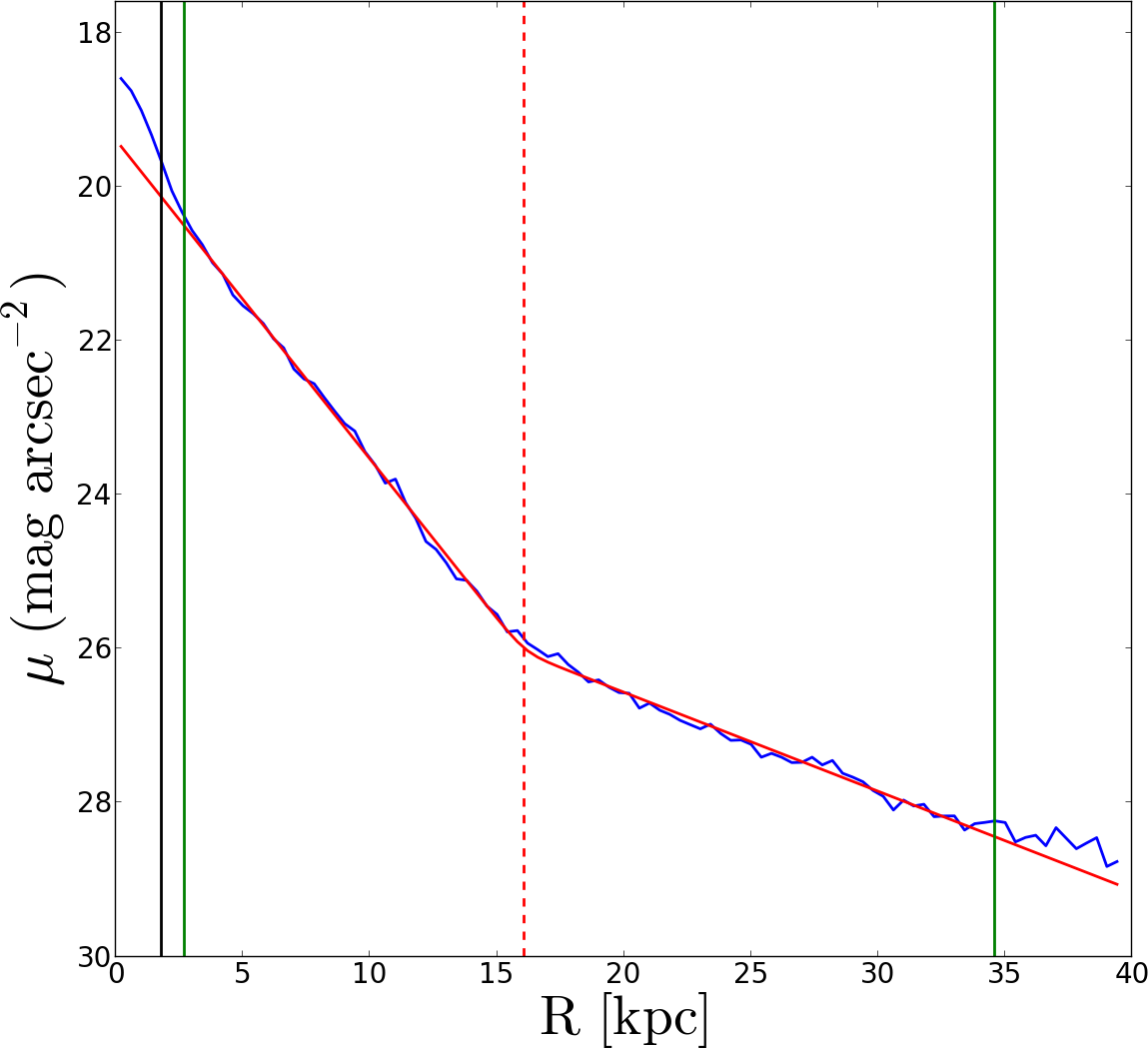} \\
\includegraphics[width=0.45\textwidth]{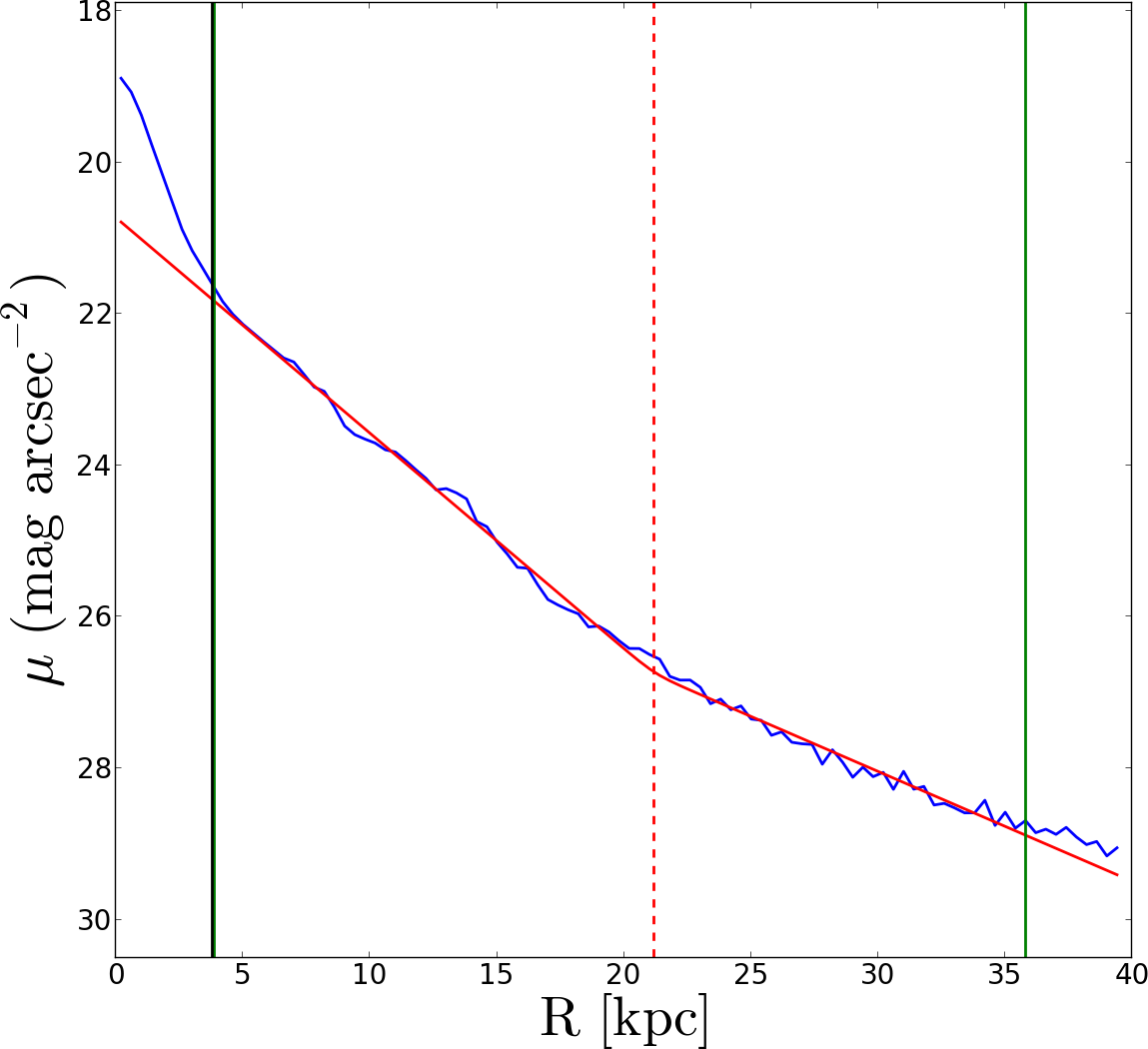} ~ 
\includegraphics[width=0.45\textwidth]{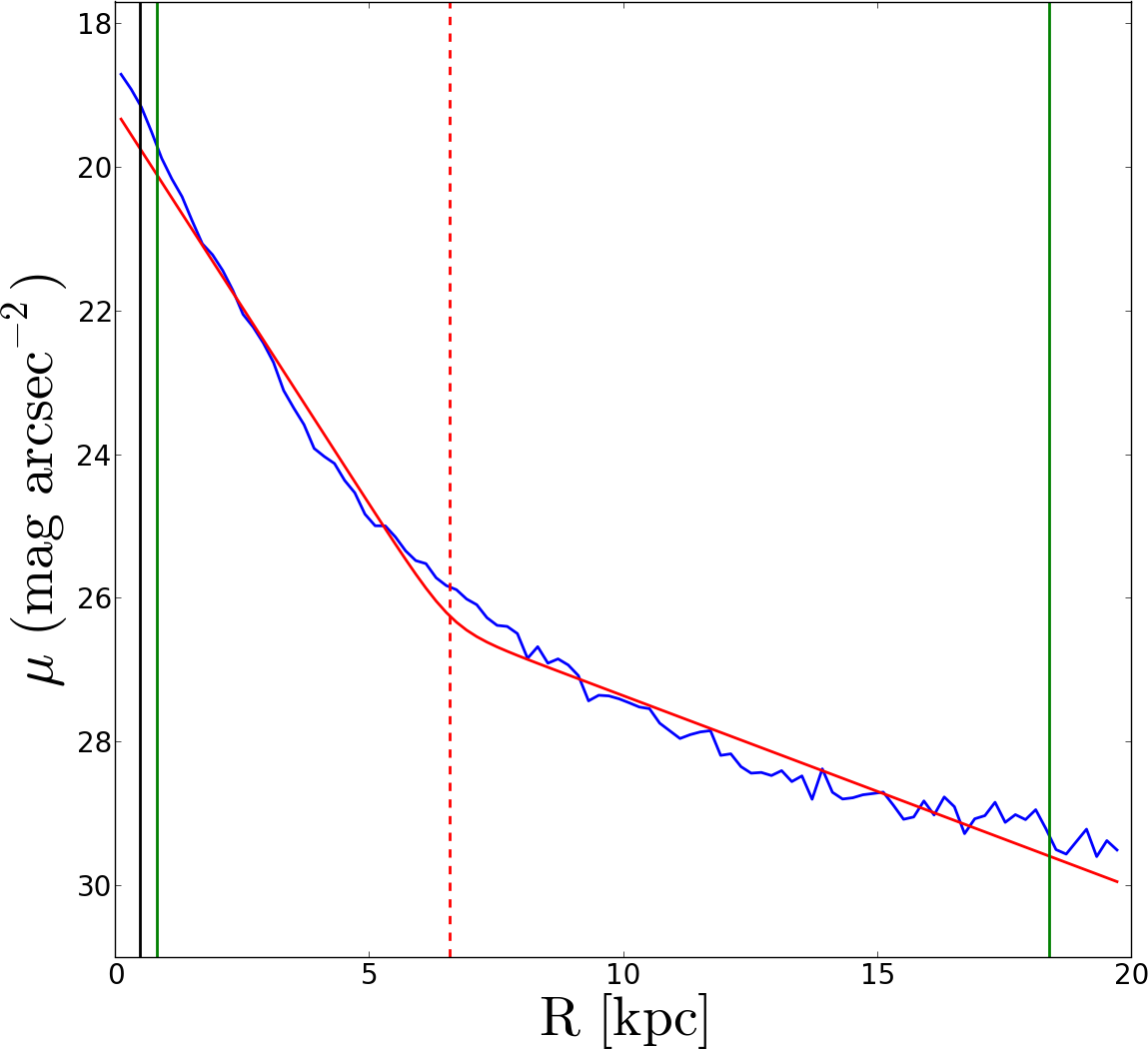} \\
\caption{Same as Fig.~\ref{SB_profiles} but for Eos (top left), Helios (top right), Hyperion (middle left), Krios (middle right), Leia (bottom left), and Leto (bottom right).}
\end{figure*}

\newpage
\begin{figure*}
\includegraphics[width=0.45\textwidth]{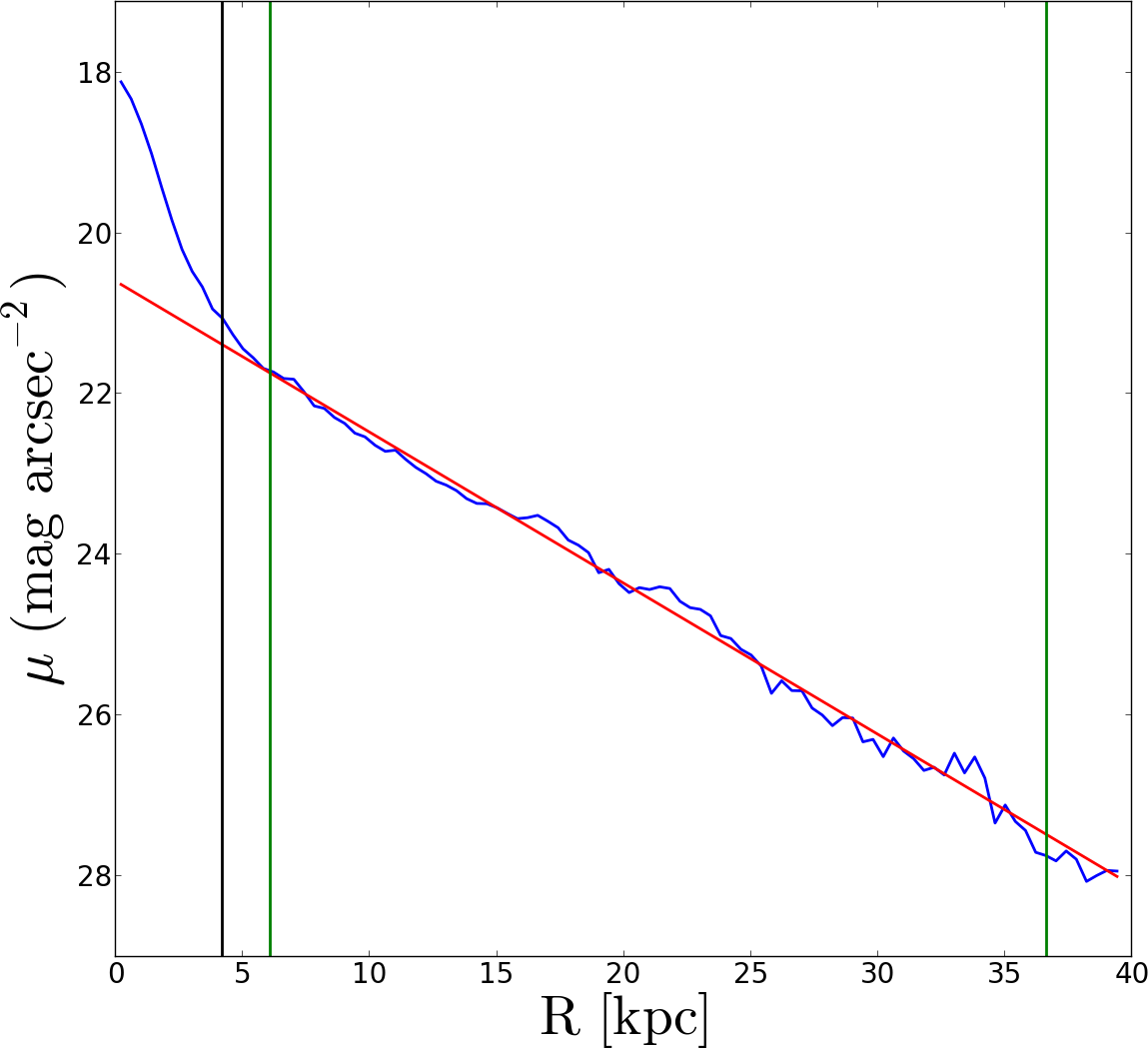} ~ 
\includegraphics[width=0.45\textwidth]{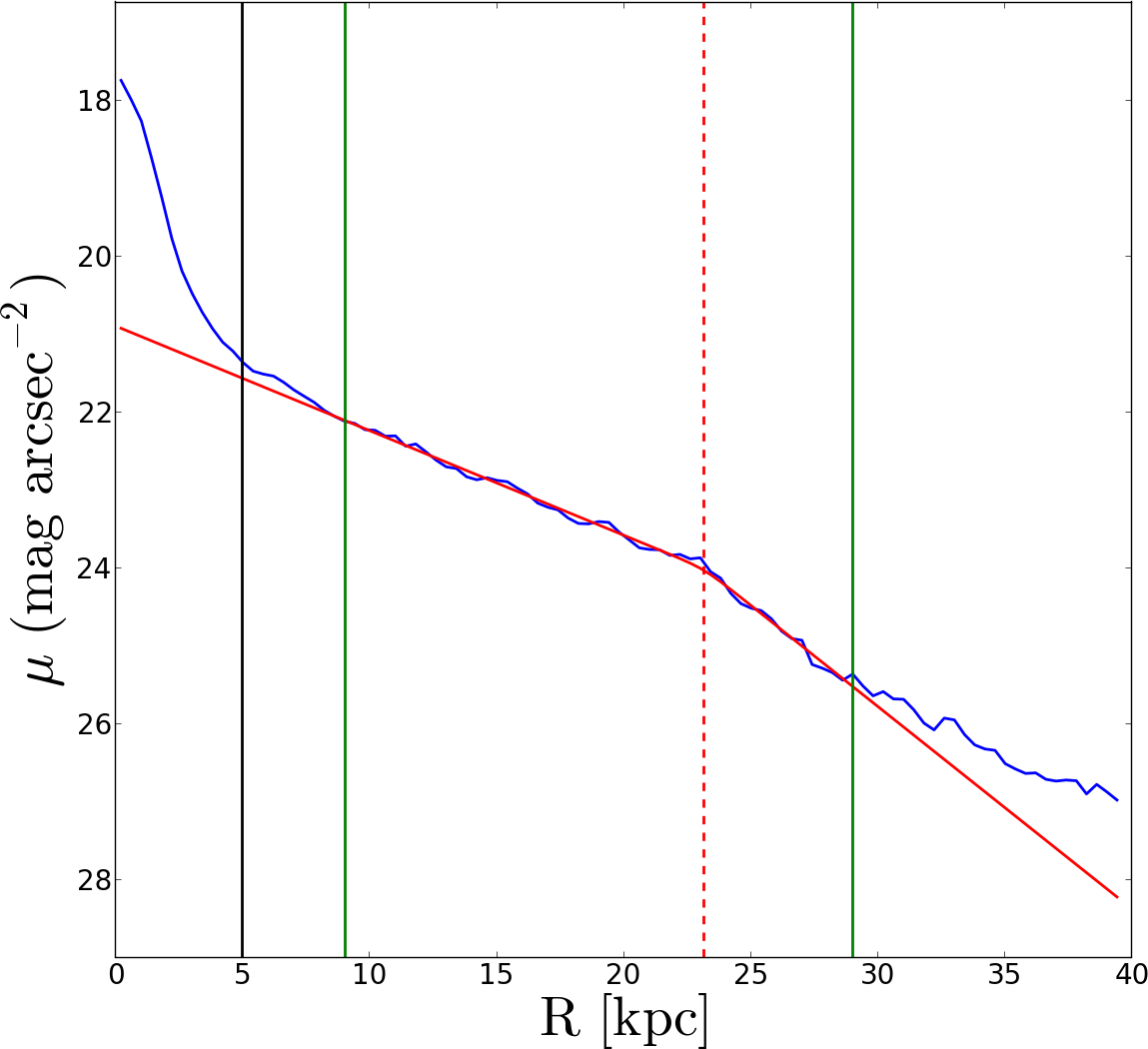} \\
\includegraphics[width=0.45\textwidth]{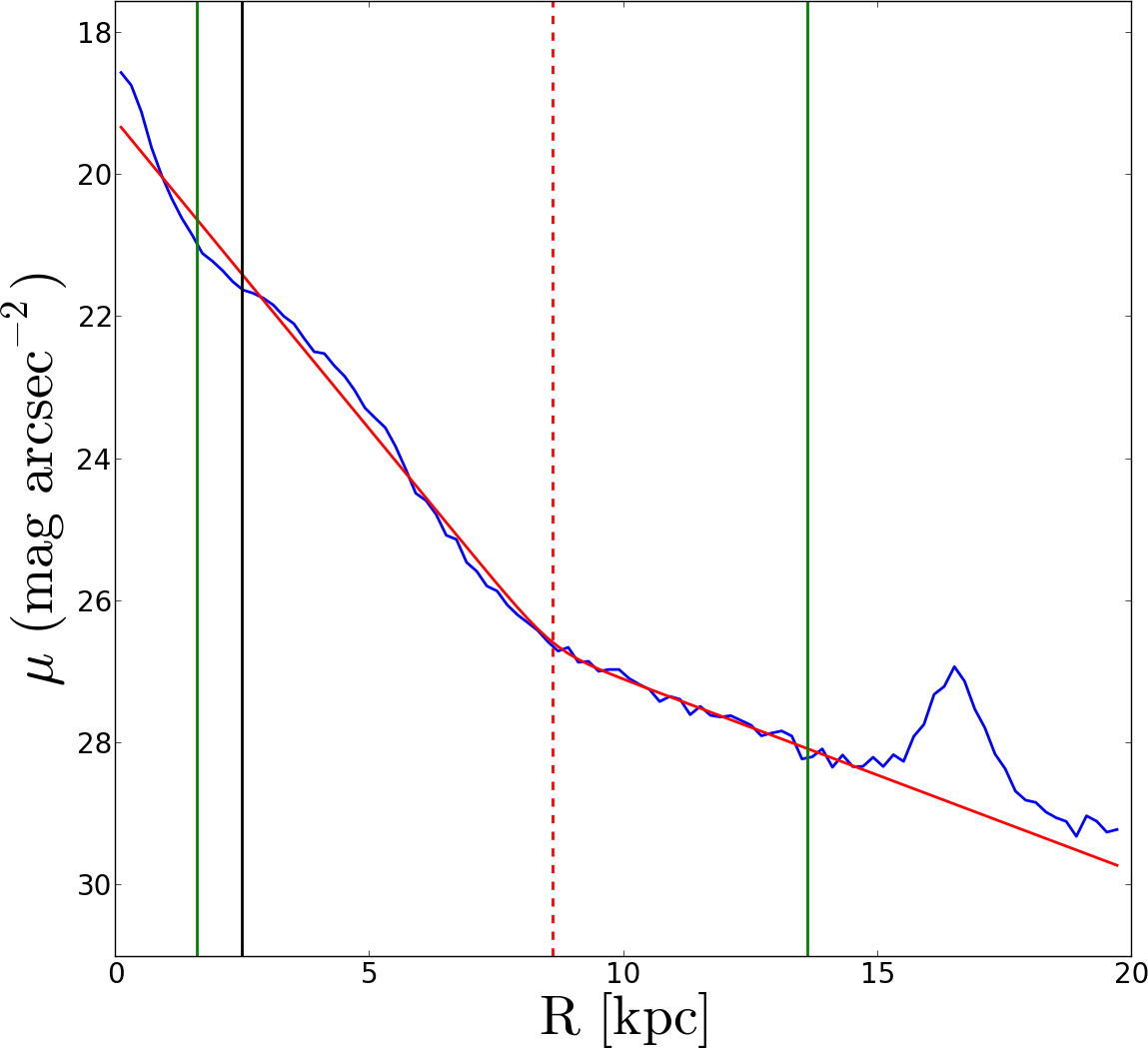} ~ 
\includegraphics[width=0.45\textwidth]{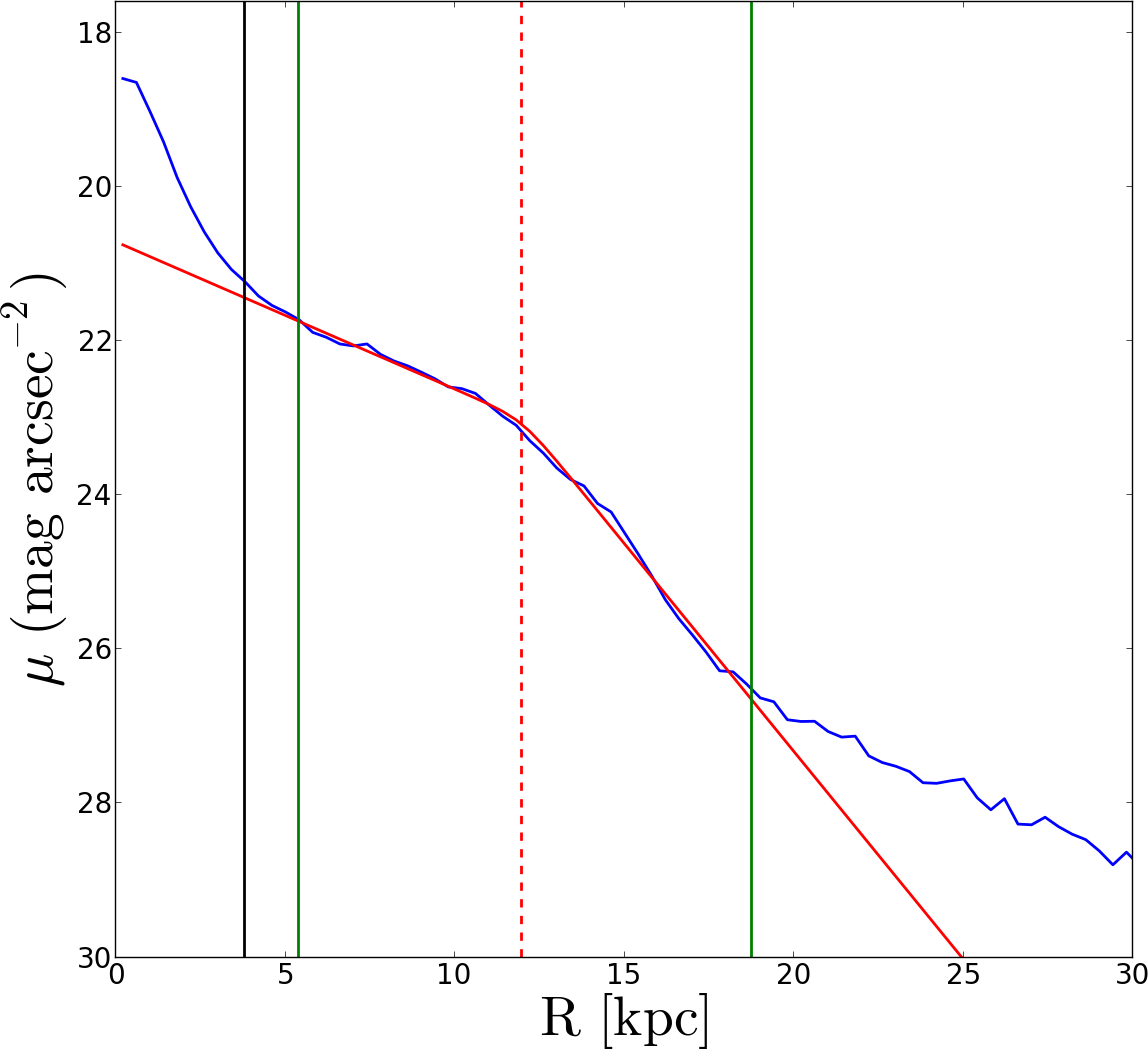} \\
\includegraphics[width=0.45\textwidth]{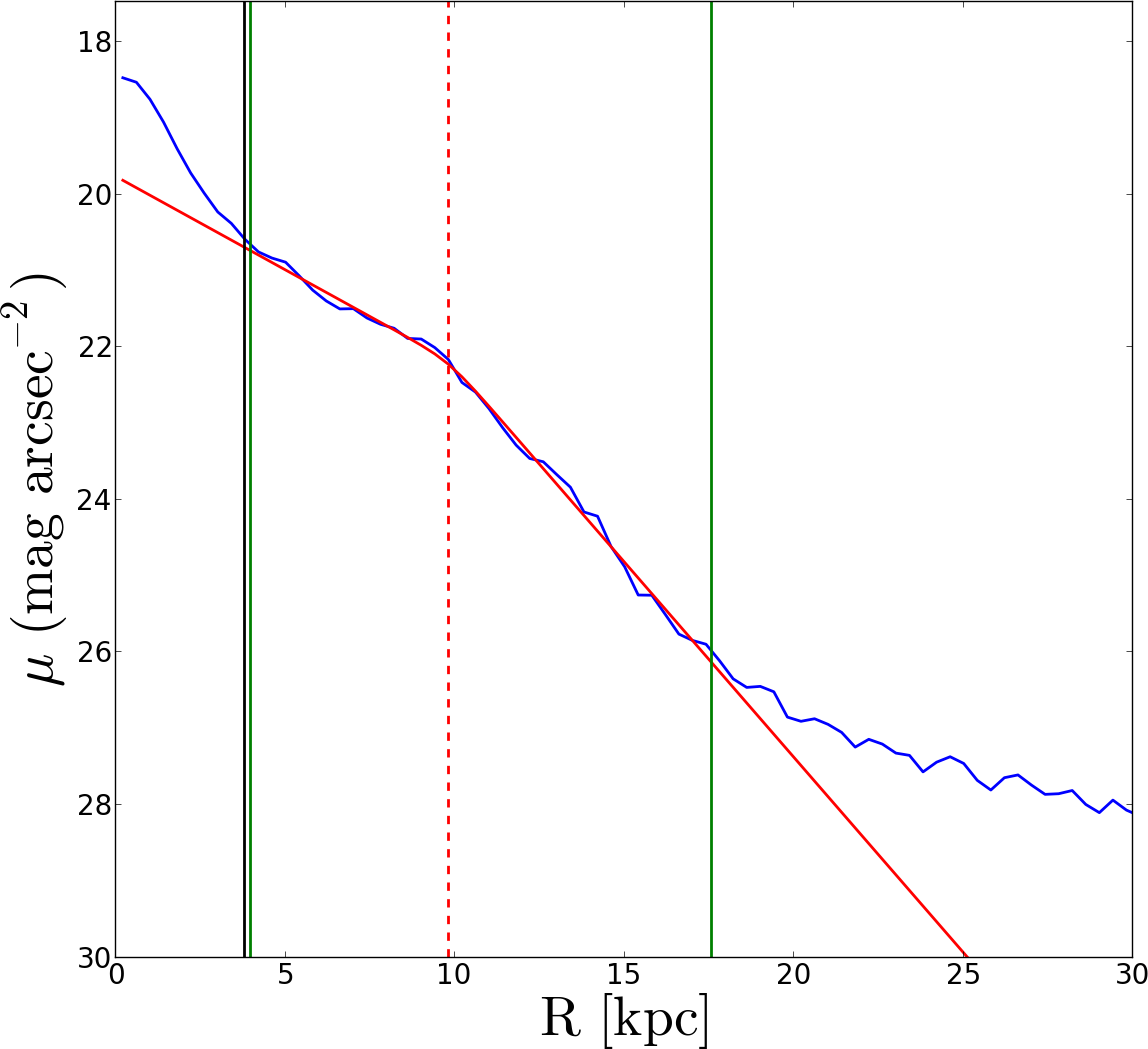} ~ 
\includegraphics[width=0.45\textwidth]{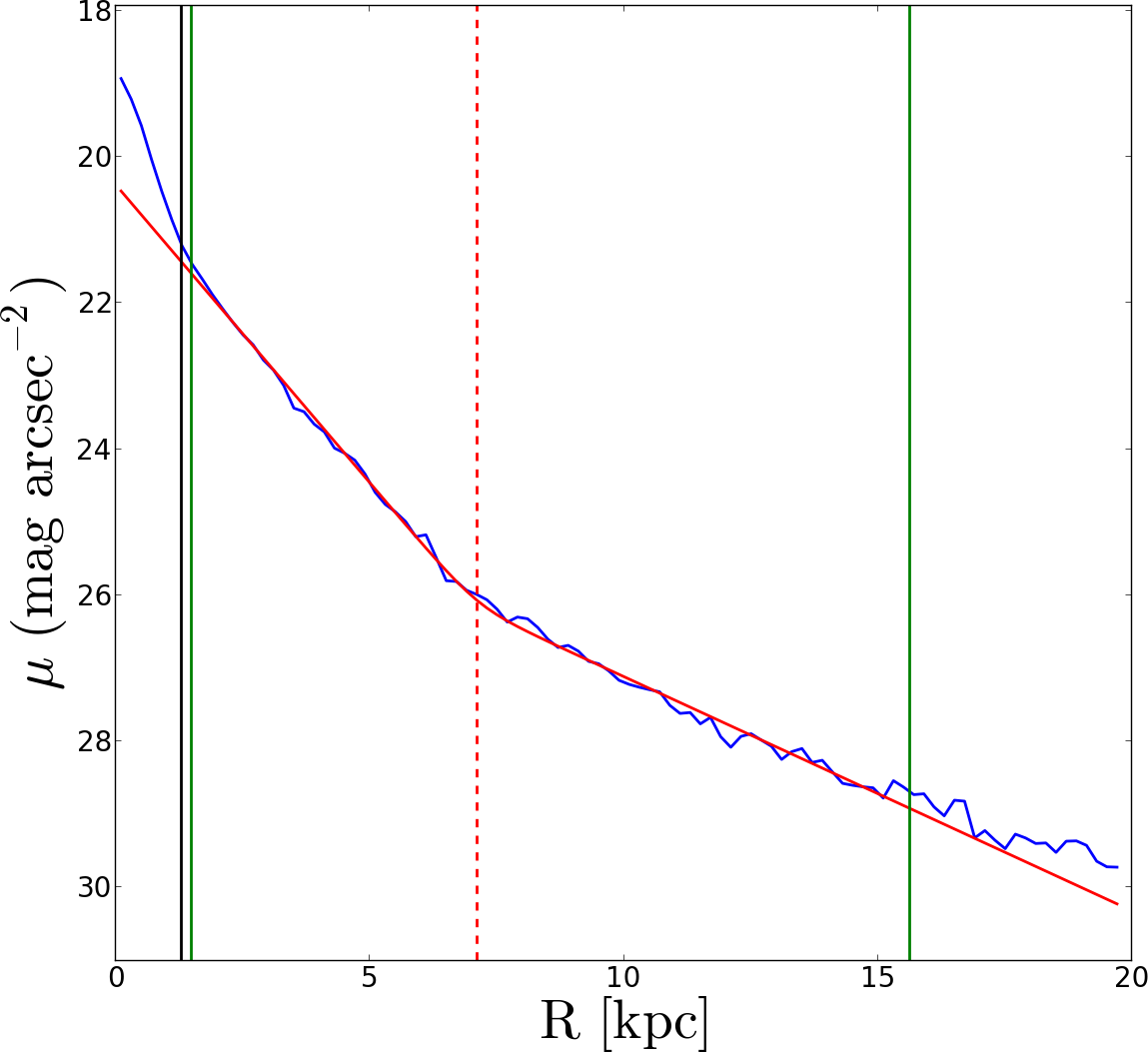} \\
\caption{Same as Fig.~\ref{SB_profiles} but for Luke (top left), Oceanus (top right), Pollux (middle left), Selene (middle right), Tethys (bottom left), and Tyndareus (bottom right).}
\end{figure*}

\newpage
\begin{figure*}
\includegraphics[width=0.45\textwidth]{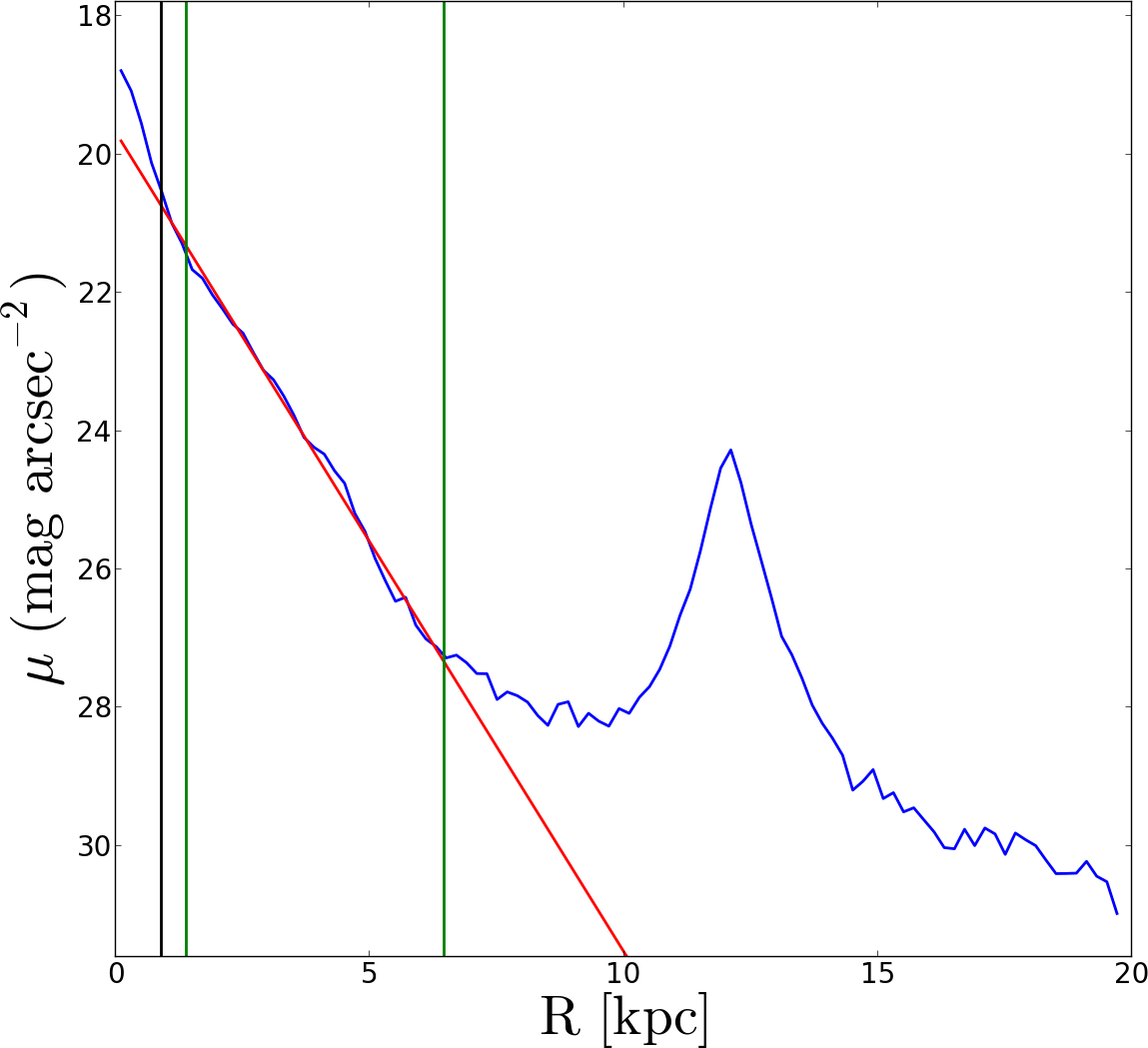} \\
\caption{Same as Fig.~\ref{SB_profiles} but for Zeus.}
\end{figure*}

\clearpage
\newpage

\end{document}